\documentstyle[manuscript,aps]{revtex}
\newcommand{\bff}[1]{{\mbox{\boldmath $#1$}}}
\begin{document}
\setlength{\baselineskip}{0.5cm}

\draft

\renewcommand\baselinestretch{1.32}

\title{Cranked Relativistic Hartree-Bogoliubov Theory:
Formalism and Application to the Superdeformed 
Bands in the $A\sim 190$ region}

\author{A. V. Afanasjev\footnote{Alexander von Humboldt fellow,
on leave of absence from the Laboratory of Radiation Physics, 
Institute of Solid State Physics, University of Latvia, LV 2169 
Salaspils, Miera str.\ 31, Latvia}, P.\ Ring and J.\ K{\"o}nig}

\address{Physik-Department der Technischen Universit{\"a}t
M{\"u}nchen, D-85747 Garching, Germany}

\date{\today}

\maketitle

\begin{abstract}
\setlength{\baselineskip}{0.5cm}
Cranked Relativistic Hartree-Bogoliubov theory without and
with approximate particle number projection by means of the
Lipkin-Nogami method is presented in detail as an extension 
of Relativistic Mean Field theory with pairing correlations 
to the rotating frame. Pairing correlations are taken into 
account by a finite range two-body force of Gogny type. The 
applicability of this theory to the description of rotating 
nuclei is studied in detail on the example of 
superdeformed bands in even-even nuclei of the 
$A\sim 190$ mass region. Different aspects such as the importance 
of pairing and particle number projection, the dependence of 
the results on the parametrization of the RMF Lagrangian and 
Gogny force etc. are investigated in detail. It is 
shown that without any adjustment of new parameters the 
best description of experimental data is obtained by using the 
well established parameter sets NL1 for the Lagrangian 
and D1S for the pairing force. Contrary to previous studies 
at spin zero it is found that the increase of the 
strength of the Gogny force is not necessary in the
framework of Relativistic Hartree-Bogoliubov theory
provided that particle number projection is performed.
\end{abstract}
\vspace{0.5cm}

\narrowtext

\input epsf

%%%%%%%%%%%%%%%%%%%%%%%%%%%%%%%%%%%%%%%%%%%%%%%%%%%%%%%%%%
\section{Introduction}
%%%%%%%%%%%%%%%%%%%%%%%%%%%%%%%%%%%%%%%%%%%%%%%%%%%%%%%%%%

  The development of self-consistent microscopic many-body 
mean field theories aimed on the description of low-energy 
nuclear phenomena provides necessary theoretical tools for 
an exploration of the nuclear chart into know and unknown regions. 
This development is motivated by theoretical and experimental 
reasons. Compared with conventional approaches such as, for 
example, the macroscopic+microscopic method, self-consistent 
theories rely on a smaller number of assumptions and start 
from a more microscopic level. For example, the starting point of 
non-relativistic mean field theories is an effective interaction 
between nucleons constituting the nucleus. Such theories based either 
on zero range Skyrme forces or finite range Gogny forces have 
been widely used starting from seventies. The next step is 
replacing the Schr\"odinger equation by the Dirac equation and 
thus considering the relativistic mean field (RMF) theory \cite{SW.86}. 
In RMF theory, the nucleus is described as a system of point-like
nucleons, Dirac spinors, which interact in a phenomenological
way by the exchange of mesons, such as the $\sigma$-meson
responsible for the large scalar attraction at intermediate
distances, the $\omega$-meson for the vector repulsion at
short distances and the $\rho$-meson for the asymmetry properties 
of nuclei with large neutron or proton excess.
Such a description has a clear advantage that the spin-orbit 
splitting, which plays an extremely important role in low-energy 
nuclear physics, emerges in a natural way as a genuine relativistic 
effect. In addition, the pseudo-spin symmetry, the origin of which was 
a long-standing puzzle, does find a natural explanation in the 
framework of RMF theory \cite{G.97}. RMF theory has been extremely 
successful also in the description of many other facets of low-energy 
nuclear physics, such as ground state properties, giant resonances, 
superdeformed rotating nuclei etc., see Ref.\ \cite{R.96} for an 
overview.   

  Since the discovery of the first superdeformed rotational (SD) 
band in $^{152}$Dy \cite{TNN.86}, the investigation of superdeformation 
at high angular momentum remains one of the most challenging topics 
of nuclear structure. The rich variety of physical phenomena at 
superdeformed shapes is based on a complicated and rather subtle
interplay of collective and single-particle properties. Although at 
the present stage, a general understanding of this phenomenon has been 
achieved, there are still many unresolved questions related, for 
example, to the phenomenon of identical bands and to the treatment of 
pairing correlations. In addition, the microscopic theoretical models 
used so far are still far from a precise quantitative description of 
SD bands which indicates the necessity of further improvements.
Different theoretical methods based mainly on the concept of the 
cranking model of Inglis \cite{Ing.54} have been employed for 
the quantitative description of various high-spin phenomena at SD 
shapes, see for example recent overviews in Refs.\ \cite{AR.99,D.99}. 
The cranked version of the RMF theory - the Cranked  Relativistic Mean
Field (CRMF) theory \cite{KR.89,KR.90,KR.93,AKR.96} is amongst the 
most successful ones. It has been applied in a systematic way to 
the description of SD bands in different mass regions such as 
$A\sim 60$ \cite{A60,Zn60,Zn68,MM.98}, $A\sim 80$ \cite{Sr83} 
and $A\sim 140-150$ \cite{KR.93,AKR.96,Hung,ALR.98,Ho153}. 
Pairing correlations are expected to be considerably 
quenched in SD bands of these regions at high spin and thus 
they have been neglected in all studies quoted above. One should 
clearly recognize that the neglect of pairing correlations 
is an approximation because pairing correlations being weak 
are still present even at the highest rotational frequencies.
Despite this a very successful description of many properties of 
SD bands, such as dynamic $J^{(2)}$ and kinematic $J^{(1)}$ 
moments of inertia, absolute and relative charge quadrupole 
moments, effective alignments $i_{eff}$, single-particle properties
in the SD minimum etc., has been obtained in these studies in an 
unpaired formalism. 

 However, the rotational properties of nuclei at low and medium 
spin are strongly affected by pairing correlations. In order to 
describe such properties within the relativistic framework, we 
have developed the Cranked Relativistic Hartree-Bogoliubov (CRHB) 
theory. This theory is an extension of CRMF theory to the description 
of pairing correlations in rotating nuclei. The brief outline of 
this theory and its application to the study of several yrast
SD bands observed in the $A\sim 190$ mass region has been reported 
in Ref.\ \cite{A190}. The present manuscript represents an extension 
of this investigation where both the theoretical formalism 
and the calculations will be presented in much greater details.

 The paper is organized in the following way: In Section
\ref{sect-CRHB} a detailed description of the CRHB theory 
without and with approximate particle number projection by 
means of the Lipkin-Nogami method and some of the specific
features of the present calculations are presented. The shell
structure in the $A\sim 190$ mass region of superdeformation
and the impact of pairing and particle number projection on 
the rotational and deformation properties of rotating nuclei
are investigated in detail on the example of the lowest SD 
bands in $^{192}$Hg and $^{194}$Pb nuclei in Section \ref{sect-pair}.
In Section \ref{par-dep}, we study the dependence of the results 
of CRHB calculations with particle number projection on the 
parametrization of the RMF Lagrangian and the Gogny force using 
SD bands in $^{194}$Hg and $^{194}$Pb as an example. The 
properties of yrast SD bands observed so far in even-even 
nuclei of the $A\sim 190$ mass region are systematically
studied in Section \ref{systematic}. Finally, Section \ref{conclusions} 
summarizes our main conclusions.

%%%%%%%%%%%%%%%%%%%%%%%%%%%%%%%%%%%%%%%%%%%%%%%%%%%%%%%%%
\section{Cranked Relativistic Hartree-Bogoliubov (CRHB) 
Theory}
\label{sect-CRHB}
%%%%%%%%%%%%%%%%%%%%%%%%%%%%%%%%%%%%%%%%%%%%%%%%%%%%%%%%%

%%%%%%%%%%%%%%%%%%%%%%%%%%%%%%%%%%%%%%%%%%%%%%%%%%%%%%%%%
\subsection{The CRHB equations}
\label{CRHBeq}
%%%%%%%%%%%%%%%%%%%%%%%%%%%%%%%%%%%%%%%%%%%%%%%%%%%%%%%%%

 In relativistic mean field (RMF) theory the nucleus is
described as a system of point-like nucleons, Dirac
spinors, coupled to mesons and to the photons. The nucleons
interact by the exchange of several mesons, namely a scalar
meson $\sigma$ and three vector particles $\omega$, $\rho$
and the photon. The isoscalar-scalar $\sigma $-mesons
provide a strong intermediate range attraction between the
nucleons. For the three vector particles we have to
distinguish the time-like components and the spatial
components. For the photons this means the Coulomb field
and possible magnetic field in the case where currents play
a role. For the isoscalar-vector $\omega$-meson the
time-like component provides a very strong repulsion at
short distances for all combinations of particles, $pp$,
$nn$ and $pn$.  For the isovector-vector $\rho$-meson the
time-like components give rise to a short range repulsion
for like particles ($pp$ and $nn$) and a short range
attraction for unlike particles ($np$). They also have a 
strong influence on the symmetry energy. In addition, the 
spatial components of the $\omega$ and $\rho$-mesons lead 
to an interaction between possible currents, which are for 
the $\omega$-meson attractive for all combinations ($pp$, $nn$ and
$pn$-currents) and for the $\rho$-meson attractive for $pp$
and $nn$-currents but repulsive for $pn$-currents. We have
to keep in mind, however, that within mean field theory
these currents only occur in cases of time-reversal
breaking mean fields as, for instance, in the case of 
the Coriolis fields.

 The starting point of RMF theory is the well known local 
Lagrangian density
\begin{eqnarray}
{\cal L}&=&\bar{\psi}(i\gamma ^\mu \partial _\mu -m)\psi
+\frac{1}{2}(\partial_\mu \sigma \partial ^\mu 
\sigma -m_\sigma ^2\sigma ^2)
-\frac{1}{3}g_2\sigma^3-\frac{1}{4}g_3\sigma^4
\nonumber \\ 
&&-\frac{1}{4}{\Omega}_{\mu\nu}{\Omega}^{\mu\nu}
  +\frac{1}{2}m_\omega^2\omega_\mu\omega^\mu 
  ~-~\frac{1}{4}{\vec R}_{\mu\nu}{\vec R}^{\mu\nu}
  +\frac{1}{2}m_\rho^2\vec{\rho}_\mu\vec{\rho\,}^\mu  
  ~-~\frac{1}{4}F_{\mu\nu}F^{\mu\nu}
\label{RMFlagr} \\
&&-g_\sigma\bar{\psi}\sigma\psi~-~
   g_\omega \bar{\psi}\gamma^\mu\omega_\mu\psi~-~
   g_\rho\bar{\psi}\gamma^\mu\vec{\tau}\vec{\rho}_\mu\psi~-~ 
   e\,\bar{\psi}\gamma^\mu\,\frac{1-\tau_3}{2}A_\mu\psi, 
\nonumber
\end{eqnarray}
where the non-linear self-coupling of the $\sigma$-field,
which is important for an adequate description of nuclear
surface properties and the deformations of finite nuclei,
is taken into account according to Ref.\ \cite{BB.77}. The 
Lagrangian (\ref{RMFlagr}) contains as parameters the 
masses of the mesons $m_{\sigma}$, $m_{\omega}$ and $m_{\rho}$, 
the coupling constants $g_{\sigma}$, $g_{\omega}$ and $g_{\rho}$ 
and the non-linear terms $g_2$ and $g_3$. The field tensors for 
the vector mesons and the photon field are: 
\begin{eqnarray}
{\Omega}_{\mu\nu}=\partial_\mu\omega_\nu-\partial_\nu\omega_\mu,\qquad 
{\vec
R}_{\mu\nu}=\partial_\mu\vec\rho_\nu-\partial_\nu\vec\rho_\mu,
\qquad F_{\mu\nu }=\partial_\mu A_\nu-\partial_\nu A_\mu.  
\label{Fieldtensor}
\end{eqnarray}
In the present state of the art of the RMF theory, the meson and 
photon fields are treated as classical fields. 

 Two approximations, namely the {\it mean-field
approximation}, in which the meson field operators are replaced
by their expectation values
\begin{eqnarray}
\langle \sigma \rangle =\sigma_0,\,\,\,\,\,\,\, 
\langle \omega_{\mu} \rangle =\delta_{\mu 0}\omega_0
\label{MFapr}
\end{eqnarray}
which is justified by the fact that the source terms are large 
\cite{SW.86}, and the {\it No-sea approximation}, in which only 
positive-energy states are taken into account \cite{RRM.86},
are employed in order to solve the Lagrangian (\ref{RMFlagr}).

 For classical fields we can restrict ourselves to the intrinsic 
symmetry violating product wave function $|\Phi \rangle$ 
which can be represented as a 
generalized Slater determinant. As long as we consider 
time-independent static or quasistatic fields, this limits the 
applicability of CRHB theory to the yrast bands and to the 
excited bands of dominantly quasiparticle nature. In particular, 
the bands based on low-lying collective vibrations, such as
for example $\gamma$-, $\beta$-vibrational bands, cannot be
described in the present formalism since they have 
(for example, in even-even nuclei) an intrinsic wave function
which is a linear superposition of many two-quasiparticle
states. In order to study such bands one should employ
time-dependent fields in the random phase approximation in 
the rotating frame: the task which has so far been resolved 
only in simple non-relativistic models with separable
forces \cite{EMR.80RPA}.

 As discussed in detail in Refs.\ \cite{KR.89,KNM.93,MM.97} in the 
unpaired formalism, the transformation to the rotating frame 
within the framework of the cranking model leads to the cranked 
relativistic mean field equations. Note that in the 
present investigation we restrict ourselves to one-dimensional 
rotation with rotational frequency $\Omega_x$ around the 
$x$-axis. Since pairing correlations work only between the 
fermions, the Bogoliubov transformation affects directly only 
the fermionic part. Thus the time-independent inhomogeneous 
Klein-Gordon equations for the mesonic fields obtained by means 
of variational principle are given in the CRHB theory by 
\cite{A190}
\begin{eqnarray}
\left\{-\Delta-({\Omega}_x\hat{L}_x)^2 + m_\sigma^2\right\}~
\sigma(\bff r) & = & 
-g_\sigma \rho_{\sl s}(\bff r) \nonumber \\
\,\,\,\,\,\,\,\,\,\,\,\,\,\,\,\,\,-g_2\sigma^2(\bff r)-g_3\sigma^3(\bff r),
\nonumber \\
\left\{-\Delta-({\Omega}_x\hat{L}_x)^2+m_\omega^2\right\}
\omega_0(\bff r)&=&
g_\omega \rho_{\sl v}^{\it i\sl s}(\bff r),
\nonumber \\
\left\{-\Delta-[{\Omega}_x(\hat{L}_x+\hat{S}_x)]^2+
m_\omega^2\right\}~
\bff\omega(\bff r)&=&
g_\omega\bff j^{\it i\sl s}(\bff r),
\nonumber \\
\left\{-\Delta-({\Omega}_x\hat{L}_x)^2+m_\rho^2\right\}
\rho_0(\bff r)&=&
g_\rho\rho_{\sl v}^{\it i\sl v}(\bff r),
\nonumber \\
\left\{-\Delta-[{\Omega}_x(\hat{L}_x+\hat{S}_x)]^2+
m_\rho^2\right\}~
\bff\rho(\bff r)&=& g_\rho \bff j^{\it i\sl v}(\bff r),
\nonumber \\
-\Delta~A_0(\bff r)&=&e\rho_{\sl v}^p(\bff r),
\nonumber \\
-\Delta~\bff A(\bff r)&=&e\bff j^p(\bff r),
\label{KGeq}
\end{eqnarray}
where the source terms are sums of bilinear products of 
baryon amplitudes 
\begin{eqnarray}
\rho_{\sl s}(\bff r) & = & \sum_{k>0} 
 [V_k^n(\bff r)]^{\dagger} \hat{\beta} V_k^n (\bff r) 
+[V_k^p(\bff r)]^{\dagger} \hat{\beta} V_k^p (\bff r), 
\nonumber \\
\rho_{\sl v}^{\it i\sl s}(\bff r) & = & \sum_{k>0} 
 [V_k^n(\bff r)]^{\dagger} V_k^n (\bff r) 
+[V_k^p(\bff r)]^{\dagger} V_k^p (\bff r), 
\nonumber \\
\rho_{\sl v}^{\it i\sl v}(\bff r) & = & \sum_{k>0} 
[V_k^n(\bff r)]^{\dagger} V_k^n (\bff r) 
-[V_k^p(\bff r)]^{\dagger} V_k^p (\bff r), 
\nonumber \\
\bff j^{\it i\sl s}(\bff r) & = & \sum_{k>0} 
[V_k^n(\bff r)]^{\dagger} \hat{\bff\alpha} V_k^n (\bff r) 
+[V_k^p(\bff r)]^{\dagger} \hat{\bff\alpha} V_k^p (\bff r), 
\nonumber \\
\bff j^{\it i\sl v}(\bff r) & = & \sum_{k>0} 
[V_k^n(\bff r)]^{\dagger} \hat{\bff\alpha} V_k^n (\bff r) 
-[V_k^p(\bff r)]^{\dagger} \hat{\bff\alpha} V_k^p (\bff r). 
\label{source}
\end{eqnarray}
The sums over $k>0$ run over all quasiparticle states corresponding
to positive energy single-particle states ({\it no-sea approximation}). 
In Eqs.\ (\ref{KGeq},\ref{source}), the indexes $n$ and $p$ indicate 
neutron and proton states, respectively, and the indexes $\it i\sl s$ 
and $\it i\sl v$ are used for isoscalar and isovector quantities.
$\rho_{\sl v}^p(\bff r)$, $\bff j^p(\bff r)$ in Eq.\ (\ref{KGeq})
correspond to $\rho_{\sl v}^{\it i\sl s}(\bff r)$ and 
$\bff j^{\it i\sl s}(\bff r)$ defined in Eq.\ (\ref{source}), 
respectively, but with the sums over neutron states neglected. 
Note that the Coriolis term for the Coulomb potential $A_0({\bff r})$
and the spatial components of the vector potential $\bff A(\bff r)$
are neglected in Eqs.\ (\ref{KGeq}) since the coupling constant of 
the electromagnetic interaction is small compared with the coupling
constants of the meson fields. On this classical level only the
direct terms of the potentials are taken into account since most
parametrizations of the RMF Lagrangian have been fitted to 
experimental data neglecting exchange terms. This means that the
exchange terms are not fully neglected, but taken into account in 
an averaged way by adjusting the parameters of the direct terms.

 The comparison of Eqs.\ (\ref{KGeq},\ref{source}) with Eqs.\ (4) 
and (5) in Ref.\ \cite{AKR.96} indicates that the pairing correlations 
between the fermions have also an impact on mesonic fields through 
the redefinition of the various nucleonic densities and currents. 
While the occupation probabilities of the single-nucleon orbitals 
are equal 1 and 0 for the orbitals below and above the Fermi level 
in the systems with no pairing, they are between 0 and 1 when the 
pairing between the fermions is taken into account.

  Contrary to the applications of the Hartree-(Fock)-Bogoliubov
theory for the ground states of even-even nuclei, the time-reversal 
symmetry of the intrinsic wave function $|\Phi \rangle$ is broken 
by the Coriolis operator $\Omega_x \hat{J}_x$ ($\hat{J_x}$ is the 
projection of total angular momentum on the rotation axis) in the 
case of rotating nuclei. Therefore, we do not know a priori the conjugate 
states in the canonical basis and thus one must solve the full
Hartree-Bogoliubov problem in the rotating frame 
\cite{RBM.70,M.75,RS.80}. The solution of this problem in the RMF
theory is to large extent similar to the one obtained earlier
in the non-relativistic case as far as we treat pairing in 
non-relativistic fashion as it is done in the CRHB theory.
Thus in the following discussion we will mainly concentrate 
on the features specific for relativistic case. 

 The CRHB equations for the fermions in the rotating frame 
are given in {\it one-dimensional cranking approximation} by 
\begin{eqnarray}
\pmatrix{ \hat{h}_D - \lambda_{\tau} - \Omega_x \hat{J}_x    & \hat{\Delta}  \cr
-\!\hat{\Delta}^* &     -\hat{h}^*_D + \lambda_{\tau} + \Omega_x \hat{J}^*_x \cr}
\pmatrix{ U_k(\bff r) \cr V_k (\bff r) } =
E_k \pmatrix{ U_k (\bff r) \cr V_k (\bff r) }
\label{CRHB}
\end{eqnarray}
where $\hat{h}_D$ is the Dirac Hamiltonian for the nucleon with mass $m$
\begin{eqnarray} 
\hat{h}_D= \bff\alpha(-i\bff\nabla-\bff V(\bff r))~+~
V_0(\bff r)~+~\beta(m+S(\bff r)) 
\end{eqnarray}
and $\lambda_{\tau}$ are the chemical potentials defined from the average 
particle number constraints for protons and neutrons ($\tau=p,n$)
\begin{eqnarray}
\langle {\Phi}_{\Omega_x} |\hat{N}_p|{ \Phi}_{\Omega_x} \rangle = Z, \qquad
\langle {\Phi}_{\Omega_x} |\hat{N}_n|{ \Phi}_{\Omega_x} \rangle =N.
\label{partnumb} 
\end{eqnarray}
The particle number expectation values 
$\langle {\Phi}_{\Omega_x} |\hat{N}_{\tau}|{\Phi}_{\Omega_x}\rangle$ are defined 
via the (normal) density matrices $\rho_p$ and $\rho_n$  
\begin{eqnarray}
\langle {\Phi}_{\Omega_x} |\hat{N}_{\tau}|{ \Phi}_{\Omega_x} \rangle
=Tr(\rho_{\tau}) \qquad {\rm where} \qquad \rho_{\tau} =V^{*}_{\tau} V^{T}_{\tau}.
\end{eqnarray} 

The Dirac Hamiltonian contains an attractive scalar potential $S(\bff r)$
\begin{eqnarray}
S(\bff r)=g_\sigma\sigma(\bff r),
\label{Spot}
\end{eqnarray}
a repulsive vector potential $V_0(\bff r)$
\begin{eqnarray}
V_0(\bff r)~=~g_\omega\omega_0(\bff r)+g_\rho\tau_3\rho_0(\bff r) 
+e \frac{1-\tau_3} {2} A_0(\bff r),
\label{Vpot}
\end{eqnarray}
a magnetic potential $\bff V(\bff r)$
\begin{eqnarray}
\bff V(\bff r)~=~g_\omega\bff\omega(\bff r)
+g_\rho\tau_3\bff\rho(\bff r)+
e\frac {1-\tau_3} {2} \bff A(\bff r),
\label{Vmag}
\end{eqnarray}
and a Coriolis term
\begin{eqnarray}
-{\Omega}_x\hat{J}_x~=~
-{\Omega}_x(\hat{L}_x+ \frac {1} {2} \hat{\Sigma}_x).
\end{eqnarray}
The latter two terms are the contributions to the mean field which 
break time-reversal symmetry and induce currents. In the Dirac 
equation the field $\bff V(\bff r)$ has the structure of a magnetic 
field.  Time-reversal symmetry 
is broken when the orbitals which are time-reversal 
counterparts are not occupied pairwise. At no rotation,
this usually takes place in one-(multi-)quasiparticle 
configurations where the magnetic potential {\bff V}(\bff r) 
breaks time-reversal symmetry. In rotating nuclei the 
time-reversal symmetry is additionally broken by the 
Coriolis field. The nuclear currents of Eq.\ (\ref{KGeq}) 
are the sources for the space-like components of the vector 
$\bff\omega(\bff r)$, $\bff\rho(\bff r)$ and $\bff A(\bff r)$ 
fields which give rise to polarization effects in the Dirac 
spinors through the magnetic potential $\bff V(\bff r)$ of 
Eq.\ (\ref{Vmag}). 
This effect is commonly referred to as {\it nuclear magnetism} 
\cite{KR.89}. It turns out that it is  very important to take 
it into account for the proper description of currents, magnetic 
moments \cite{HR.88} and moments of inertia \cite{KR.93}.
In the present calculations the spatial components of the vector 
mesons are properly taken into account in a fully self-consistent 
way.  

In Eq.\ (\ref{CRHB}), the rotational frequency $\Omega_x$ 
along the $x$-axis is defined from the condition \cite{Ing.54}
\begin{eqnarray}
J = \langle{\Phi}_{\Omega_x}\mid\hat{J}_x
\mid {\Phi}_{\Omega_x}\rangle~=~\sqrt{I(I+1)}.  
\label{eq3}
\end{eqnarray}
where $I$ is total nuclear spin. $U_k (\bff r)$ and 
$V_k (\bff r)$ are quasiparticle Dirac spinors and 
$E_k$ denotes the quasiparticle energies. 

The pairing potential (field) $\hat{\Delta}$ in Eq.\ (\ref{CRHB}) 
is given by
\begin{eqnarray}
\hat{\Delta} \equiv \Delta_{ab}~=~\frac{1}{2}\sum_{cd} V^{pp}_{abcd} 
\kappa_{cd}
\label{gap}
\end{eqnarray}
where the indices $a,b,\dots$ denote quantum numbers which specify 
the single-particle states with the space coordinates
$\bff r$ as well as the Dirac and isospin indices $s$ and 
$\tau$. It contains the pairing tensor \footnote{This quantity is
sometimes called as {\it abnormal density}.} $\kappa$
\begin{eqnarray}
\kappa = V^{*}U^{T} 
\label{kappa}
\end{eqnarray}
and the matrix elements $V^{pp}_{abcd}$ of the effective
interaction in the $pp$-channel. 
 
 The matrix elements $V_{abcd}^{pp}$ in the pairing channel can, 
in principle, be derived as a one-meson exchange interaction by
eliminating the mesonic degrees of freedom in the model 
Lagrangian as discussed in detail in the relativistic 
Hartree-Bogoliubov model developed in Ref.\ \cite{KurcR.91}. 
However, the resulting pairing matrix elements obtained with 
the standard sets of the RMF theory are unrealistically 
large. In nuclear matter the strong repulsion produced by the 
exchange of vector mesons at short distances results in a 
pairing gap at the Fermi surface that is by factor of 3 too 
large. In addition,
standard RMF parameters do not reproduce scattering data
in the $S_0$-channel, which is necessary for a reasonable 
description of pairing correlations. 
On the other hand, since we are using effective forces, there 
is no fundamental reason to have the same interaction both in 
the particle-hole and particle-particle channel \cite{RS.80}. 
In a first-order approximation,
the effective interaction contained in the mean field $\hat{\Gamma}$
is a $G$ matrix, i.e. the sum over all ladder diagrams. 
On the contrary, the effective force in the $pp$ channel (pairing
potential $\hat{\Delta}$) should be the $K$ matrix, the sum
of all diagrams irreducible in $pp$ direction. Although encouraging 
results have been reported in applications to nuclear matter 
\cite{GCF.96,MFD.97}, a microscopic and fully relativistic derivation 
of the pairing force starting from the Lagrangian of quantum
hadrodynamics  still cannot be applied to realistic nuclei. 
Thus we follow the prescription of Ref.\ \cite{GELR.96} and use a 
phenomenological interaction of Gogny type with finite range 
(see Eq.\ (\ref{Vpp}) below) in the particle-particle channel. 
Such a procedure provides both the automatic cutoff of 
high-momentum components and, as follows from 
non-relativistic and relativistic studies, a reliable 
description of pairing properties in finite nuclei.

  Although this procedure formally breaks the Lorentz structure 
of the RMF equations, one has to 
keep in mind that pairing itself is a completely non-relativistic 
phenomenon. Relativistic effects such as the nuclear saturation
mechanism due to the cancellation between attractive scalar and 
repulsive vector potentials, spin-orbit splitting and
the admixture of small components through the kinetic term 
$-i \bff\alpha \bff\nabla$ of the Dirac equation \cite{R.96} 
are only important for the mean field part of Hartree-Bogoliubov 
theory and have only a negligible counterpart in the pairing field. 
The pairing density $\kappa$ and the corresponding pairing field 
result from the scattering of pairs in the vicinity of the Fermi surface. 
The pairing density is concentrated in an energy window of a few
MeV around the Fermi level, i.e. the contributions from the small 
components of the wave functions to the pairing tensor $\kappa$ 
are very small. Thus in the present version of the CRHB theory, 
pairing correlations are only considered between the positive 
energy states. Consequently, it is justified to approximate the pairing 
force by the best currently available non-relativistic interaction: 
the pairing part of the Gogny force. At present, this is certainly 
more realistic than the use of a one-meson exchange force in the 
pairing channel, since this type of interactions has never been 
optimized for the description of pairing properties in finite 
nuclei. Thus the word {\it relativistic} in CRHB applies only to 
the Hartree particle-hole channel of this theory.

 The phenomenological Gogny-type finite range interaction is
given by 
\begin{eqnarray}
V^{pp}(1,2) = f \sum_{i=1,2} e^{-[({\bff r}_1-{\bff r} _2)/\mu_i]^2} 
\times (W_i+B_i P^{\sigma}- H_i P^{\tau} - M_i P^{\sigma} P^{\tau})
\label{Vpp}
\end{eqnarray}
where $\mu_i$, $W_i$, $B_i$, $H_i$ and $M_i$ $(i=1,2)$ are the 
parameters of the force and $P^{\sigma}$ and $P^{\tau}$ are the
exchange operators for the spin and isospin variables, respectively. 
This interaction is density-independent. In most 
of the applications of the CRHB theory in the present article, the 
parameter set D1S \cite{D1S-a} (see also Table \ref{table-Gog}) is 
employed for the Gogny force. 
Note also that an additional factor $f$ affecting the strength 
of the Gogny force is introduced in Eq.\ (\ref{Vpp}). This
is motivated by the fact that previous studies of different
nuclear phenomena within the Relativistic Hartree-Bogoliubov
theory followed the prescription of Ref.\ \cite{GELR.96} 
where $f=1.15$ was employed. It turns out, however, that the 
rotational properties of the SD bands are very well described
with $f=1.0$ if the approximate particle number projection
is performed by means of the Lipkin-Nogami method. Thus 
the value $f=1.0$ is used in the following if no other value
is specified.  

 In Hartree-(Fock)-Bogoliubov calculations the size of the 
pairing correlations is usually measured in terms of the 
pairing energy\footnote{This quantity is sometimes called 
as particle-particle correlation energy, see for 
example Ref.\ \protect\cite{VER.99}} defined as 
\begin{eqnarray}
E_{pairing}~=~-\frac{1}{2}\mbox{Tr} (\Delta\kappa).
\label{Epair}
\end{eqnarray}
This is not an experimentally accessible quantity, but 
it is a measure for the size of the pairing correlations 
in the theoretical calculations. An alternative way to 
look for the size of pairing correlations is to use the 
BCS-like pairing energies defined as a difference between 
the total energies obtained at a given spin $I$ in 
the calculations with ($E_{\rm CRHB}(I)$) and without 
($E_{\rm CRMF}(I)$) pairing 
\begin{eqnarray}
E_{\rm BCS}(I) = E_{\rm CRHB}(I)-E_{\rm CRMF}(I)
\label{Ebcs}
\end{eqnarray}
The quantities $E_{\rm BCS}$ and $E_{pairing}$ coincide 
only in the BCS approximation. The disadvantage of the 
use of the BCS-like pairing energies is related to the 
fact that the calculations with and without pairing
should be performed in order to define this quantity
in a proper way.

The total energy of system in the laboratory frame is given
by
\begin{eqnarray}
E_{\rm CRHB} = E^{\rm F}+E^{\rm B}
\end{eqnarray}
where $E^{\rm F}$ and $E^{\rm B}$ are the contributions from 
fermionic and mesonic (bosonic) degrees of 
freedom. Fermionic energies $E^{\rm F}$ are given by
\begin{eqnarray}
E^{\rm F}=E_{part}  + \Omega_x J + E_{pairing}+ E_{cm} 
\end{eqnarray}
where
\begin{eqnarray}
E_{part}=Tr (h_D \rho),\qquad \qquad J=Tr (j_x \rho),
\end{eqnarray} 
are the particle energy and the expectation value of the total 
angular momentum along the rotational axis and
\begin{eqnarray} 
E_{cm}=-\frac 34\hbar \omega _0=-\frac 3441A^{-1/3}\,\,\,\, {\rm MeV}
\end{eqnarray}
is the correction for the spurious center-of-mass motion 
approximated by its value in a non-relativistic harmonic 
oscillator potential.

  The bosonic energies $E^{\rm B}$ in the laboratory frame are
given by
\begin{eqnarray}
E^{\rm B}= && -~\frac{1}{2}\int d {\bff r} \,[g_\sigma\,\sigma (\bff r) 
\rho_{\sl s} (\bff r) + \frac{1}{3}g_2\sigma^3(\bff r) 
+\frac{1}{2}g_3\sigma^4(\bff r)]  
\nonumber\\
&&-~\frac {1}{2}\, g_\omega \int d {\bff r}\,[\omega_0(\bff r) 
\rho_{\sl v}^{\it i\sl s} (\bff r) 
          -\bff\omega (\bff r) \bff j^{\it i\sl s}(\bff r)]
\nonumber\\
&&-~\frac{1}{2}\, g_\rho \int d {\bff r}\,[\rho_0(\bff r) 
\rho_{\sl v}^{\it i\sl v} (\bff r)
-\bff\rho (\bff r) \bff j^{\it i \sl v} (\bff r)]  
\nonumber\\
&&-~\frac{1}{2}\, e \int d {\bff r} \left[A_0(\bff r) \rho_{\sl v}^p (\bff r) 
+ \bff A (\bff r) \bff j^p (\bff r) \right]  
\nonumber\\
&&+~{\Omega}_x^2\int d {\bff r} \,
[\sigma(\bff r) \hat{L}_x^2\sigma(\bff r)- 
\omega_0(\bff r) \hat{L}_x^2\omega_0(\bff r) 
+\bff\omega(\bff r) (\hat{L}_x+\hat{S}_x)^2
\bff\omega (\bff r)
\nonumber \\
&&\qquad\qquad\qquad
-\rho_0(\bff r)\hat{L}_x^2\rho_0(\bff r)
+\bff\rho(\bff r)(\hat{L}_x+\hat{S}_x)^2\bff\rho(\bff r)]
\end{eqnarray}
In the systems with broken time-reversal symmetry the spatial 
components of the $\omega$-mesons give larger contributions to 
the total energy than the ones of the $\rho$-mesons because of 
the isovector nature of the $\rho$-meson. One should also note 
that the contribution of the terms proportional to ${\Omega}_x^2$
to the total energy is very small being typically in the range 
of $\sim 10-20$ keV at the highest frequencies of interest 
in the $A\sim 190$ mass region and thus, in general, can 
be neglected.

%%%%%%%%%%%%%%%%%%%%%%%%%%%%%%%%%%%%%%%%%%%%%%%%%%%%%%%%%%%%%%%%%
\subsection{The Lipkin-Nogami Method}
%%%%%%%%%%%%%%%%%%%%%%%%%%%%%%%%%%%%%%%%%%%%%%%%%%%%%%%%%%%%%%%%%

 In recent years it has become clear that a proper treatment
of pairing correlations is required to describe many nuclear
properties. One should note that the Bogoliubov transformation 
is not commutable with the nucleon number operator and 
consequently the resulting wave function does not 
correspond to a system having a definite number of 
protons and neutrons. The best way to deal with this problem 
would be to perform an exact particle number projection 
before the variation \cite{RS.80}. However, for heavy nuclei
this has been realized so far only for non-relativistic separable
models \cite{ER.82} in part due to the fact that such 
calculations are expected to be extremely time-consuming
for realistic interactions. As a result,
an approximate particle number projection by means of the
Lipkin-Nogami (LN) method \cite{L.60,N.64,NZ.64,PNL.73} 
(further APNP(LN)) is most widely used just because of its 
simplicity. As illustrated by the results of non-relativistic 
\cite{GBDFH.94,THBDF.95,GDBPL.97,VER.97,VE.97,VER.99,WS.94,SWM.94} 
and recent relativistic \cite{A190,Rare} calculations, the 
application of this method considerably improves an agreement 
with experiment especially for rotational properties of nuclei
compared with the calculations without particle number
projection.

  Our derivation of the Lipkin-Nogami method is based on the 
Kamlah expansion \cite{Ka68} which is an approximation to
the exact projection methods. It is applied to the two-body
interaction $V^{pp}$ in the pairing channel. Note that the 
Gogny force in the pairing channel is density independent 
in the CRHB theory. For the derivation of 
the Lipkin-Nogami method with density dependent forces see 
Refs.\ \cite{VER.97,VER.99}.

 In the following discussion, we will limit ourselves  
to one type of particles. In the numerical calculations
we use this approach for protons and neutrons separately.
The eigenstates $|\Psi^N\rangle$ of the particle number 
operator $\hat{N}$ can be constructed from particle number 
violating Bogoliubov wave functions $|\Phi\rangle$ by 
means of the particle number projection operator 
$\hat{P}^N$ 
\begin{eqnarray}
|\Psi^N\rangle = \hat{P}^N|\Phi\rangle= \frac{1}{2\pi} \int_0^{2\pi}\,d\phi
e^{i(\hat{N}-N)\phi}|\Phi\rangle
\end{eqnarray}
where $N$ is actual particle number. With this wave function one obtains 
the total particle number projected energy 
\begin{eqnarray}
  E^N_{proj} =
  \frac{\langle\Phi| \hat{H}\hat{P}^N |\Phi\rangle}
       {\langle\Phi| \hat{P}^N        |\Phi\rangle}
  = 
  \frac{\int_0^{2\pi}\,d\phi e^{-i\phi N} h(\phi)}
  {\int_0^{2\pi}\,d\phi e^{-i\phi N} n(\phi)}.
\end{eqnarray}
In the case of large particle number and strong deformations in 
gauge space one expects that the Hamiltonian and norm 
overlap integrals
\begin{eqnarray}
  h(\phi)= \langle \Phi | \hat{H} e^{i\phi\hat{N}}|\Phi\rangle
  \mbox{~~~~and~~}
  n(\phi)= \langle \Phi | e^{i\phi\hat{N}}|\Phi\rangle,
\end{eqnarray}
are sharply peaked at $\phi=0$ and are very small elsewhere. 
In addition, the quotient $h(\phi)/n(\phi)$ should be a rather 
smooth function. In such a case one can make an expansion of 
$h(\phi)$ in terms of $n(\phi)$ in the following way
\begin{eqnarray}
  h(\phi)= \sum_{m=0}^{M} \lambda_m \hat{\cal{N}}^m n(\phi)
\label{hphi}
\end{eqnarray}
where the Kamlah operator
\begin{eqnarray}
  \hat{\cal{N}}=\frac{1}{i}\frac{\partial}{\partial \phi}
  -\langle \Phi |\hat{N}|\Phi\rangle
\end{eqnarray}
represents the particle number operator in the space of the gauge
angle 
$\phi$ and $\lambda_m$ are constants. Mathematically, Eq.\ (\ref{hphi}) 
essentially corresponds to a Taylor expansion of the Fourier 
transformed function $h(\phi)/n(\phi)$ \cite{RS.80}. $M$ in 
Eq.\ (\ref{hphi}) represents the order of expansion, for 
$M\rightarrow \infty$ this equation is exact. Since we need $h(\phi)$ 
only in the vicinity of $\phi=0$ we can already hope to get a very 
good approximation for this function with a few non-vanishing 
constants $\lambda_0, \lambda_1, \lambda_2, ...$, if they are 
properly adjusted. The number $M$ certainly depends on the widths 
of the overlap integral $n(\phi)$ which reflects the symmetry 
violation. The larger the symmetry violation, the better the 
approximation will already be for low $M$-values.

  Inserting Kamlah operator into Eq.\ (\ref{hphi}) one gets
\begin{eqnarray}
  h(\phi)=  \sum_{m=0}^{M} \lambda_m 
  \langle \Phi | (\Delta\hat{N})^m e^{i\phi\hat{N}}|\Phi\rangle
  \mbox{~~~~with ~~~~~} 
  \Delta \hat{N}=  \hat{N}- \langle\Phi| \hat{N} |\Phi\rangle  .
\end{eqnarray}
The expansion coefficients $\lambda_m$ are determined by 
applying the operators $\hat{\cal N}^k$ $(k=0,1,...M)$ on 
Eq.\ (\ref{hphi})
\begin{eqnarray}
\hat{\cal{N}}^k h(\phi)= \sum_{m=0}^{M} \lambda_m \hat{\cal{N}}^{m+k} n(\phi)
\,\,, 
\mbox{~~~~~}k=0\ldots M\,.
\end{eqnarray}
This yields
\begin{eqnarray}
  \langle \Phi | (\Delta \hat{N})^{k} \hat{H} e^{i\phi\hat{N}}|\Phi\rangle =
  \sum_{m=0}^{M} \lambda_m
  \langle \Phi | (\Delta \hat{N})^{m+k} e^{i\phi\hat{N}}|\Phi\rangle\,.
\end{eqnarray}
Taking the limit $\phi \rightarrow 0$ this results in a system 
of the $M+1$ linear equations
\begin{eqnarray}
  \langle \Phi |
  [\hat{H} - \sum_{m=0}^{M} \lambda_m (\Delta \hat{N})^{m}] 
(\Delta \hat{N})^{k} |\Phi\rangle = 0
\label{kamlahset}
\end{eqnarray}
for the constants $\lambda_m$. The case of $M=1$ corresponds
to CRHB theory discussed in Sect.\ \ref{CRHBeq}. In the following
we will use the shorthand notation $\langle \hat{O}\rangle=
\langle \Phi | \hat{O} |\Phi \rangle$.
In the case of 
$M=2$, the solution of the system of Eqs.\ (\ref{kamlahset}) 
under the particle number constraint $\langle \Delta\hat{N} \rangle = 0$
leads to the following values of $\lambda_0,\lambda_1,\lambda_2$:
\begin{eqnarray}
\lambda_0&=& \langle \hat{H} \rangle -
            \lambda_2 \langle (\Delta\hat{N})^2 \rangle\,,
\\
\lambda_1 &=&  \frac{\langle\hat{H} \Delta \hat{N} \rangle 
 -\lambda_2 \langle (\Delta \hat{N})^3 \rangle }
{\langle(\Delta\hat{N})^2\rangle}\,, 
\\
\lambda_2&=& \frac{\langle\hat{H}[ (\Delta \hat{N})^2 -
\langle(\Delta\hat{N})^2\rangle ]\rangle 
      -\langle\hat{H}\Delta \hat{N}\rangle
      \langle(\Delta\hat{N})^3\rangle / \langle(\Delta\hat{N})^2\rangle}
               {\langle(\Delta\hat{N})^4\rangle-
                \langle(\Delta\hat{N})^2\rangle^2 
                -
\langle(\Delta\hat{N})^3\rangle^2/\langle(\Delta\hat{N})^2\rangle}
\;,
\label{Lambda2}
\end{eqnarray}
where the moments of the operator $\Delta\hat{N}$ are given by
\begin{eqnarray}
\langle(\Delta\hat{N})^2\rangle &=& 2 Tr [\chi]\,, 
\\
\langle(\Delta\hat{N})^3\rangle &=& 4 Tr [\gamma \chi]\,,
\\
\langle(\Delta\hat{N})^4\rangle &=& \langle(\Delta\hat{N})^2\rangle
    + 8 Tr [\chi(1-6\chi)]\,,
\end{eqnarray}
with
\begin{eqnarray}
\chi=\rho (1-\rho)\,\,\,\,\,\,\, {\rm and} \,\,\,\,\,\,\, \gamma=1-2\rho\,.
\end{eqnarray}

Because $\hat{\cal{N}}$ is a hermitian operator with respect
to the integration on $\phi$ between 0 and $2\pi$, one gets a
very simple result for the particle number projected energy 
to order $M$
\begin{equation}
  E_{proj}^N = \sum_{m=0}^{M} \lambda_m (N -\langle \hat{N} \rangle)^m.
\end{equation}
In a full variation after projection method one should vary $E^N_{proj}$.
The Lipkin-Nogami method consists in treating $\lambda_2$ as a constant
during the variation, with its value defined according to 
Eq.\ (\ref{Lambda2}) being readjusted self-consistently at each 
iteration. 

This leads to the variational equation 
\begin{eqnarray}
  \frac{\delta}{\delta \Phi} 
  \langle \Phi | \hat{H} -\lambda_2 (\Delta \hat{N})^{2} |\Phi\rangle -
  \lambda \frac{\delta}{\delta \Phi} \langle \Phi |\hat{N}|\Phi \rangle = 0\,,
\end{eqnarray} 
which means that we have to minimize the expectation value
of the particle number projected energy within the set of 
product wave functions $|\Phi\rangle$ under subsidiary 
condition that $\lambda$ is determined by the particle 
number constraint 
\begin{eqnarray}
\langle \Phi | \hat{N}|\Phi\rangle = N\,, 
\end{eqnarray}
provided that the condition $\lambda=\lambda_1$ is
fulfilled and the terms proportional to 
$\frac{\delta}{\delta \Phi} \lambda_2$ are neglected. 
This means in particular that the Lipkin-Nogami
method violates the variational principle. An extension of
the Lipkin-Nogami method to a full variation of the Kamlah
expansion of the projected energy up to second order is very
complicated. So far it has only been carried out in
non-relativistic theories \cite{VER.99Kam}.

  The formulation of the Lipkin-Nogami method presented above
is given for one kind of nucleons (protons or neutrons). In
realistic calculations we perform it simultaneously for protons 
and neutrons. It can easily be shown that for $M=2$ there 
is no coupling between protons and neutrons.

The evaluation of the term $\lambda_2$ for a general two-particle 
interaction $V^{pp}$ is given as
\begin{eqnarray}
  \lambda_2=\frac{1}{4} \;
 \frac{
   2 \,Tr_1 Tr_1(\kappa \kappa^+\, \overline{v}\, \kappa \kappa^+)
  - Tr_2 Tr_2(\kappa^*\rho \,\overline{v}\,\sigma \kappa )}
      {
        [Tr(\kappa\kappa^+)]^2
        -2Tr(\kappa\kappa^+\kappa\kappa^+) }\,,
\label{lambda2-gen}
\end{eqnarray}
where $\sigma=1-\rho$ and 
$\overline{v}_{abcd}=\langle ab | V^{pp} | cd-dc\rangle$
is antisymmetrized matrix element of the two-particle interaction
$V^{pp}$. The trace $Tr_1$ represents the summation in the
particle-hole channel, while $Tr_2$ in particle-particle
channel \cite{RS.80}. In Relativistic Hartree-Bogoliubov theory 
with the Gogny force in the pairing channel, the sum over the 
particle-hole part is zero, since according to our definition 
the two-particle interaction $V^{pp}$ acts only between the 
fermions. In the particle-hole
channel we have only classical fields conserving the particle 
number. As a result, the $\lambda_2$ value used in the CRHB
calculations with APNP(LN) is given by
\begin{eqnarray}
  \lambda_2=-\frac{1}{4} \;
 \frac{Tr_2 Tr_2(\kappa^*\rho \,\overline{v}\,\sigma \kappa )}
      {
        [Tr(\kappa\kappa^+)]^2
        -2Tr(\kappa\kappa^+\kappa\kappa^+) }\;.
\label{lambda2-Gog}
\end{eqnarray}
Note that in the harmonic oscillator basis used presently in the 
CRHB(+LN) calculations, the density matrix $\rho$ and pairing 
tensor $\kappa$ entering into Eq.\ (\ref{lambda2-Gog}) are real. 

  In Ref.\ \cite{JKthesis} it is shown in detail that 
Eq.\ (\ref{lambda2-Gog}) can be represented by 
\begin{eqnarray}
  \lambda_2=\frac{1}{4} \;
 \frac{ E_p[\gamma \kappa]- E_p[\kappa] 
}{
        [Tr(\kappa\kappa^+)]^2
        -2Tr(\kappa\kappa^+\kappa\kappa^+) 
}\,,
\label{lambda2Ep}
\end{eqnarray}
where the unprojected pairing energy functional is given by
\begin{eqnarray}
  E_p[ \kappa]= 
  \frac{1}{4}\,Tr_2 Tr_2( \kappa^* \,\overline{v}\,\kappa ) \,.
\end{eqnarray}
Since the modified pairing tensor  $\gamma \kappa$ is much 
smaller than the pairing tensor $\kappa$, the main contribution 
to $\lambda_2$ comes from the pairing energy.

 The application of the Lipkin-Nogami method leads to a modification 
of the Hartree-Bogoliubov equations for the fermions, while the
mesonic part
of the CRHB theory is not affected. This modification is obtained by
the restricted variation of $\lambda_2 \langle (\Delta N)^2 \rangle$,
namely, $\lambda_2$ is not varied and its value is calculated 
self-consistently using Eq.\ (\ref{lambda2Ep}) in each step of
the iteration. One should note that the
form of the CRHB+LN equations is not unique (see Refs.\ 
\cite{PNL.73,SWM.94,GBDFH.94} for details). In the general case, the 
CRHF+LN equation contains a parameter $\eta\, (\eta=0,\pm1)$ and is 
given by
\begin{eqnarray}
\pmatrix{ 
  \hat{h}_D(\eta) -\lambda(\eta)-\Omega_x \hat{J_x} & \hat{\Delta}(\eta)        \cr
 -\hat{\Delta}^*(\eta) & -\hat{h}^*_D(\eta) +\lambda(\eta)-\Omega_x \hat{J_x}^*   \cr}
  \pmatrix{ U(\bff r) \cr V(\bff r)}_k
  = E_k(\eta) \pmatrix{ U(\bff r) \cr V(\bff r)}_k
\end{eqnarray}
where
\begin{eqnarray}
  \hat{h}_D(\eta) &= & \hat{h}_D  + 2 
\lambda_2\,[(1+\eta)\rho - Tr(\rho)]
\,,  \\
  \hat{\Delta}(\eta) & =& \hat{\Delta} - 
2\lambda_2(1-\eta)\kappa\,,   \\
\lambda(\eta) &=&  \lambda_1+\lambda_2\,[1 +\eta]\,, \\ 
\label{equasi}
E_k(\eta) &=&  E_k -\eta \lambda_2\,.
\end{eqnarray}
With these definitions and neglecting for simplicity
$2\lambda_2 Tr ( \rho)$ in $ \hat{h}_D(\eta)$ in the 
following discussion it is clear that the case 
of $\eta=+1$ corresponds to the shift of whole modification into 
the particle-hole channel of the CRHB+LN theory: 
$\hat{h}_D \rightarrow \hat{h}_D + 4\lambda_2 \rho$ leaving 
pairing potential $\hat{\Delta}$ unchanged. The case of 
$\eta=-1$ correspond to the shift of the modification into 
the particle-particle channel $\hat{\Delta} \rightarrow 
\hat{\Delta} - 4\lambda_2 \kappa$ leaving $\hat{h}_D$ 
unchanged. An intermediate situation is obtained in the 
case of $\eta=0$: $\hat{h}_D \rightarrow \hat{h}_D + 2\lambda_2 \rho$,
$\hat{\Delta} \rightarrow \hat{\Delta} - 2\lambda_2 \kappa$. One must 
note that the eigenvalues $E_k(\eta)$ of the CRHB+LN equations are 
identical to the quasiparticle energies $E_k$ only in the case of 
$\eta=0$. In the present calculations we are using the case of 
$\eta=+1$ which provides reasonable numerical stability of
the CRHB+LN equations.

%%%%%%%%%%%%%%%%%%%%%%%%%%%%%%%%%%%%%%%%%%%%%%%%%%%%%%%%%%%%%%%%%%
\subsection{Physical observables and details of the calculations}
%%%%%%%%%%%%%%%%%%%%%%%%%%%%%%%%%%%%%%%%%%%%%%%%%%%%%%%%%%%%%%%%%%

  Because, with a few exceptions, the observed SD bands are not linked 
to the low-spin level schemes, the absolute angular momentum quantum 
numbers $I$ are not known experimentally. Thus the dynamic moment of 
inertia $J^{(2)}$ which contains only the differences $\Delta I=2$
plays an important role in our understanding of their structure.  In 
the calculations, the rotational frequency ${\Omega}_x$, the kinematic 
moment of inertia $J^{(1)}$ and the dynamic moment of inertia
$J^{(2)}$ are defined by
\begin{eqnarray}
{\Omega}_x=\frac{dE}{dJ},\qquad 
J^{(1)}({\Omega}_x)=J\left\{\frac{dE}{dJ}\right\}^{-1}
=\frac{J}{\Omega_x},\qquad 
J^{(2)}({\Omega}_x)=\left\{\frac{d^2E}{dJ^2}\right\}^{-1}
=\frac{dJ}{d\Omega_x}.
\label{}
\end{eqnarray}
Experimental quantities such as a the rotational frequency, 
the kinematic and dynamic moments of inertia are extracted from 
the observed 
energies of $\gamma$-transitions within a band according to the
prescription given in Sect.\ 4.1 of Ref. \cite{PhysRep}. One 
should note that the kinematic moment of inertia depends on the 
absolute values of the spin which in most cases are not know in 
SD bands.

 The charge quadrupole $Q_0$ and mass hexadecupole $Q_{40}$
moments are calculated by using the expressions
\begin{eqnarray}
Q_0&=&e\sqrt{\frac{16\pi}{5}}
\sqrt{\left\langle r^2Y_{20}\right\rangle_p^2
+2\left\langle r^2Y_{22}\right\rangle_p^2}
\label{Q_0}\\
Q_{40}&=&\left\langle r^4Y_{40}\right\rangle _p+\left\langle
r^4Y_{40}\right\rangle _n
\label{Q_40}
\end{eqnarray}
where the labels $p$ and $n$ are used for protons and neutrons, 
respectively and $e$ is the electrical charge. The transition 
quadrupole moment $Q_t$ for a triaxially deformed nucleus is 
calculated by the following expression \cite{RHHEG.82,NR.96}
\begin{eqnarray}
Q_t = e \sqrt{\frac{16\pi}{5}} \left\langle r^2Y_{20}\right\rangle_p
\,\,\frac{\cos{(\gamma +30^{\circ})}}{\cos(30^{\circ})}
\end{eqnarray}
where the $\gamma$-deformation (of the proton subsystem) is defined 
by
\begin{eqnarray}
\tan \gamma = \frac{\left\langle r^2Y_{22}\right\rangle_p}
{ \left\langle r^2Y_{20}\right\rangle_p}
\end{eqnarray}
In the limit $\gamma \rightarrow 0$ which is typical for SD bands
in the $A\sim 190$ region, $Q_t$ is almost identical to $Q_0$. The 
comparison between calculated transition quadrupole moments and available 
experimental data has been performed in Ref.\ \cite{A190} and thus 
it will not be repeated here. They agree with each other if the 
uncertainties due to stopping powers are taken into account in the
experimental data. In addition, it was concluded that more accurate 
and consistent experimental data on $Q_t$ is needed in order to 
carry out such a comparison in more detail.

   In the absence of experimentally known spins, an effective 
alignment approach \cite{Rag91,Rag93,ALR.98} plays an extremely 
important role in the definition of the structure and relative 
spins of SD bands. The effective alignment of two bands (A and B) 
is simply the difference between their spins at constant rotational 
frequency $\Omega_x$ \cite{Rag91,Rag93,ALR.98}:
\begin{eqnarray} 
i_{eff}^{\rm B,A}(\Omega_x)= I^{\rm B}(\Omega_x) - I^{\rm A}(\Omega_x)
\label{effalign}
\end{eqnarray}
 Experimentally, $i_{eff}$ includes the effects associated
with the change of the number of the particles and the relevant 
changes in the alignments of the single-particle orbitals, 
in the deformation, in pairing etc. between two bands. This physical
observable has been used frequently in unpaired calculations
for the configuration and spin assignments of SD bands in 
the $A\sim 60$ and $A\sim 140-150$ mass regions (see Refs.\ 
\cite{A60,ALR.98} and references therein) and for the study of 
relative properties of smooth terminating bands in the $A\sim 110$ mass 
region \cite{Ieff}. The effective 
alignment approach exploits the fact that the spin is quantized, 
integer for even nuclei and half-integer for odd nuclei and that 
it is furthermore constrained by signature. One should note 
that with the configurations and specifically the signatures 
fixed, the relative spins of observed bands can only change 
in steps of $\pm 2 \hbar \cdot n$, where $n$ is an integer 
number.

  The excitation energies of SD bands relative to the ground 
state are known definitely in $^{194}$Hg and $^{194}$Pb and 
tentatively in $^{192}$Pb. The RMF theory with a somewhat simplistic 
pairing and no particle number projection excellently reproduces 
these data \cite{LR.98}. It would be interesting to see how this result
will be changed when the Gogny force is used in the pairing 
channel and APNP(LN) is performed. Considering, however, that 
such an investigation will require a constraint on the quadrupole
moment and thus will be extremely time consuming within
a three-dimensional CRHB computer code, we will leave this question 
open for future studies.

 The CRHB-equations are solved in the basis of an anisotropic 
three-dimensional harmonic oscillator in Cartesian coordinates. 
The same basis deformation $\beta_0=0.5$, $\gamma=0^{\circ}$ and oscillator
frequency $\hbar \omega_0=41$A$^{-1/3}$ MeV have been used
for all nuclei. All fermionic and bosonic states belonging 
to the shells up to $N_F=14$ and $N_B=16$ are taken into 
account in the diagonalization of the Dirac equation
and the matrix inversion of the Klein-Gordon equations, 
respectively. The detailed investigations performed for
$^{194}$Hg indicate that this truncation scheme provides 
reasonable numerical accuracy. The values of the kinematic 
moment of inertia $J^{(1)}$ and the transition quadrupole 
moment $Q_t$ obtained with such a truncation of the basis
are different from the ones obtained in the fermionic
basis with $N_F=17$ by less than 1\%. The accuracy of
the calculated mass hexadecupole moment $Q_{40}$ is a 
bit lower being $\sim 1.5$\%. The numerical accuracy of the
calculations of the physical observables within the employed 
basis has also been tested using different combinations
of $\beta_0$ and $\hbar \omega_0$, namely $\beta_0=0.4$, 
0.5 and 0.6 as well as $\hbar \omega_0=38A^{-1/3}$,  
$41A^{-1/3}$, $44A^{-1/3}$, $47A^{-1/3}$, $50A^{-1/3}$, 
$53A^{-1/3}$ MeV. A similar level of accuracy has been 
obtained for $J^{(1)}$,  but the new estimations of  
accuracy for the calculation of $Q_t$ and $Q_{40}$ are 
around 2\% and 3\%, respectively. The numerical errors 
for the total energy (in \%) are even smaller.

At $\Omega_x=0.0$ MeV, the single-particle orbitals are 
labeled by means of the asymptotic quantum numbers 
$[N n_z \Lambda]\Omega$ (Nilsson quantum numbers) of the
dominant component of the wave function and superscripts to 
the orbital labels (e.g. $[651]1/2^+$) are used to indicate 
the sign of the signature $r$ for that orbital $(r=\pm i)$.

%%%%%%%%%%%%%%%%%%%%%%%%%%%%%%%%%%%%%%%%%%%%%%%%%%%%%%%%%%%%%%
\section{The impact of pairing and particle number projection}
\label{sect-pair}
%%%%%%%%%%%%%%%%%%%%%%%%%%%%%%%%%%%%%%%%%%%%%%%%%%%%%%%%%%%%%%

  In the present section, the impact of pairing and approximate
particle number projection will be studied in detail on the example 
of the yrast SD bands observed in $^{192}$Hg and $^{194}$Pb. In order 
to differentiate different types of calculations, the following 
abbreviations are introduced: 
\begin{itemize}
\item
CRMF - cranked relativistic mean 
field calculations without pairing correlations, 

\item
CRHB - cranked 
relativistic Hartree+Bogoliubov calculations without particle 
number projection, 

\item
CRHB+LN - cranked relativistic Hartree+Bogoliubov 
calculations with approximate particle number projection by means 
of the Lipkin-Nogami method (APNP(LN)). 
\end{itemize}
In the following the lines 
representing the results of different CRHB(+LN) calculations will 
be labelled in the figures by 
\begin{eqnarray}
   {\rm \bf RMFset} + f*{\rm \bf Gogset} + {\rm \bf LN}  
\nonumber
\end{eqnarray}
where {\bf RMFset} and {\bf Gogset} are the RMF and Gogny 
forces used in the calculations, $f$ is the scaling factor
for the Gogny force (see Eq.\ (\ref{Vpp})). Note that $f$ will
in general be omitted when $f=1.0$. {\bf LN} is used only
in the cases when APNP(LN) is performed in the calculations. 
The lines showing the results of the CRMF calculations will 
be labelled by {\bf RMFset [CRMF]}. 

\begin{figure}
\epsfxsize 12.0cm
\epsfbox{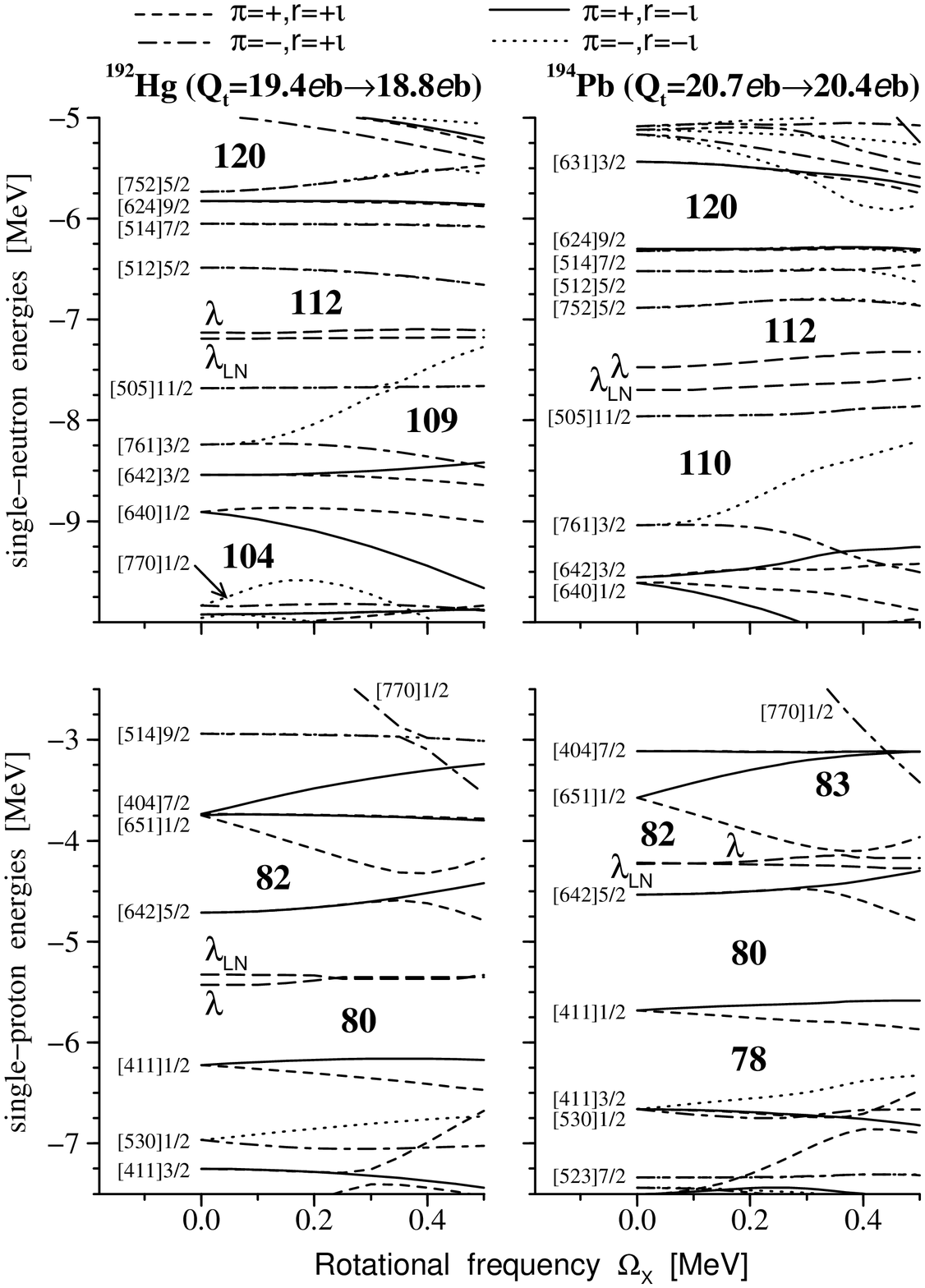}
\caption{Neutron (top) and proton (bottom) single-particle
energies (routhians) in the self-consistent rotating potential
as a function of rotational frequency $\Omega_x$. They are 
given along the deformation path of the lowest SD configurations 
in $^{192}$Hg (left panels) and $^{194}$Pb (right panels) obtained 
in the calculations neglecting pairing and with the NL1 
parametrization of the RMF Lagrangian. The calculated transition 
quadrupole moments $Q_t$ are shown on the top of the figure as 
$(Q_t= Q_t(\Omega_x=0.0) \rightarrow Q_t(\Omega_x=0.5))$.
The notation of lines used for routhians is also indicated on 
the top of the figure. Chemical potentials $\lambda$ 
($\lambda_{\rm LN}$) obtained in the paired calculations
without (with) particle number projection by means of the
Lipkin-Nogami method are shown by long-dashed lines.}
\label{routh-np}
\end{figure}

\begin{figure}
\epsfxsize 16.0cm
\epsfbox{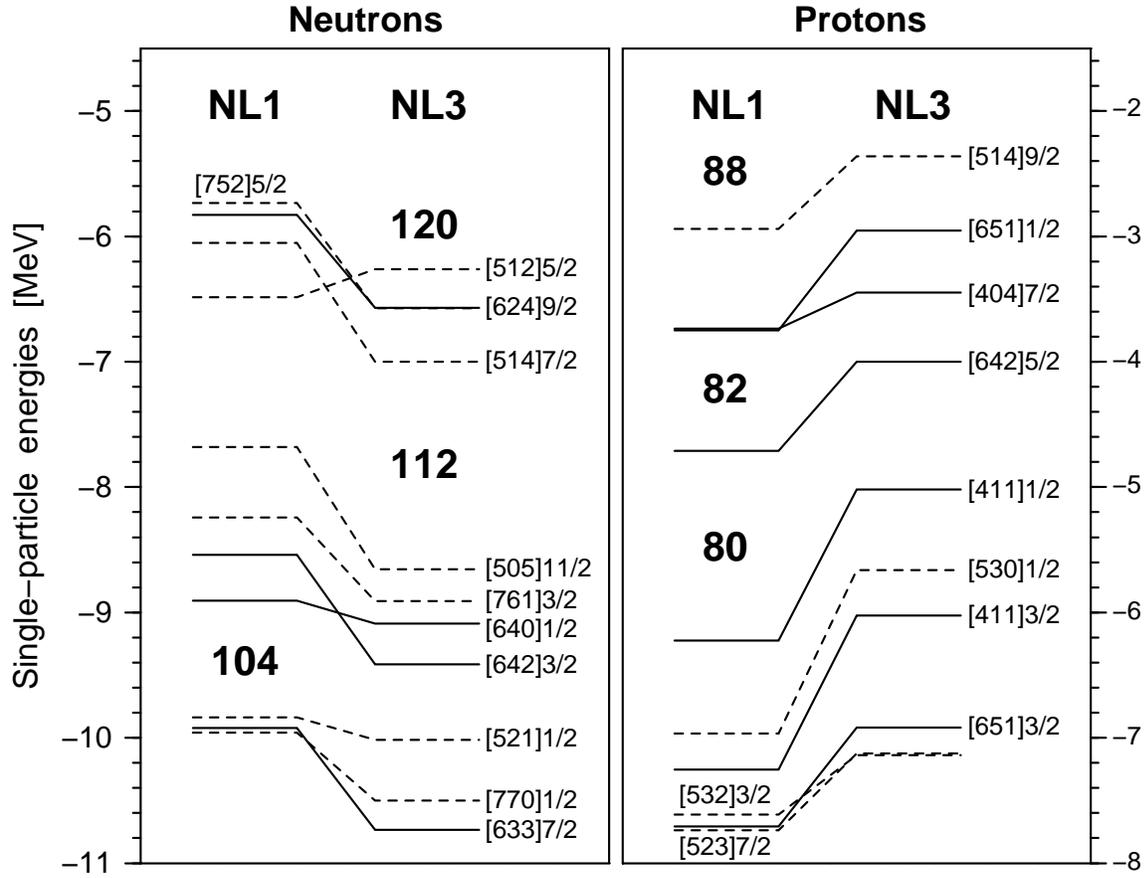}
\caption{The single-particle states around the $Z=80$ and
$N=112$ SD shell gaps obtained in the calculations with NL1 
and NL3 and with no pairing at the corresponding equilibrium 
deformations of the lowest SD configuration in $^{192}$Hg. 
The single-particle orbitals are labeled by means of the 
asymptotic quantum numbers $[N n_z \Lambda]\Omega$ (Nilsson 
quantum numbers) of the dominant component of the wave 
function. Solid and dashed lines are used for the positive 
and negative parity states, respectively. The calculated
charge quadrupole $Q_0$ and mass hexadecupole $Q_{40}$
moments are $Q_0=19.44$ $e$b, $Q_{40}=17.49 \times 10^3$ fm$^4$
(NL1) and $Q_0=19.11$ $e$b, $Q_{40}=17.04 \times 10^3$ fm$^4$.
}
\label{nl1-vr-nl3}
\end{figure}

 We start from the results of the CRMF calculations. Neutron and 
proton single-routhian diagrams calculated with the parameter 
set NL1 (see Table \ref{RMFsets}) and drawn along the deformation 
path of the yrast SD configurations in these two nuclei are shown 
in Fig.\ \ref{routh-np}. The shell structure of $^{192}$Hg is 
dominated by the large $Z=80$ and $N=112$ SD shell gaps. The same 
shell gaps appear also in the calculations with the set NL3, see 
Fig.\ \ref{nl1-vr-nl3}, but compared with the set NL1 the $Z=80$ SD 
shell gap is somewhat smaller while the $N=112$ SD shell gap is 
more pronounced. Similar shell gaps are seen also in $^{194}$Pb
(see right panels of Fig.\ \ref{routh-np}). Considering that
the $Z=82$ SD shell gap decreases with increasing rotational 
frequency due to down-sloping $\pi [651]1/2^+$ orbital, it
is more reasonable to consider the nucleus $^{192}$Hg as a doubly
magic in this mass region.  
While the NL1 and NL3 forces give the same set of single-particle 
orbitals in the vicinity of magic SD shell gaps, their relative 
ordering (energies) is somewhat different. As studied in detail 
in Ref.\ \cite{ALR.98} (Section 6.4), the difference in the 
single-particle energies at superdeformation when different RMF 
parametrizations are used is to a great extend connected with 
their energy differences  at spherical shape. Note that the proton
single-particle spectra in the vicinity of the $Z=80$ SD shell
gap in the $A\sim 190$ mass region reveal large similarities
with the neutron single-particle spectra in the vicinity of the 
$N=80$ SD shell gap in the $A\sim 140-150$ mass region, 
see Fig.\ \ref{nl1-vr-nl3} in the present article and Fig. 17
in Ref.\ \cite{ALR.98}. 

 Figs.\ \ref{routh-np} and \ref{nl1-vr-nl3} show some similarities 
and differences with the results obtained in the non-relativistic
calculations with the Wood-Saxon potential (see Fig.\ 3 in Ref.\ 
\cite{Ndiag-WS}), Skyrme forces (see Figs.\ 6 and 7 in 
Ref.\ \cite{GBDFH.94}) and Gogny forces (see Fig.\ 5 in 
Ref.\ \cite{LGD.99}). Both in relativistic and non-relativistic 
calculations a similar shell structure and a similar set of
single-particle states (although their relative energies
are different) appear in the vicinity of the SD shell gaps.
The main difference with non-relativistic calculations is 
related to the larger size of the SD shell gaps and to the 
lower level density in the vicinity of these gaps. These features
are connected with low effective mass of RMF theory, see the
discussion in Ref.\ \cite{AKR.96} for details.

\begin{figure}
\epsfxsize 16.0cm
\epsfbox{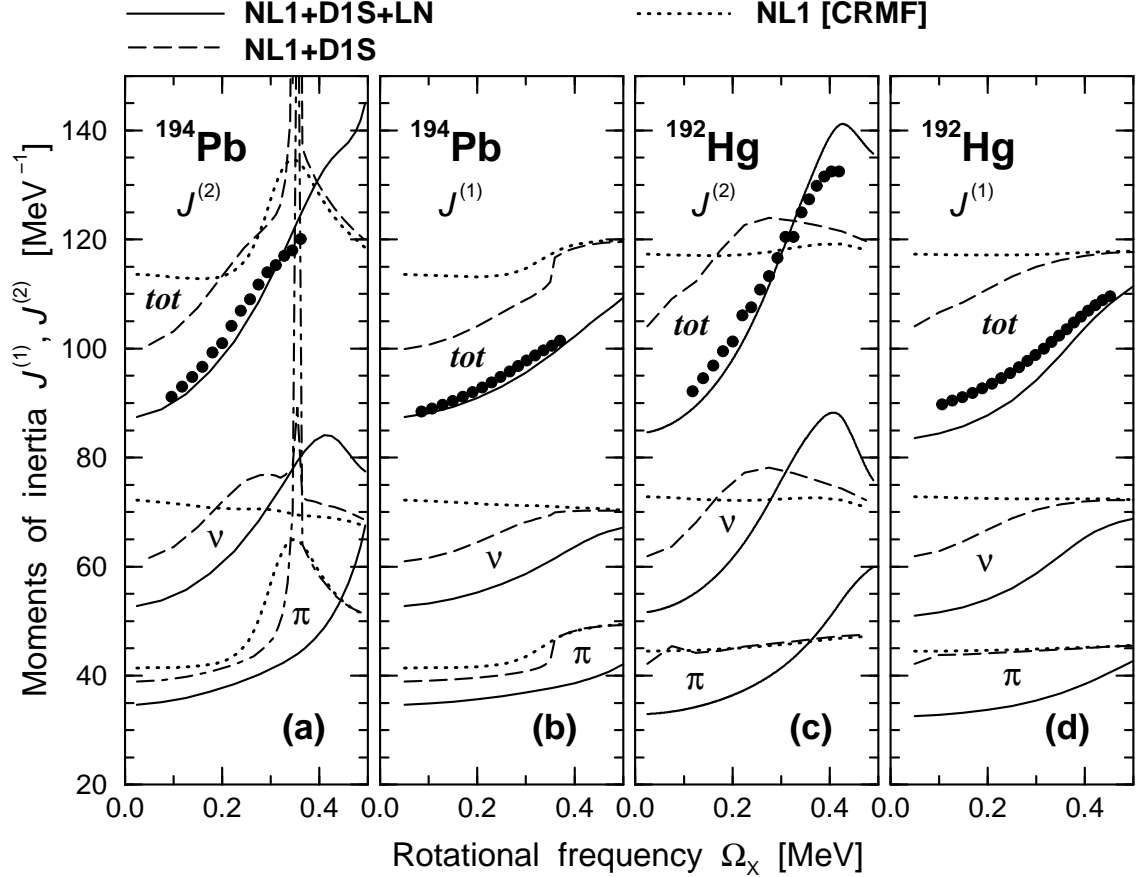}
\vspace{0.0cm}
\caption{The dependence of calculated kinematic ($J^{(1)}$)
and dynamic ($J^{(2)}$) moments of inertia on the pairing
and particle number projection illustrated on the example
of the lowest SD configurations in $^{194}$Pb and
$^{192}$Hg. Experimental data are shown by solid unlinked
symbols.  The calculations have been performed with
no pairing (indicated as NL1 [CRMF]), without particle number 
projection (NL1+D1S) and with approximate particle number 
projection by means of the Lipkin-Nogami method (NL1+D1S+LN).
Proton and neutron contributions to the kinematic and dynamic
moments of inertia are indicated. Note that in   
panel (a) the proton contribution to $J^{(2)}$ is shown by 
a dash-dotted line for the case of NL1+D1S.}
\label{j2j1-unpr}
\end{figure}

As can be seen in Fig.\ \ref{j2j1-unpr}, the CRMF calculations 
do not reproduce the experimental kinematic and dynamic 
moments of inertia. In the case of $^{192}$Hg, the calculated values 
of $J^{(1)}$ and $J^{(2)}$ are almost constant, while in the case 
of $^{194}$Pb the unpaired proton band crossing is clearly seen. 
It originates from the interaction between $\pi[642]5/2^+$ and 
$\pi[651]1/2^+$ orbitals, see bottom panels in Fig.\ \ref{routh-np}. 
The orbital $\pi[642]5/2^+$ is occupied before band crossing in the
lowest SD configuration, while the orbital $\pi[651]1/2^+$ is occupied
after band crossing. Note that in the $A\sim 150$ mass region, the 
crossings observed in some bands of the nuclei around $^{147}$Gd 
have been attributed to the unpaired interaction of these two orbitals, 
but in the neutron subsystem, see Refs.\ \cite{AKR.96,Haas}.

\begin{figure}
\epsfxsize 16.0cm
\epsfbox{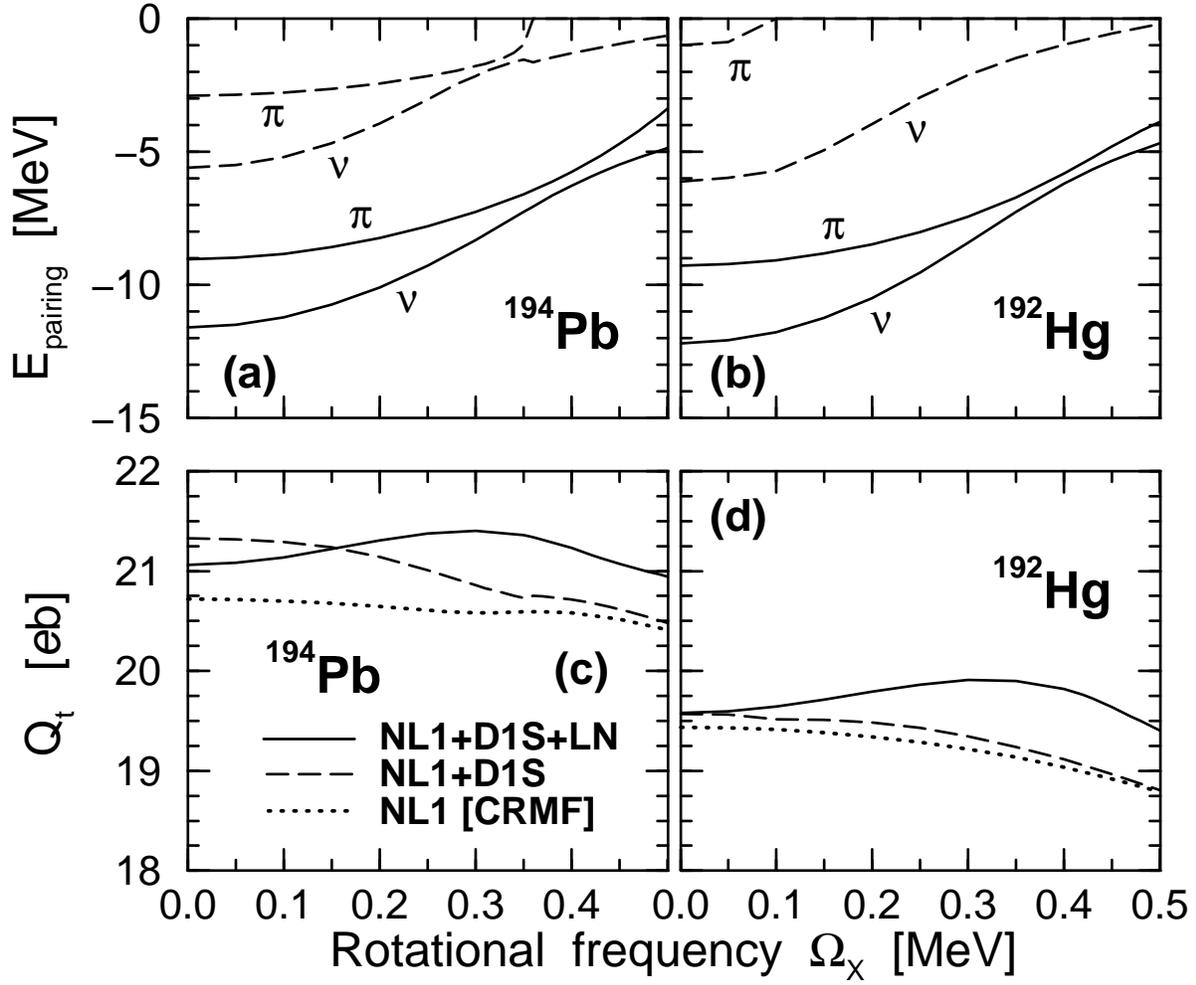}
\caption{Transition quadrupole moments $Q_t$ and neutron 
and proton pairing energies $E_{pairing}$ of the lowest 
SD configurations in $^{194}$Pb and $^{192}$Hg, see 
caption of Fig.\ \protect\ref{j2j1-unpr} for details.
In panels (a) and (b), the letters $\nu$ and $\pi$ are 
used in order to indicate neutron and proton pairing 
energies, respectively.}
\label{qtep-unpr}
\end{figure}

  The inclusion of pairing without particle number projection 
somewhat improves the agreement with experiment, see Fig.\ 
\ref{j2j1-unpr}. Indeed, the slope of the calculated kinematic
and dynamic moments of inertia at low rotational frequencies
is coming closer to experiment, but still the disagreement
is considerable. In addition, a proton pairing collapse
takes place in the CRHB calculations. This collapse is calculated 
in $^{194}$Pb at $\Omega_x \sim 0.35$ MeV and in $^{192}$Hg at 
$\Omega_x \sim 0.10$ MeV, see Figs.\ \ref{qtep-unpr}a and 
\ref{qtep-unpr}b. In the case of $^{194}$Pb
it is correlated with the alignment of the $\pi [651]1/2^+$
orbital which reveals itself in a sharp increase of $J^{(2)}$,
see Fig.\ \ref{j2j1-unpr}a. This crossing coincides with
the one seen in the CRMF calculations. At higher 
frequencies, the calculated kinematic and dynamic moments of 
inertia for the proton subsystem and the transition quadrupole moments 
obtained in the CRMF and CRHB calculations are very close, see 
Figs.\ \ref{j2j1-unpr} and \ref{qtep-unpr}c,d, respectively. 
However, at lower frequencies proton pairing is still 
important and the results of the calculations with and without 
pairing for proton $J^{(1)}$ and $J^{(2)}$ values are different. 
Contrary to the proton subsystem, the pairing correlations 
in the neutron subsystem do not collapse in the rotational frequency 
range under consideration. Neutron pairing energies $E_{pairing}^{\nu}$, 
being larger in absolute value than the proton ones at 
$\Omega_x=0.0$ MeV, start around $-6$ MeV and then smoothly 
decrease in absolute value with increasing rotational frequency 
coming close to 0 MeV at $\Omega_x=0.5$ MeV, see Figs.\ 
\ref{qtep-unpr}a,b. At low and medium rotational frequencies, 
there is a considerable difference between the corresponding neutron 
moments of inertia ($J^{(1)}$ or $J^{(2)}$) obtained in the 
CRMF and CRHB calculations, see Fig.\ \ref{j2j1-unpr}. This 
difference, however, is  small at the highest rotational 
frequencies, where the magnitude of neutron pairing 
correlations is negligible. 

 Note that neutron pairing correlations alone do not lead to 
a considerable modification of transition quadrupole moments 
$Q_t$ compared with the CRMF calculations, see an example 
of $^{192}$Hg (Figs.\ \ref{qtep-unpr}b,d). On the contrary,
in $^{194}$Pb, where both neutron and proton pairing persist
up to $\Omega_x \sim 0.35$ MeV, the transition quadrupole 
moment $Q_t$ obtained in the CRHB calculations is larger
than the one of the CRMF calculations.

  The inclusion of approximate particle number projection by 
means of the Lipkin-Nogami method leads to a very good agreement 
between the calculated and experimental kinematic and dynamic 
moments of inertia, see Fig.\ \ref{j2j1-unpr}. The correlations 
induced by the Lipkin-Nogami prescription compared with CRHB 
produce the desired effects, namely, to diminish the kinematic 
moments of inertia at all frequencies, to diminish the dynamic 
moments of inertia at low and medium rotational frequencies 
and to delay proton and neutron alignments to higher frequencies. 
These effects are due to stronger pairing correlations seen in 
CRHB+LN compared with CRHB (Figs.\ \ref{qtep-unpr}a,b). Contrary 
to the CRHB approach where the proton pairing collapses at some 
frequencies, no such collapse appears in the CRHB+LN calculations 
in the rotational frequency range under consideration. Thus the 
proton and neutron subsystems remain correlated even at 
$\Omega_x=0.5$ MeV. The decrease of the 
magnitude of pairing correlations with increasing rotational 
frequency seen both in CRHB and CRHB+LN is due to a strong Coriolis 
anti-pairing effect. The transition quadrupole moments $Q_t$ 
calculated in CRHB+LN show a different behavior as a function of 
rotational frequency compared with CRMF and CRHB, see 
Figs.\ \ref{qtep-unpr}c,d. While in the latter approaches, the 
$Q_t$ values decrease with increasing rotational frequency, 
they initially increase and then decrease in the CRHB+LN 
approach. The calculated decrease of $Q_t$ is due to an
anti-stretching effect caused by the Coriolis force, but the 
differences in the behavior of $Q_t$ as a function of 
$\Omega_x$  should be attributed to different strengths of 
the pairing correlations in the CRHB and CRHB+LN calculations.

\begin{figure}
\epsfxsize 14.0cm
\epsfbox{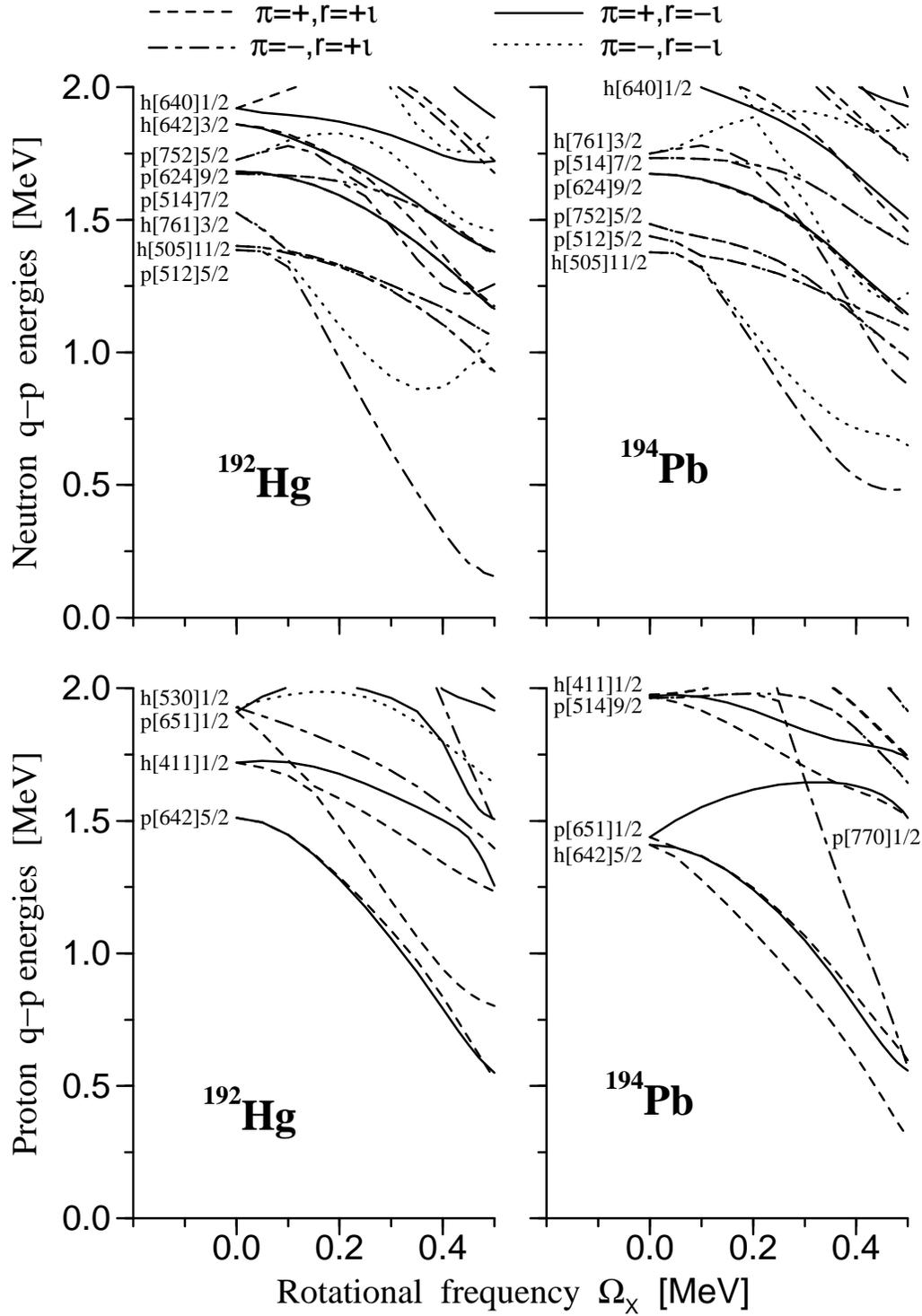}
\caption{Neutron (top panels) and proton 
(bottom panels) quasiparticle energies corresponding to the 
lowest SD configurations in $^{192}$Hg and $^{194}$Pb. The 
calculations have been performed in the NL1+D1S+LN scheme. 
The notation of the lines is given in the figure. The letters
'p' and 'h' before the Nilsson labels are used to indicate
whether a given routhian is of particle of hole type.}
\label{qpe-routh}
\end{figure}

  The increase of the kinematic and dynamic moments of inertia
is a complex effect which predominantly includes
the gradual alignments of the pairs of $j_{15/2}$ neutrons
and $i_{13/2}$ protons and the decreasing pairing correlations
with increasing rotational frequency. The quasiparticle
routhian diagrams of Fig.\ \ref{qpe-routh} shows the
typical quasiparticle spectra obtained in the CRHB+LN
calculations and the alignments in the above discussed pairs. 
Note that the alignment of the proton pair takes place at a 
higher rotational frequency than the one of the neutron pair. 
It is reasonable to expect that the behavior of the total 
dynamic moment of inertia will sensitively depend on the 
balance of the alignments in the proton and neutron
subsystems. Indeed, the total $J^{(2)}$ in $^{194}$Pb
does not show the decrease above the frequency of full
alignment of the neutron pair, while such a decrease is 
seen in $^{194}$Hg (see Figs.\ \ref{j2j1-unpr}a,c).
 
\begin{figure}
\epsfxsize 10.0cm
\epsfbox{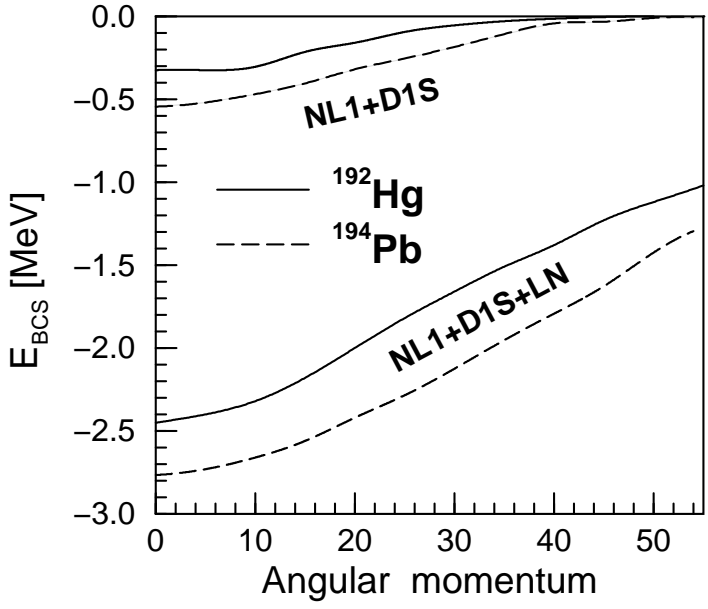}
\caption{Total (proton + neutron) BCS-like pairing energies 
obtained for the lowest SD configurations in $^{192}$Hg and
$^{194}$Pb in the calculations with (NL1+D1S+LN) and without 
(NL1+D1S) approximate particle number projection.}
\label{bcs}
\end{figure}

   Total (proton + neutron) BCS-like pairing energies $E_{\rm BCS}$ 
(see Eq.\ (\ref{Ebcs})) obtained for the lowest SD configurations in 
$^{192}$Hg and $^{194}$Pb in the calculations with and without 
APNP(LN) are shown in Fig.\ \ref{bcs}. These quantities have 
more physical content than $E_{pairing}$ defined in Eq.\ (\ref{Epair})
since they directly show the gain in binding energy due to the
pairing correlations. The comparison with Figs.\ \ref{qtep-unpr}a,b
allows to conclude that $E_{\rm BCS}$ is smaller by roughly 
an order of magnitude than 
$E_{pairing}^{tot}=E_{pairing}^{\nu}+E_{pairing}^{\pi}$.
The CRHB calculations show rather
small BCS-like pairing energies $E_{\rm BCS}$ which slowly 
converge to zero with increasing spin and almost vanish 
already at $I\sim 30-35\hbar$. Similar to $E_{pairing}$,
APNP(LN) significantly (by factor $5-10$) increases the
size of the  BCS-like pairing energies. Although these energies
decrease with increasing spin reflecting the quenching
of the pairing correlations, they do not vanish even at
highest calculated spins.

\begin{figure}
\epsfxsize 16.0cm
\epsfbox{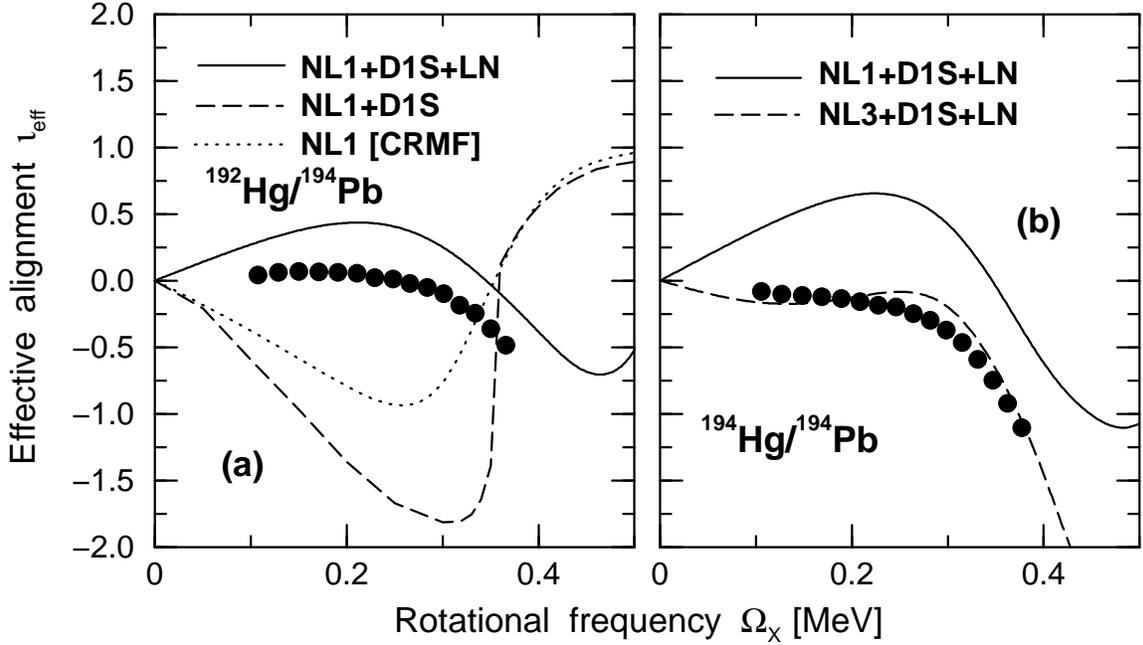}
\vspace{0.0cm}
\caption{The dependence of the effective alignment $i_{eff}$ on the 
pairing and the particle number projection (panel (a)) and 
on the parametrization of the RMF Lagrangian (panel (b)). In 
both panels, the experimental data are shown by unlinked 
solid circles, while the results of the calculations by 
lines of different type. The experimental effective alignment 
between bands A and B is indicated as ``A/B''. Band A in the 
lighter nucleus is taken as a reference, so the effective 
alignment measures the effect of 
the additional particle(s). In panel (a), the results
of the calculations with no pairing (indicated as NL1 
[CRMF]), without particle number projection (NL1+D1S) and 
with approximate particle number projection by means of the
Lipkin-Nogami method (NL1+D1S+LN) are compared with the
experimental effective alignment in the $^{192}$Hg/$^{194}$Pb
pair. Panel (b) compares effective alignments in the 
$^{194}$Hg/$^{194}$Pb pair obtained with NL1 and NL3 forces 
with experimental data.}
\label{efaldep}
\end{figure}

  The effective alignments $i_{eff}$ between the lowest SD 
configurations in $^{192}$Hg and $^{194}$Pb calculated in 
CRMF, CRHB and CRHB+LN are shown in Fig.\ \ref{efaldep}a and 
are compared with experiment. It is clearly seen that the CRMF 
and especially the CRHB results deviate considerably from 
experiment in absolute value. In addition, the considerable change of the 
slope of $i_{eff}$ obtained in these calculations at 
$\Omega_x\approx 0.3-0.35$ MeV, which is due to proton 
band crossing, contradicts to experimental data. 
Similar to the moments of inertia, APNP(LN) considerably 
improves the agreement between calculations and experiment 
also for the effective alignments. Although the values 
$i_{eff}$ calculated in CRHB+LN deviate by $0.4-0.5\hbar$ 
from experiment, this deviation should not be considered
as crucial since the compared bands differ by 2 protons.
Moreover, the slope of $i_{eff}$ as a function of 
$\Omega_x$ is rather well reproduced in the 
calculations.  

%%%%%%%%%%%%%%%%%%%%%%%%%%%%%%%%%%%%%%%%%%%%%%%%%
\section{The dependence of the results on the
parametrization of the mean field and pairing 
force.}
\label{par-dep}
%%%%%%%%%%%%%%%%%%%%%%%%%%%%%%%%%%%%%%%%%%%%%%%%%

%%%%%%%%%%%%%%%%%%%%%%%%%%%%%%%%%%%%%%%%%%%%%%%%%
\subsection{The dependence of the results on the 
parametrization of the RMF Lagrangian.}

%%%%%%%%%%%%%%%%%%%%%%%%%%%%%%%%%%%%%%%%%%%%%%%%%

\begin{figure}
\epsfxsize 16.0cm
\epsfbox{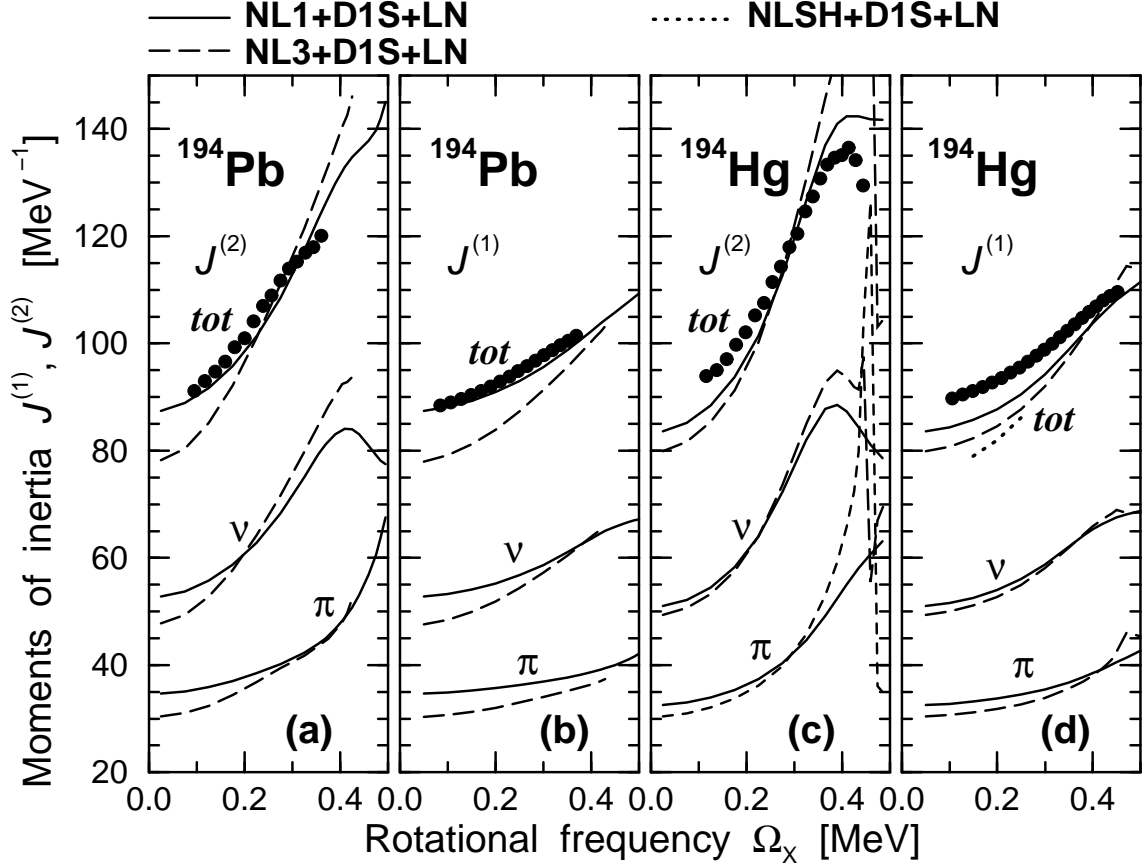}
\vspace{0.0cm}
\caption{The dependence of the kinematic $(J^{(1)})$ and 
dynamic $(J^{(2)})$ moments 
of inertia of the lowest SD configurations in $^{194}$Pb and 
$^{194}$Hg on the parametrization of the RMF Lagrangian. The 
calculations have been performed with the sets NL1, NL3 and 
NLSH. In all calculations, the D1S set of parameters has been
used for the Gogny force and approximate particle number projection
has been performed by means of the Lipkin-Nogami method.
Due to slow convergence the calculations 
with the NLSH set have been done only in the case of $^{194}$Hg 
and only in the rotational frequency range $0.15-0.25$ MeV. In 
this frequency range the dynamic moment of inertia calculated
with NLSH merge together with the ones obtained with NL3
and NL1 and thus it is not shown. No convergence is obtained
in the calculations with NL3 above $\Omega_x >0.43$ MeV in the
case of $^{194}$Pb. Note that in the case of $^{194}$Hg, 
the calculations with the NL3 force considerably overestimate
the experimental dynamic moment of inertia in the band crossing
region: the calculated value of $J^{(2)}$ reaches 202 MeV$^{-1}$ 
at 0.445 MeV. In panel (c), the proton contribution to 
$J^{(2)}$ is shown by a short-dashed line in order to make it more 
visible.}
\label{j2j1-nl3}
\end{figure}

\begin{figure}
\epsfxsize 16.0cm
\epsfbox{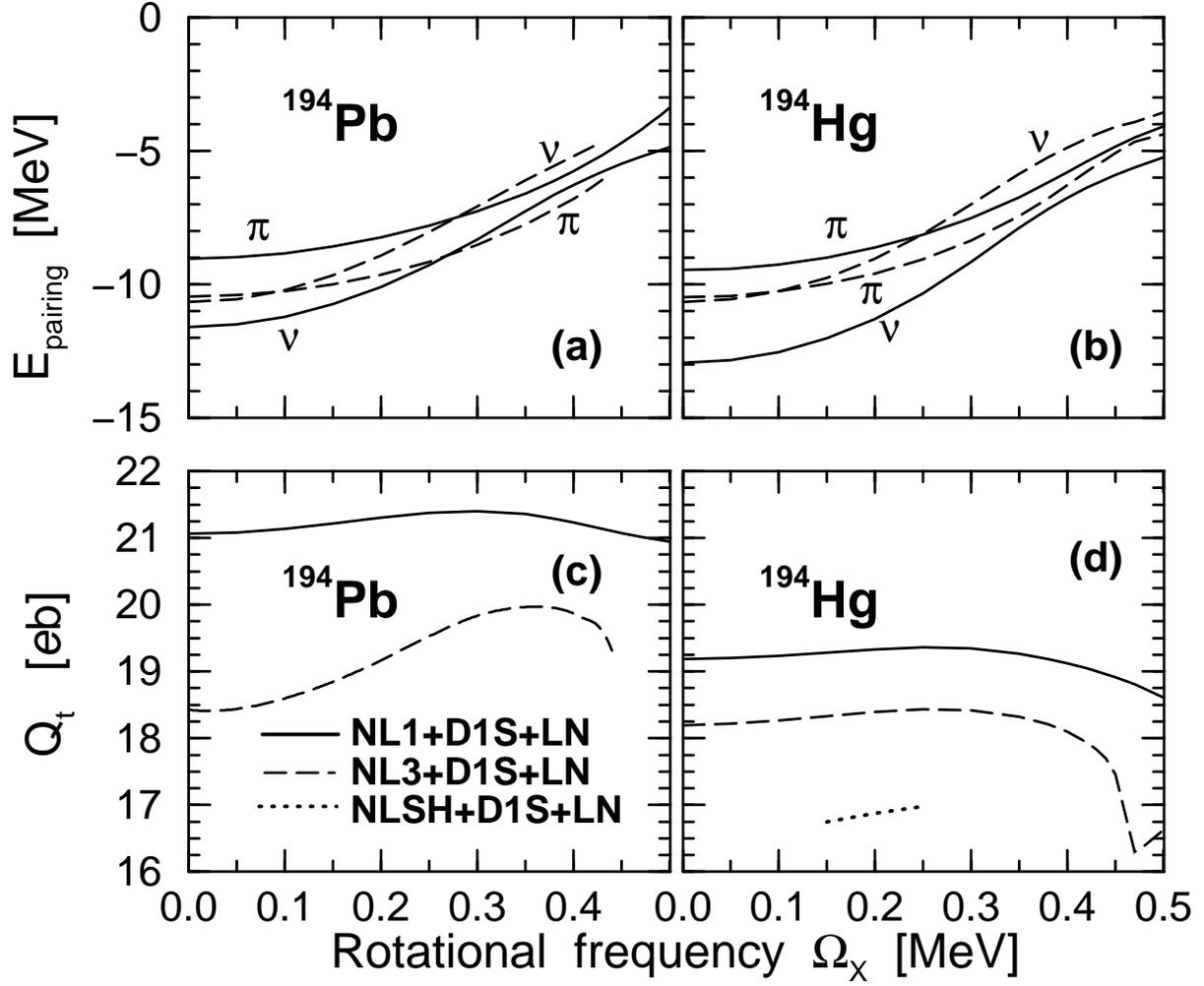}
\caption{Transition quadrupole moments $Q_t$ and neutron 
and proton pairing energies $E_{pairing}$ of the lowest 
SD configurations in $^{194}$Hg and $^{194}$Pb shown in
Fig.\ \protect\ref{j2j1-nl3}. The notation of lines is 
given there.}
\label{qtep-nl3}
\end{figure}

 The pairing correlations depend not only on the properties of the 
effective pairing force, but also on the single-particle level
density. A full relativistic Hartree-Bogoliubov calculation is 
therefore only meaningful if the Hartree field yields a reasonable 
single-particle spectrum \cite{RS.80}. In order to investigate the 
dependence of the results on the parametrization of the RMF Lagrangian,
the CRHB+LN calculations have been performed with the NL3 force 
\cite{NL3} for the 
lowest SD configurations in $^{194}$Pb and $^{194}$Hg, see Figs.\ 
\ref{j2j1-nl3} and \ref{qtep-nl3}. In addition, the NLSH force \cite{NLSH} 
has been employed in $^{194}$Hg, but due to slow convergence it was 
used only in a short frequency range. In all these calculations, 
the D1S set has been used for the Gogny force.

 At low rotational frequencies, total, neutron and proton
kinematic and dynamic moments of inertia obtained in the 
calculations with NL3 are smaller than the corresponding 
quantities calculated with NL1, see Fig.\ \ref{j2j1-nl3}.
However, the increase of $J^{(2)}$ and $J^{(1)}$ as a function 
of rotational frequency is larger in the calculations with NL3 
compared with the ones employing NL1. Thus at some frequencies, 
they reach each other. At even higher frequencies, the $J^{(2)}$ and 
$J^{(1)}$ values calculated with NL3 become larger than the ones 
calculated with NL1. It is also clear that the NL1 force provides 
better agreement with experimental kinematic and dynamic moments
of inertia than NL3. The results of the calculations with NLSH 
(see Fig.\ \ref{j2j1-nl3}d) are in even larger disagreement with 
experiment than the ones with NL3. One should note however that
the NL3 force provides better reproduction of effective alignment 
in the $^{194}$Hg/$^{194}$Pb pair compared with the NL1 force,
see Fig.\ \ref{efaldep}b. 

 In the NL1 parametrization, the $Q_t$ values are nearly 
constant as a function of $\Omega_x$ both in $^{194}$Hg and 
$^{194}$Pb nuclei. The $Q_t$($^{194}$Pb) is approximately 2$e$b 
larger than $Q_t$($^{194}$Hg). The situation is different when 
NL3 force is used in the calculations. At $\Omega_x \sim 0$ MeV, 
$Q_t$($^{194}$Pb) is approximately equal to $Q_t$($^{194}$Hg). 
This fact is most likely related to the $N=112$ SD shell gap 
which is more pronounced in the NL3 parametrization 
(see Fig.\ \ref{nl1-vr-nl3} and discussion in 
Sect.\ \ref{sect-pair}).
However, with increasing rotational frequency $Q_t$($^{194}$Pb) 
increases considerably with a maximum gain of $\approx 1.5$ $e$b at
$\Omega_x\sim 0.35$ MeV (Fig.\ \ref{qtep-nl3}c). On the contrary, 
with exception of the band crossing region the evolution of 
$Q_t$($^{194}$Hg) as a function of $\Omega_x$ is similar to the 
one seen in the calculations with NL1. Comparing different
parametrizations of the RMF Lagrangian, it is clear that 
NL1 produces the largest values of $Q_t$, while NLSH the 
smallest (Fig.\ \ref{qtep-nl3}d)). This tendency has already 
been seen in the $A\sim 60$ and $A\sim 140-150$ regions of 
superdeformation \cite{A60,ALR.98}.

  In the calculations with NL3, proton and neutron pairing 
energies are similar at low rotational frequencies (see
Figs.\ \ref{qtep-nl3}a,b). At $\Omega_x\geq 0.15$ MeV,
proton pairing energies are larger than neutron ones. 
On the other hand, in the calculations
with NL1 neutron pairing energies are larger than 
proton ones at all rotational frequencies. The comparison 
of single-particle energies at $\Omega_x=0.0$ MeV 
obtained with the NL1 and NL3 parametrizations of the RMF
theory (see Fig.\ \ref{nl1-vr-nl3}) suggests that this 
is due to the larger $N=112$ and the smaller $Z=80$ SD shell 
gaps in the calculations with NL3. This leads to an 
additional quenching of neutron pairing correlations
and to an increase of proton pairing correlations as
compared with the case of the NL1 force.

%%%%%%%%%%%%%%%%%%%%%%%%%%%%%%%%%%%%%%%%%%%%%%%%%
\subsection{The dependence of the results on the 
parametrization of the Gogny force.}
%%%%%%%%%%%%%%%%%%%%%%%%%%%%%%%%%%%%%%%%%%%%%%%%%

  In the present section, we will study how the results 
of the calculations depend on the parametrization and the
strength of the Gogny force. In all calculations given in
this section, the set NL1 is used for the RMF Lagrangian
and approximate particle number projection is performed
by means of the Lipkin-Nogami method. Such a study is motivated 
by the fact that a precise quantitative information on the 
pairing correlations is not easy to extract 
in nuclei. There are no simple physical processes allowing 
one to isolate completely the 
pairing effects from the rest and to use them for the 
fit of the interaction in the particle-particle channel.
Different sets of the Gogny force have been obtained by 
the fit to the properties of finite nuclei. As a result,
these fits are more sensitive to the properties in the 
particle-hole channel than in the particle-particle channel 
since the pairing energies represent only a small portion of 
the total binding energies. In addition, 
apriori it is not clear that existing parametrizations of 
the Gogny force should be reliable in conjuction with the RMF 
theory. One should note that the moments of inertia of rotating 
nuclei are very sensitive to the properties of the pairing 
interaction \cite{RS.80,NilMI}. Considering that the rotation-vibration 
coupling is small in strongly deformed nuclei \cite{RS.80,GDBPL.97}, 
one can try to use this fact for the definition of the best
parametrization of the Gogny force which has to be used in the 
particle-particle channel in conjuction with the RMF theory.

 The parameter set D1 has been defined in Ref.\ \cite{D1}
based on the features of the $^{16}$O, $^{48}$Ca, 
$^{90}$Zr nuclei and the pairing properties of the Sn 
isotopes. It turns out that in non-relativistic calculations 
with the Gogny force this set overestimates the strength of
the pairing correlations. The D1' set  \cite{DG.80} 
differs from the D1 set only in the strength of the spin-orbit 
interaction and thus it will not be employed in the present 
calculations since only the central force with two-finite range 
Gaussian terms (see Eq.\ (\ref{Vpp})) of the original Gogny force
\cite{RS.80} is used in the CRHB theory. The D1S set \cite{D1S-a,D1S} 
differs, as follows from non-relativistic studies, from D1 by improved
surface properties and by producing smaller pairing correlations. It 
is used in most of the calculations in the present
manuscript. Recently another set of the Gogny force has been 
suggested in Ref.\ \cite{D1P}. So far it has not been applied 
for detailed studies of finite nuclei in non-relativistic approaches. 
However, comparing it with D1 and D1S (see Table \ref{table-Gog}), one
can conclude that it has larger similarities with D1 than with D1S.

\begin{figure}
\epsfxsize 16.0cm
\epsfbox{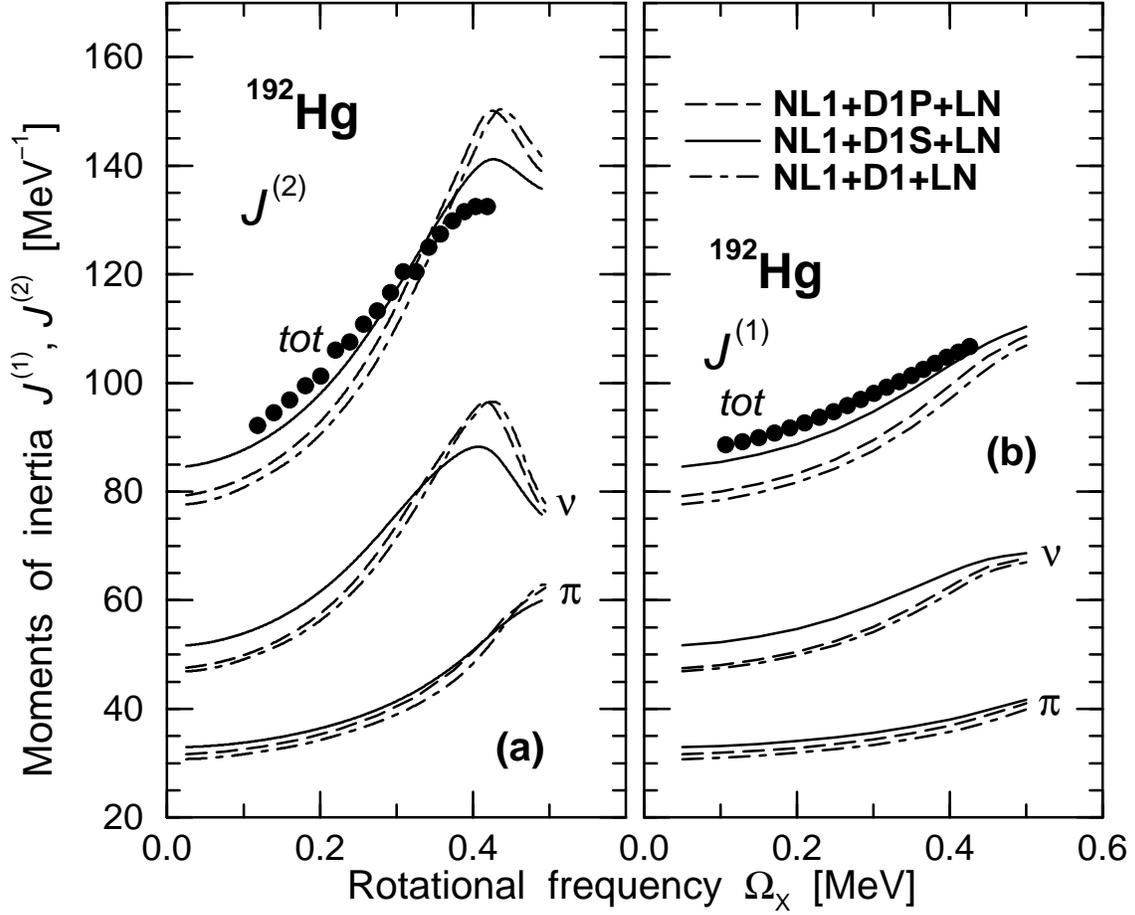}
\vspace{0.0cm}
\caption{The dependence of the kinematic and dynamic moments 
of inertia of the lowest SD configuration in $^{192}$Hg on 
the parametrization of the Gogny force. The results with the
sets D1, D1P and D1S for the Gogny force are shown.}
\label{j2j1-gog}
\end{figure}

 The dynamic and kinematic moments of inertia of the lowest SD 
configuration in $^{192}$Hg calculated with D1, D1S and D1P sets 
are compared with experiment in Fig.\ \ref{j2j1-gog}. One can see 
that only the set D1S provides good agreement with experimental 
data. The values of $J^{(1)}$ calculated with D1 and D1P are 
appreciable below both the experiment and the results obtained 
with D1S, see Fig.\ \ref{j2j1-gog}b. This is mainly due to smaller 
$J^{(1)}$ values for neutrons. At low rotational frequencies, the 
$J^{(2)}$ values obtained with D1 and D1P are lower than both
experimental data and the ones obtained with D1S. However, the 
increase of $J^{(2)}$ as a function of rotational frequency is 
larger in the calculations with D1 and D1P and thus at 
$\Omega_x \sim 0.32$ MeV they become larger than both experiment 
and the values calculated with D1S. In the neutron band crossing 
region at $\Omega_x \sim 0.42$ MeV, the $J^{(2)}$ values calculated 
with D1 and D1P are larger than both the ones obtained with D1S and 
the experiment. It is interesting to note that all three sets give
very similar neutron band crossing frequencies. Similar to the case 
of $J^{(1)}$, the differences between $J^{(2)}$'s calculated with 
D1 and D1P on the one side and with D1S on the other side are 
connected mainly with different alignment patterns in neutron 
subsystem.   

\begin{figure}
\epsfxsize 16.0cm
\epsfbox{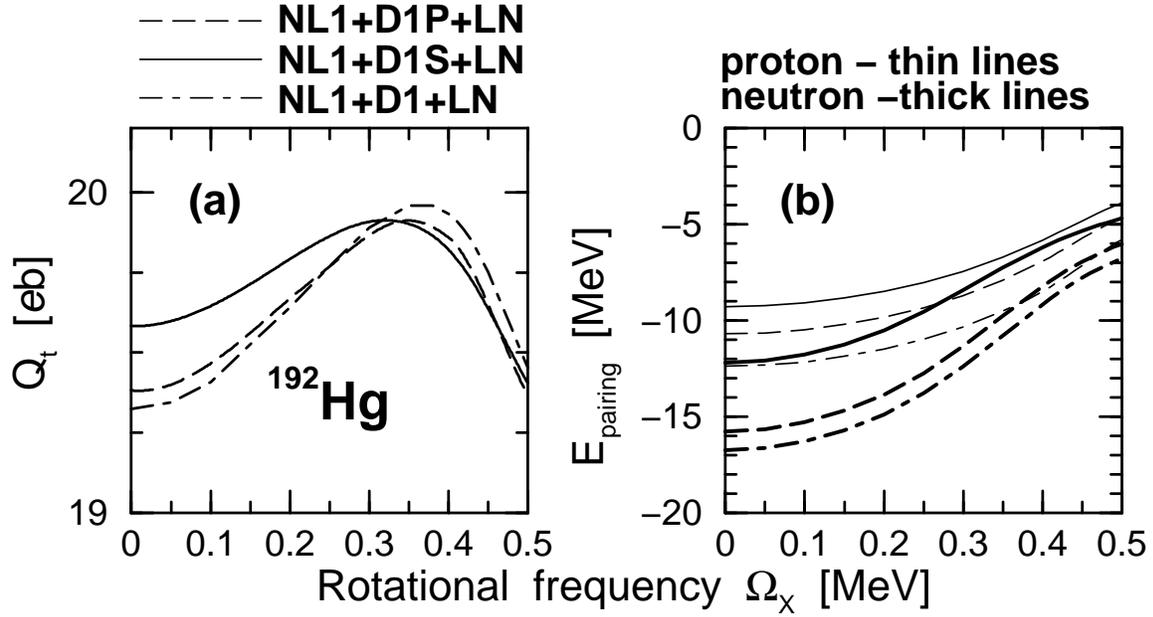}
\vspace{0.0cm}
\caption{The dependence of the transition quadrupole moment
$Q_t$ and neutron and proton pairing energies $E_{pairing}$
of the lowest SD configuration in $^{192}$Hg on the 
parametrization of the Gogny force. The results with the sets 
D1, D1P and D1S for the Gogny force are shown.}
\label{qtep-gog}
\end{figure}

  Calculated neutron and proton pairing energies $E_{pairing}^{\nu,\pi}$
are displayed in Fig.\ \ref{qtep-gog}b. The results of the
calculations with different sets show a similar behavior as a function 
of rotational 
frequency, but the absolute values of pairing energies strongly 
depend on the parametrization of the Gogny force. The strongest pairing 
correlations are provided by the D1 set, while the smallest ones are
obtained with D1S. The pairing energies calculated with D1P are in 
between the ones obtained with D1S and D1. Comparing transition 
quadrupole moments $Q_t$ obtained with different sets (see Fig.\
\ref{qtep-gog}a), one can see that (i) the D1 and D1P sets provide very 
similar values of $Q_t$ and (ii) at high rotational frequencies the
values of $Q_t$ calculated with all three sets come very close to
each other reflecting the decrease of pairing correlations. 

\begin{figure}
\epsfxsize 16.0cm
\epsfbox{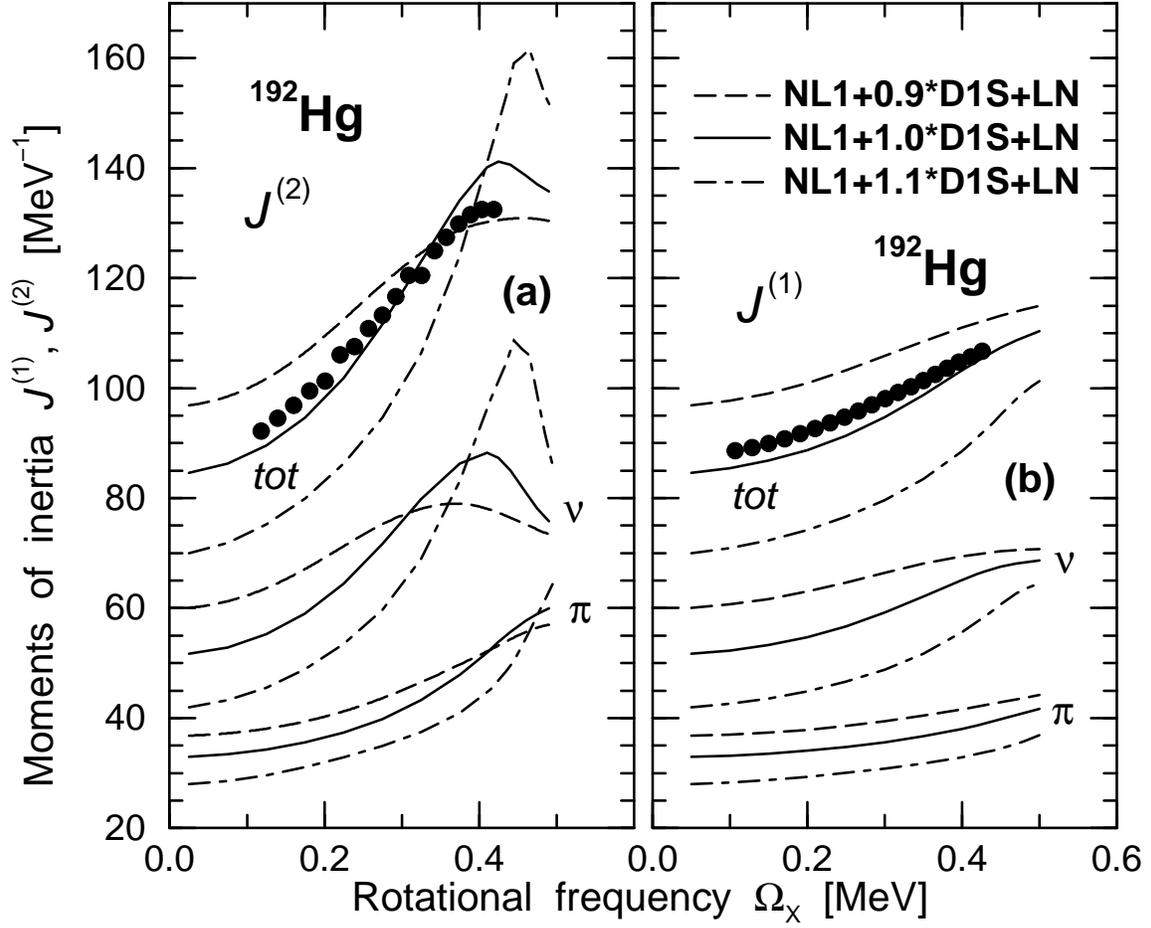}
\vspace{0.0cm}
\caption{Calculated and experimental dynamic and kinematic 
moments of inertia for yrast SD band in $^{192}$Hg. 
Experimental data are shown by solid unlinked symbols.  
The calculations have been performed with scaling of the
strength of the Gogny force: scaling factors 0.9, 1.0 and 1.1
have been used. Neutron and proton contributions into 
kinematic and dynamic moments of inertia are indicated
by the letters $\nu$ and $\pi$, respectively}
\label{j2j1-scal}
\end{figure}

\begin{figure}
\epsfxsize 16.0cm
\epsfbox{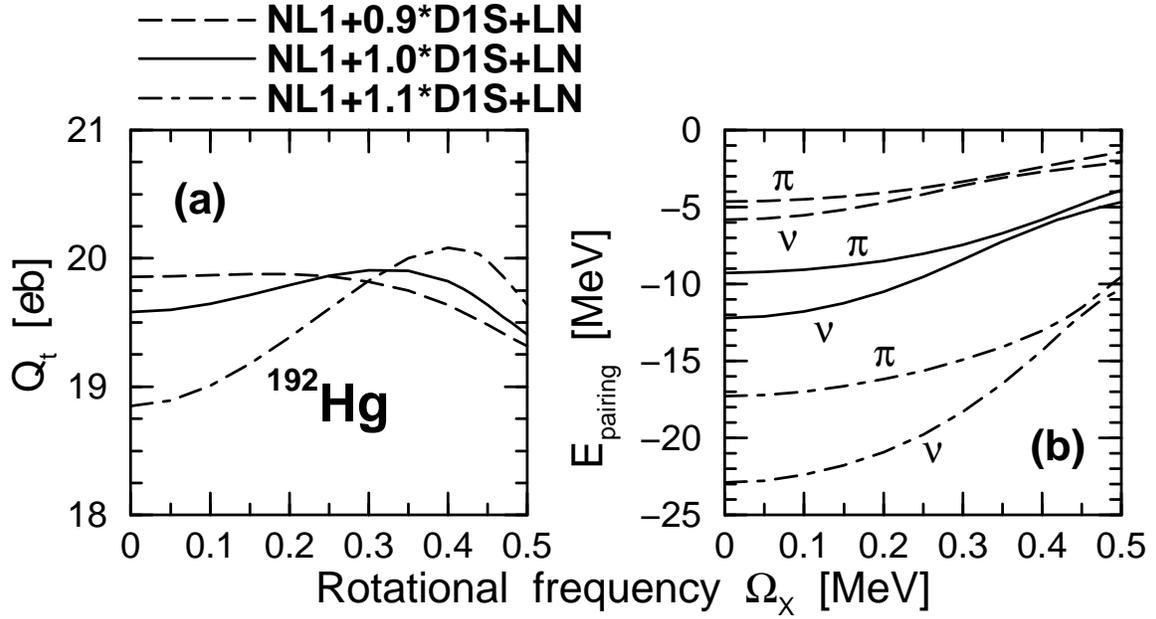}
\caption{Transition quadrupole moments $Q_t$ and neutron 
and proton pairing energies $E_{pairing}$ of the lowest SD 
configuration in $^{192}$Hg calculated with different scaling 
of the strength of the Gogny force. The notation of lines is 
given in the figure.}
\label{qtep-scal}
\end{figure}

 An alternative way to study the role of pairing correlations for 
different physical observables is to look how the results of the 
calculations are affected when the strength of the Gogny force is 
changed. This is done by introducing the 
scaling factor $f$ into the Gogny force (see Eq.\ (\ref{Vpp})). 
Such an investigation is in part motivated by the fact that in many 
studies performed within the relativistic Hartree-Bogoliubov theory 
with Gogny forces in the pairing channel and with no particle number 
projection, the strength of the Gogny force is increased by the 
factor 1.15, see for example Ref.\ \cite{GELR.96}. The results of the 
CRHB calculations for the 
lowest SD configuration in $^{192}$Hg with the scaling factors 
$f=0.9$, $f=1.0$ and $f=1.1$ for the D1S Gogny force are shown  
in Figs.\ \ref{j2j1-scal} and \ref{qtep-scal}. The effect of the 
scaling of the Gogny force is especially drastic on the pairing 
and rotational properties of the configuration under investigation 
(see Figs.\ \ref{qtep-scal}b and \ref{j2j1-scal}). The increase 
(decrease) of the strength of the Gogny force by 10\% leads to 
approximately twofold increase (decrease) of the absolute 
values of proton and neutron pairing energies, see Fig.\
\ref{qtep-scal}b. On the contrary, the increase (decrease) of 
the strength of the Gogny force leads to a significant decrease 
(increase) of the kinematic moment of inertia (Fig.\
\ref{j2j1-scal}b). One should note that the impact of the 
change of the strength of the Gogny force on the proton and 
neutron $J^{(1)}$ values is different. The effect of the scaling 
of the Gogny force on the dynamic moment of inertia is more 
complicated. While at low rotational frequencies it is similar 
to the one seen for the kinematic moment of inertia, it is 
completely different in the paired band crossing regions as 
seen in the proton, neutron and the total dynamic moments of 
inertia (Fig.\ \ref{j2j1-scal}a). Contrary to the low 
frequency range, stronger pairing correlations lead to larger 
$J^{(2)}$ values. In addition, as it is seen on the example 
of the neutron subsystem, stronger pairing correlations (a larger 
strength of the Gogny force) lead to the shift of the neutron 
paired band crossing to higher frequencies and make it sharper. 
A larger strength of the Gogny force leads also to smaller 
values of $Q_t$ at low $\Omega_x$ and to larger $Q_t$ at high 
$\Omega_x$, see Fig.\ \ref{qtep-scal}a. 

 In addition, different strengths and different sets of the 
Gogny force lead to different quasiparticle spectra in the 
vicinity of the Fermi level. In part, this effect is caused by 
the different calculated equilibrium deformations. It is 
reasonable to expect that this dependence of the quasiparticle 
spectra upon the parametrization and the strength of the Gogny 
force will have a pronounced impact on the band crossing 
frequencies in the SD configurations in odd and odd-odd 
nuclei.

%%%%%%%%%%%%%%%%%%%%%%%%%%%%%%%%%%%%%%%%%%%%%%%%
\subsection{Concluding remarks.}
%%%%%%%%%%%%%%%%%%%%%%%%%%%%%%%%%%%%%%%%%%%%%%%%

  The difference between the transition quadrupole moments 
$Q_t$ calculated with different sets of the RMF forces 
using the same set of the Gogny force 
is appreciable (Figs.\ \ref{qtep-nl3}c,d). On the 
contrary, the difference between $Q_t$'s obtained 
with three different sets (Fig.\ \ref{qtep-gog}) or 
with three different scalings of the strength 
(Fig.\ \ref{qtep-scal}a) of the Gogny force is 
smaller reflecting the fact that the equilibrium 
deformations are mostly defined by the properties of 
effective interaction in the particle-hole channel. The 
experimental uncertainties on transition quadrupole 
moments (see discussion in Ref.\ \cite{A190}) prevent 
the use of calculated $Q_t$'s for the selection 
of a better parametrization of the RMF Lagrangian for
the particle-hole channel and the Gogny force for the
particle-particle channel. Thus this selection can be 
based only on the rotational properties of the SD bands
under study such as the kinematic and dynamic moments of 
inertia which are very accurately defined in experiment. 

  Comparing different parametrizations of the Gogny force, 
it is clear that the D1 and D1P sets in connection with 
the NL1 force provide too strong pairing correlations and give 
very similar results for kinematic and dynamic moments of 
inertia, which deviate appreciable from experiment. The 
results of the calculations with the NL3 and NLSH forces in 
conjuction with the D1S set of the Gogny force also lead to 
considerable deviations from experiment for $J^{(1)}$ and 
$J^{(2)}$. It is then reasonable to expect that the use 
of the D1 and D1P sets of the Gogny force in conjuction with 
NL3 and NLSH will lead to even larger deviations from 
experiment for kinematic and dynamic moments of inertia. 
Thus one can conclude (see also Ref.\ \cite{A190}) that 
only the NL1 force in conjuction with the D1S set of the 
Gogny force and APNP(LN) will lead to a reasonable description 
of rotational properties of SD bands in the $A\sim 190$
mass region.

%%%%%%%%%%%%%%%%%%%%%%%%%%%%%%%%%%%%%%%%%%%%%%%%%%%%%%%%%
\section{Properties of yrast SD bands in even-even nuclei}
\label{systematic}
%%%%%%%%%%%%%%%%%%%%%%%%%%%%%%%%%%%%%%%%%%%%%%%%%%%%%%%%%

   It is our believe that the strong and the weak points
of a theoretical approach can be determined only by a
systematic comparison between experiment and theory. 
In the present Section the results of a systematic 
investigation of yrast SD bands in even-even nuclei 
of the $A\sim 190$ mass region will be presented. 
This investigation covers all even-even nuclei in this
region in which 
SD bands have been observed so far, namely, $^{190,192,194}$Hg, 
$^{192,194,196,198}$Pb and $^{198}$Po. First, the results 
for Hg and Pb isotopes with $N=110,112$ and 114 will be 
presented in detail. Partial results for these nuclei have 
been already presented in Ref.\ \cite{A190}. Then the results 
for $^{198}$Pb and $^{198}$Po will be discussed for each 
nucleus separately. All the calculations presented in this 
Section have been performed with the force NL1 for the RMF 
Lagrangian and the set D1S for the Gogny force in the 
$pp$-channel and approximate particle number projection
by means of the Lipkin-Nogami method (APNP(LN)) has been
used.

%%%%%%%%%%%%%%%%%%%%%%%%%%%%%%%%%%%%%%%%%%%%%%%%%%%%%%
\subsection{The $N=110,112$ and $114$ Hg and Pb nuclei}
\label{PbHg}
%%%%%%%%%%%%%%%%%%%%%%%%%%%%%%%%%%%%%%%%%%%%%%%%%%%%%%

\begin{figure}
\epsfxsize 12.0cm
\epsfbox{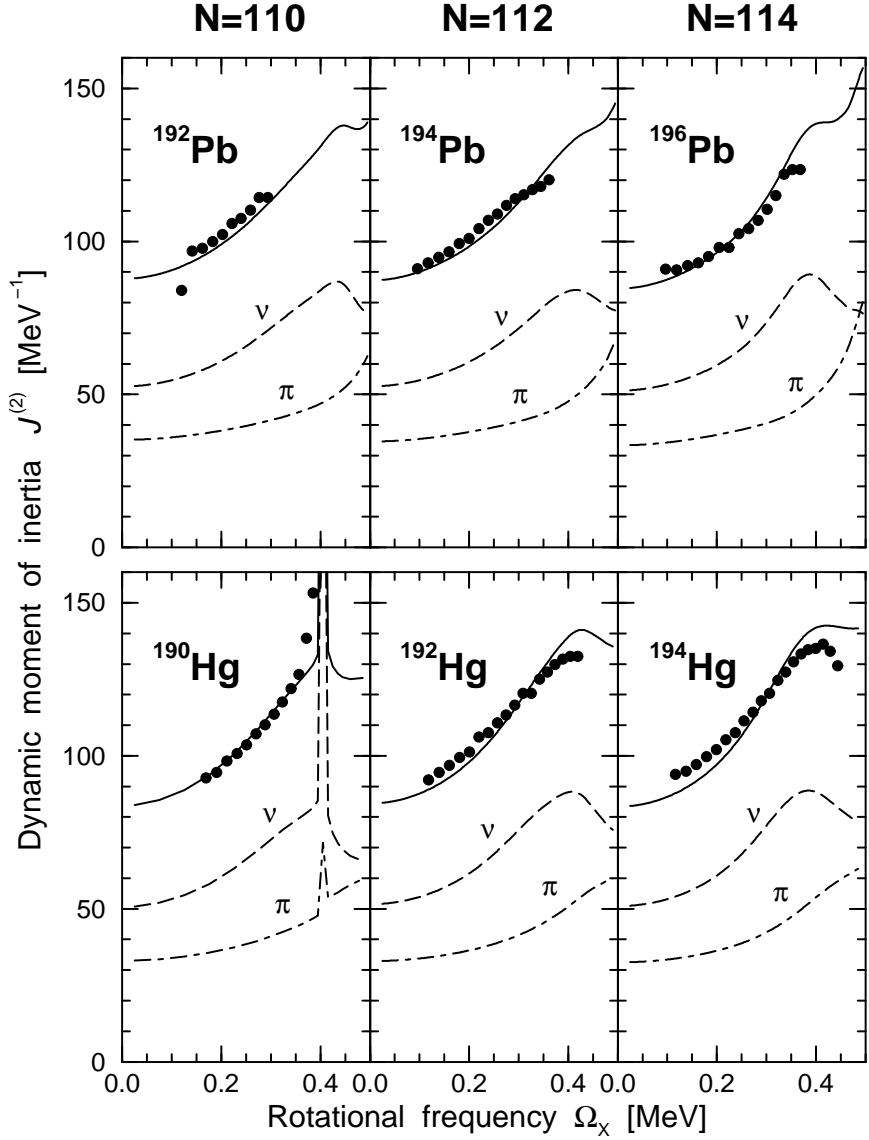}
\caption{Dynamic moments of inertia $J^{(2)}$ of observed
(solid circles) yrast SD bands in the $^{190,192,194}$Hg and
$^{192,194,196}$Pb nuclei versus the ones of calculated lowest
in energy SD configurations. Solid lines show the total calculated
dynamic moments of inertia $J^{(2)}$, while long-dashed and
dash-dotted lines show the contribution in $J^{(2)}$ from neutron
and proton subsystems. The experimental data are taken from
Refs.\ \protect\cite{Pb192c} ($^{192}$Pb),
\protect\cite{Pb194a,Pb194b,Pb194c} ($^{194}$Pb),
\protect\cite{Pb196a} ($^{196}$Pb), \protect\cite{Hg190}
($^{190}$Hg), \protect\cite{Hg192b} ($^{192}$Hg) and
\protect\cite{Hg194a,Hg194b} ($^{194}$Hg).}
\label{sys-j2}
\end{figure}

\begin{figure}
\epsfxsize 12.0cm
\epsfbox{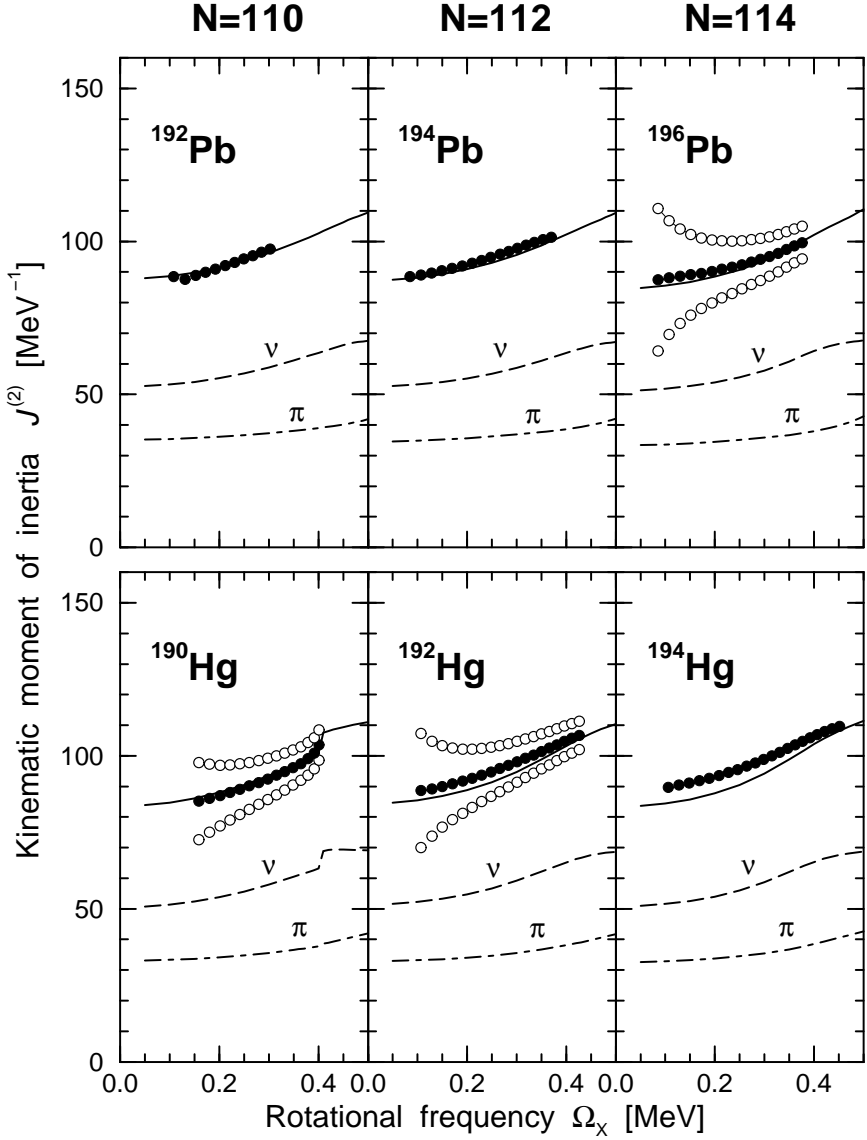}
\caption{The same as Fig.\ \protect\ref{sys-j2} but for kinematic
moments of inertia $J^{(1)}$. The experimental moments of inertia
of linked or tentatively linked SD bands are shown by solid
circles. In other cases ($^{196}$Pb, $^{190,194}$Hg), the 
'experimental' kinematic moments of inertia $J^{(1)}$ are shown 
for three different spin values of the lowest state $I_0$ in the 
SD band, see text for details. The values being in best agreement 
with the calculations are indicated by solid circles, while open 
circles are used for alternatives.}
\label{sys-j1}
\end{figure}

  The calculated total, neutron and proton dynamic and 
kinematic moments of inertia are shown in Figs.\ \ref{sys-j2},
\ref{sys-j1} and compared with experiment. One can see that 
a very successful description of dynamic moments of inertia 
of experimental bands is obtained in the calculations without 
adjustable parameters (Fig.\ \ref{sys-j2}). When comparing 
calculated and experimental kinematic moments of inertia, one 
should keep in mind that only yrast SD bands in $^{194}$Pb 
and $^{194}$Hg are definitely linked to the low-spin level scheme 
\cite{Pb194b,Pb194c,Hg194b}. In addition, there is a tentative
linking of the SD band in $^{192}$Pb \cite{Pb192c}. On the 
contrary, at present the yrast SD bands in $^{190,192}$Hg 
and $^{196}$Pb are not linked to the low-spin level scheme yet. 
Thus some spin values consistent with the signature of the 
calculated lowest SD configuration should be assumed for the 
experimental bands when a comparison is made with respect 
of the kinematic moment of inertia $J^{(1)}$. Taking into
account that kinematic moments of inertia of linked SD 
bands in $^{194}$Pb and $^{194}$Hg and tentatively linked 
SD band in $^{192}$Pb are well described
in the calculations, the spin values of unlinked bands
can be obtained by comparing the calculated values of 
$J^{(1)}$ with experimental ones under different spin
assignments. Such a comparison (see Ref.\ \cite{A190}
for detailed figures) leads to the spin values $I_0$ for the
lowest states in SD bands listed in Table \ref{table-I0}. Under
these spin assignments, the CRHB+LN calculations rather
well describe 'experimental' kinematic moments of inertia
of SD bands in $^{192,196}$Pb and $^{194}$Hg 
(Fig.\ \ref{sys-j1}). Alternative spin assignments, which are 
different from the ones given in Table \ref{table-I0} by $\pm 2\hbar$,
can be ruled out since they (open circles in Fig.\ \ref{sys-j1}) lead
to considerable deviations from the results of the calculations. 

 The increase of kinematic and dynamic moments of inertia in this 
mass region can be understood in the framework of the CRHB+LN theory
as emerging predominantly from a combination of three effects: the 
gradual alignment of a pair of $j_{15/2}$ neutrons, the alignment 
of a pair of $i_{13/2}$ protons at a somewhat higher frequency, 
and decreasing pairing correlations with increasing rotational 
frequency, see also Sect.\ \ref{sect-pair}. The interplay of 
alignments of neutron and proton pairs is more clearly seen in Pb 
isotopes where the calculated $J^{(2)}$ values show either a small 
peak (for example, at $\Omega_x \sim 0.44$ MeV in $^{192}$Pb, see 
Fig.\ \ref{sys-j2}) or a  plateau (at $\Omega_x \sim 0.4$ MeV in 
$^{196}$Pb, see Fig.\ \ref{sys-j2}). With increasing rotational 
frequency, the $J^{(2)}$ values determined by the alignment in 
the neutron subsystem decrease. The maximum in $J^{(2)}$ of the
neutron subsystem is reached at $\Omega_x \approx 0.44$, 0.41,
0.39, 0.425 and 0.39 MeV in $^{192,194,196}$Pb and $^{194,196}$Hg, 
respectively. The decrease in these frequencies with increasing 
$N$ within each isotope chain correlates with the decrease of
the transition quadrupole moments $Q_t$ (Fig.\ 4 in Ref.\ \cite{A190}). 
However, the fact that the maximum in neutron $J^{(2)}$ is 
obtained at the same frequencies in Hg and Pb isotones, which have 
$Q_t$ values differing by $1-1.5$ $e$b (Fig.\ 4 in Ref.\ 
\cite{A190}), indicates that the alignment process is very 
complicated and depends not only on the equilibrium deformation 
but also on the position of the Fermi level.  
  
  The decrease of neutron $J^{(2)}$ at frequencies higher than
$\Omega_x \sim 0.4$ MeV is in part compensated by the increase 
of proton $J^{(2)}$ due to the alignment of the $i_{13/2}$ proton 
pair. This leads to the increase of the total $J^{(2)}$-value at 
$\Omega_x \geq 0.45$ MeV in the Pb isotopes, 
while no such increase has been found in the calculations after 
the peak up to $\Omega_x=0.5$ MeV in $^{192}$Hg, see Fig.\ 
\ref{sys-j2}. Thus one can conclude that the shape of the peak 
(plateau) in total $J^{(2)}$ in the band crossing region is 
determined by a delicate balance between alignments in the proton 
and neutron subsystems which depends on deformation, rotational 
frequency and Fermi energy. It is also of interest to mention that 
the sharp increase in $J^{(2)}$ of the yrast SD band in $^{190}$Hg 
is also reproduced in the present calculations. In the calculations,
this increase is due to a two-quasiparticle alignment associated with 
the $\nu [761]3/2$ orbital. One should note that 
the calculations slightly overestimate the magnitude of $J^{(2)}$ at 
the highest observed frequencies. Possible reasons could be 
the deficiencies either of the Lipkin-Nogami method \cite{Mag93} 
or of the cranking model in the band crossing region or both of 
them. In addition, the calculations do not reproduce the sudden 
decrease in $J^{(2)}$ at the bottom of the SD band in $^{192}$Pb, 
the origin of which is not understood so far, see 
Ref.\ \cite{Pb192c}. 

\begin{figure}
\epsfxsize 14.0cm
\epsfbox{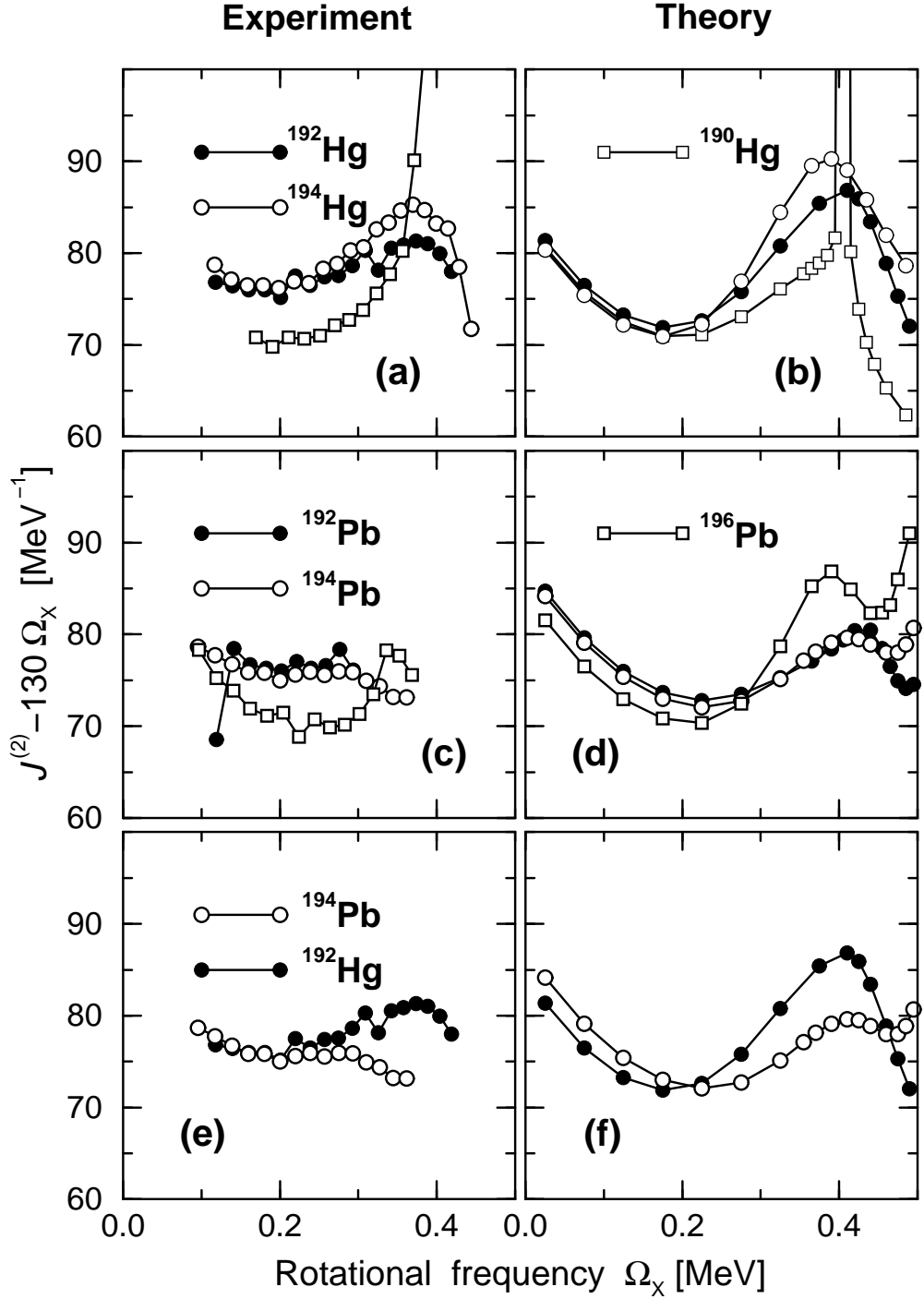}
\caption{Comparison of experimental and calculated dynamic 
moments of inertia $J^{(2)}$ in the Hg (top panels) and the 
Pb (middle panels) isotopes as well as in the $N=112$ 
isotones (bottom panels). The same type of symbols is
used for experimental bands (left panels) and their 
theoretical counterparts (right panels). In order to
show the differences between different bands in more
detail, the frequency dependent term is extracted
from $J^{(2)}$, see text for details.}
\label{rel-j2}
\end{figure}

\begin{figure}
\epsfxsize 16.0cm
\epsfbox{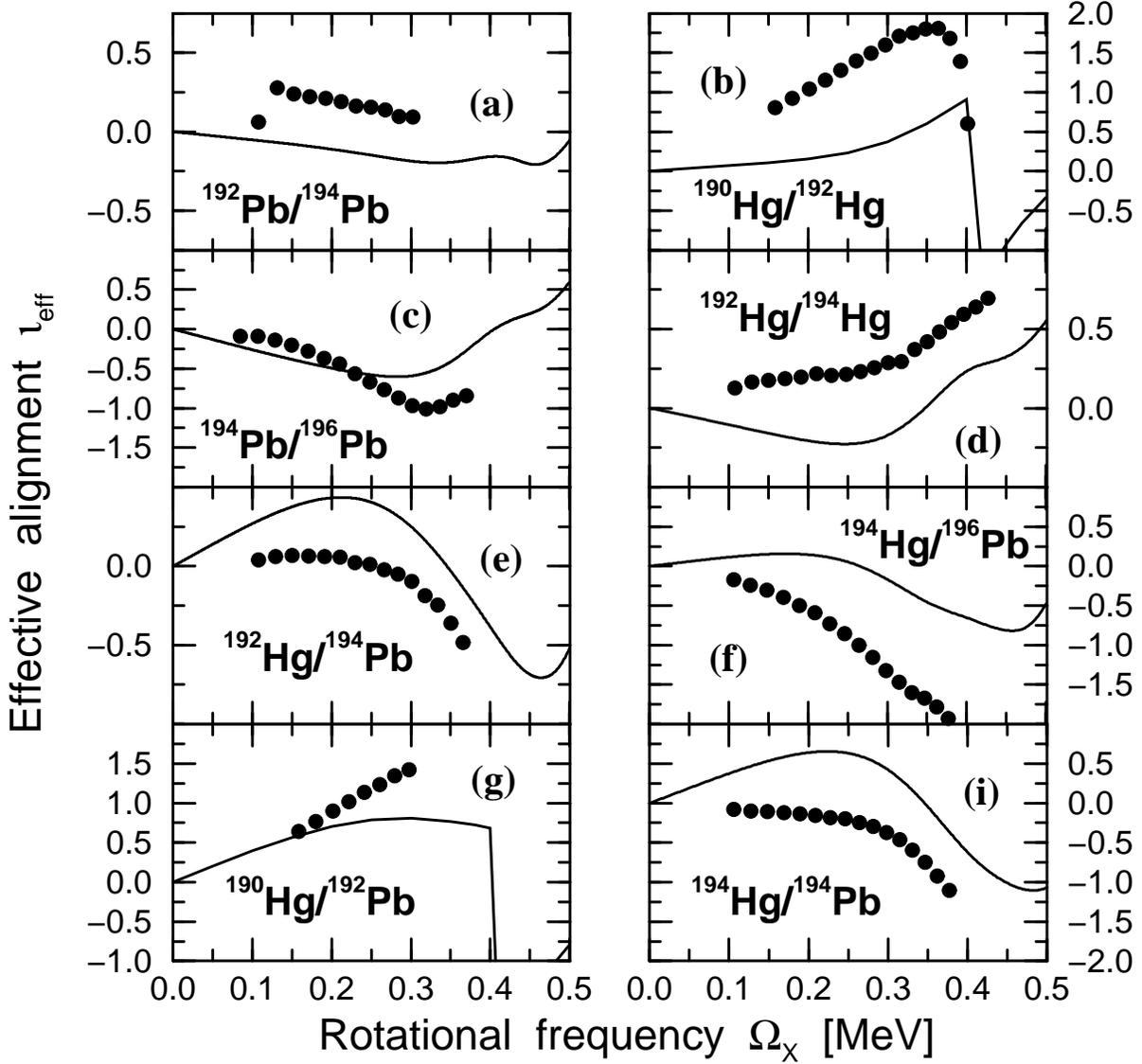}
\caption{Effective alignments, $i_{eff}$ (in units $\hbar$),
extracted from experiment (unlinked solid circles) are compared
with those extracted from the corresponding calculated
configurations (solid lines). Note that two absolute scales 
are used for the vertical axis. Panels b,c,f,g and i use
the same scale with the difference between the lowest and 
the highest $i_{eff}$ values being $3\hbar$. On the other
hand, the panels a,d and e use a scale with a difference
of $1.5\hbar$.}
\label{efalign}
\end{figure}

  In order to compare relative properties of the calculated and 
experimental dynamic moments of inertia in more detail we tilt 
them by extracting the frequency dependent term from $J^{(2)}$. The 
resulting quantities ($J^{(2)}-130\Omega_x$) are shown in 
Fig.\ \ref{rel-j2} for Hg (top panel) and Pb (middle panel) 
isotopes as well as for $N=112$ (bottom panel) isotones. 
Since the difference of the dynamic moments of inertia of two 
bands is proportional to the derivative of the effective (relative) 
alignment $i_{eff}$ of these bands (see Refs.\ \cite{BHN.95,ALR.98}) 
for details), the effective alignments of compared bands will also 
be presented here. The relative properties 
of the dynamic moments of inertia
of the yrast SD bands in $^{192,194}$Hg nuclei are
very well reproduced in the calculations. Both in
calculations and in experiment, 
$J^{(2)}$($^{192}$Hg)$\approx$$J^{(2)}$($^{194}$Hg)
at $\Omega_x\leq 0.3$ MeV, while at higher frequencies 
$J^{(2)}$($^{194}$Hg)$\geq$$J^{(2)}$($^{192}$Hg)
(Figs.\ \ref{rel-j2}a,b). This result correlates with the 
fact that effective alignment $i_{eff}$ in the 
$^{192}$Hg/$^{194}$Hg pair and especially its slope is 
also reproduced rather well in the calculations (see 
Fig.\ \ref{efalign}d). On the contrary, while the 
properties of $J^{(2)}$($^{190}$Hg) relative to 
$J^{(2)}$($^{192}$Hg) are reasonably well reproduced 
at frequencies $\Omega_x\geq 0.3$ MeV, the situation
is different at lower frequencies. There the 
difference between $J^{(2)}$ values stays almost constant 
around 6 MeV$^{-1}$ in experiment, while it is decreasing 
to zero at $\Omega_x \sim 0.2$ MeV in the calculations
(see Figs.\ \ref{rel-j2}a,b). This feature correlates with
the fact that the effective alignment in this pair of SD
bands is not well reproduced in the calculations (see 
Fig.\ \ref{efalign}b). 
 
 The dynamic moments of inertia of yrast SD bands in 
$^{192,194}$Pb are almost identical in experiment. This 
feature (Figs.\ \ref{rel-j2}c,d) and the effective alignment 
in the $^{192}$Pb/$^{194}$Pb pair (Figs.\ \ref{efalign}a) 
are rather well reproduced in the calculations. In experiment,
the dynamic moments of inertia of SD bands in $^{194,196}$Pb 
are almost identical at $\Omega_x\sim 0.1$ MeV. With 
increasing $\Omega_x$, $J^{(2)}$($^{196}$Pb) decreases 
below $J^{(2)}$($^{194}$Pb) with a maximum difference 
between them of $\approx 5$ MeV$^{-1}$ being reached 
at $\Omega_x \sim 0.22$ MeV and then this difference
becomes smaller up to $\Omega_x\sim 0.32$ MeV where
the dynamic moments of inertia of both bands 
coincide (Fig.\ \ref{rel-j2}c). At even higher frequencies, 
$J^{(2)}$($^{196}$Pb)$\geq$$J^{(2)}$($^{194}$Pb). The
calculations reasonably well reproduce the general features, 
however, somewhat underestimate the difference between 
$J^{(2)}$'s of these two bands at medium frequencies and 
overestimate this difference at highest observed 
frequencies. The experimental effective alignment
in the $^{194}$Pb/$^{196}$Pb pair is very well reproduced 
at low frequencies, while the difference between 
experiment and calculations is somewhat larger at high 
frequencies (Fig.\ \ref{efalign}c). 

  The experimental dynamic moments of inertia of SD bands 
in $^{194}$Pb and $^{192}$Hg are identical at $\Omega_x\leq 0.2$ 
MeV, while at higher frequencies the condition  
$J^{(2)}$($^{194}$Pb)$\geq$$J^{(2)}$($^{192}$Hg) holds
(Fig.\ \ref{rel-j2}e). The relative properties of dynamic moments of
inertia of these two bands at $\Omega_x\geq 0.2$ MeV are rather 
well reproduced in the calculations (Fig.\ \ref{rel-j2}f).
At lower frequencies, there is however some difference 
between the calculated $J^{(2)}$ values for these two
bands of $\approx 2$ MeV$^{-1}$. The
effective alignment in the $^{192}$Hg/$^{194}$Pb pair and 
especially its slope as a function of $\Omega_x$ is well 
reproduced in the calculations (Fig.\ \ref{efalign}e). 

  The effective alignments between other pairs of SD bands which 
have not been discussed before are also shown in Fig.\ 
\ref{efalign}g,f,i. It is seen that the effective alignment in 
the $^{194}$Hg/$^{194}$Pb pair is reasonably well reproduced in 
the calculations, although they somewhat overestimate the 
absolute value of $i_{eff}$. Since the spins of these two bands 
are determined experimentally, this result can be considered
as a first direct justification of the reliability of the effective 
alignment approach used frequently for the configuration and spin 
assignments of SD bands in different mass regions, see Refs.\ 
\cite{ALR.98,A60} for details. Comparing the experimental and
effective alignments (Fig.\ \ref{efalign}) and taking into account 
that compared bands differ by at least 2 particles, one can conclude
that considerable deviations from experiment are seen only 
in the cases of the effective alignments in the $^{190}$Hg/$^{192}$Pb, 
$^{194}$Hg/$^{196}$Pb and $^{190}$Hg/$^{192}$Hg pairs 
(Figs.\ \ref{efalign}g,f,b). In order to find are observed deviations 
from experiment related to the particle-hole or the particle-particle 
channel of the CRHB theory the investigation of SD bands in odd nuclei 
of this mass region are needed. Such an investigation is in progress and 
its results will be reported later. {\bf The} investigation of
relative alignments of SD bands in this mass region has been performed 
in non-relativistic approaches such as the total routhian surface 
Strutinsky-type approach based on the Woods-Saxon potential and 
Skyrme-Hartree-Fock-Bogoliubov approach in Ref.\ \cite{FHS.99}. 
Similar to our case, some discrepancies between theory and experiment 
have been found.

\begin{figure}
\epsfxsize 16.0cm
\epsfbox{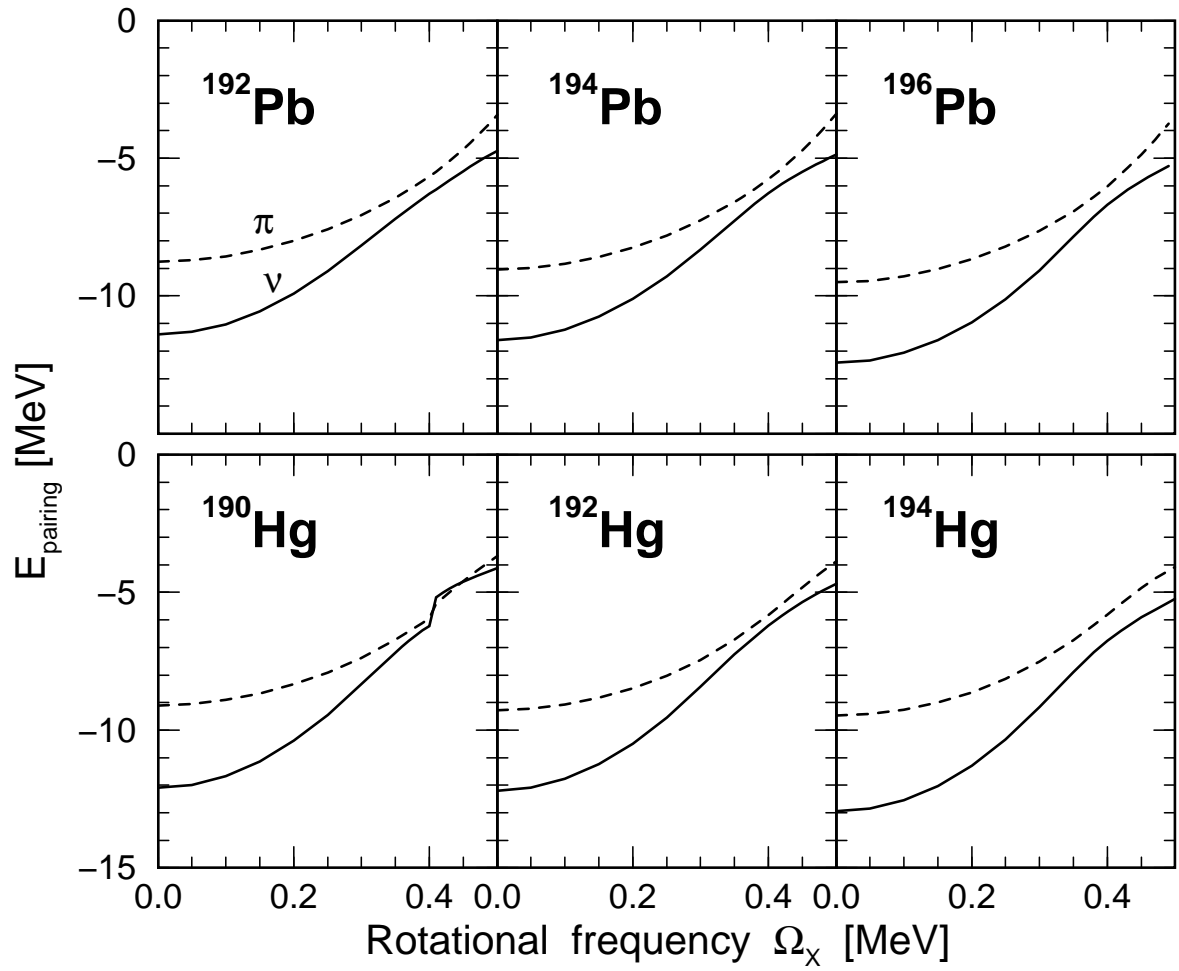}
\vspace{0.0cm}
\caption{Calculated neutron and proton pairing energies
for the lowest SD configurations in $^{192.194,196}$Pb
and $^{190,192,194}$Hg. Solid and dashed lines are 
used for neutrons and protons, respectively.}
\label{sys-pair}
\end{figure}

  The calculated neutron and proton pairing energies 
$E^{\nu,\pi}_{pairing}$ are summarized in Fig.\ \ref{sys-pair}. 
In all nuclei, they decrease with increasing rotational frequency 
reflecting the quenching of pairing correlations due to the 
Coriolis antipairing effect. In addition, neutron pairing is 
stronger than proton pairing. Neutron pairing energies in 
$^{190,192}$Hg almost coincide up to the band crossing seen in 
$^{190}$Hg. A similar situation exists also in $^{192,194}$Pb where 
the difference between calculated neutron pairing energies does 
not exceed 0.2 MeV. These energies are smaller than the ones in 
$^{190,192}$Hg by $\approx 0.5$ MeV at $\Omega_x=0.0$ MeV, while 
at high rotational frequencies neutron pairing energies are similar 
in these Pb and Hg nuclei. On the contrary, neutron 
pairing energies in the $N=114$ Hg and Pb nuclei are larger 
than the ones in the nuclei with $N=112$ by 0.5-0.8 MeV 
dependent on the rotational frequency and they do coincide at 
$\Omega_x \geq 0.3$ MeV. A similar trend is seen also for
proton pairing energies where, however, the difference 
between the pairing energies in different nuclei is smaller 
than in the case of neutron pairing energies and it does not exceed 
0.5 MeV.

%%%%%%%%%%%%%%%%%%%%%%%%%%%%%%%%%%%%%
\subsection{The nucleus $^{198}$Pb}
%%%%%%%%%%%%%%%%%%%%%%%%%%%%%%%%%%%%%

\begin{figure}
\epsfxsize 14.0cm
\epsfbox{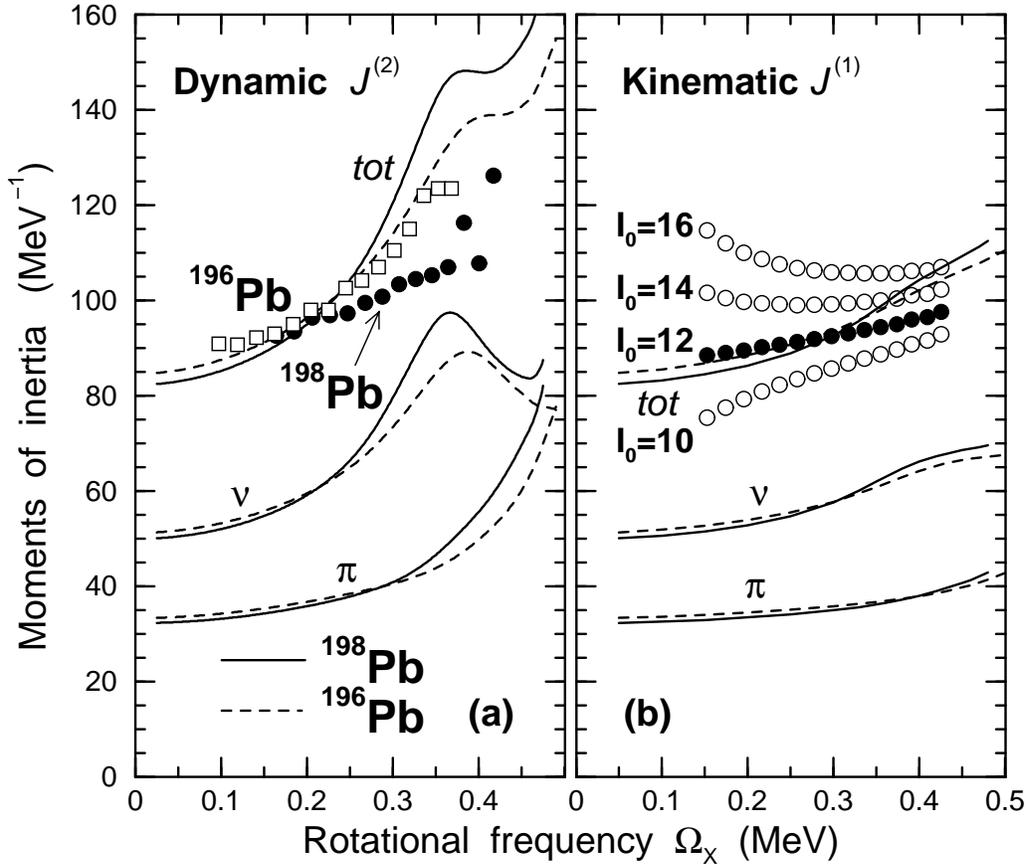}
\vspace{2.0cm}
\caption{Calculated and experimental dynamic and kinematic 
moments of inertia for yrast SD band in $^{198}$Pb. The 
results of the calculations are shown by solid lines. 
Experimental data for $^{198}$Pb band is taken from Ref.\ 
\protect\cite{Pb198} and 'experimental' kinematic moments
of inertia are shown for 4 different spin values for lowest
state $I_0$. The $J^{(1)}$ values being in best agreement
with results of the calculations at low spin are shown by
solid circles. For comparison, the same quantities obtained 
for the lowest SD configuration in $^{196}$Pb are shown by 
dashed lines. The experimental values of $J^{(2)}$ for yrast 
SD band in $^{196}$Pb are shown by open squares.}
\label{pb98-j2j1}
\end{figure}

\begin{figure}
\epsfxsize 16.0cm
\epsfbox{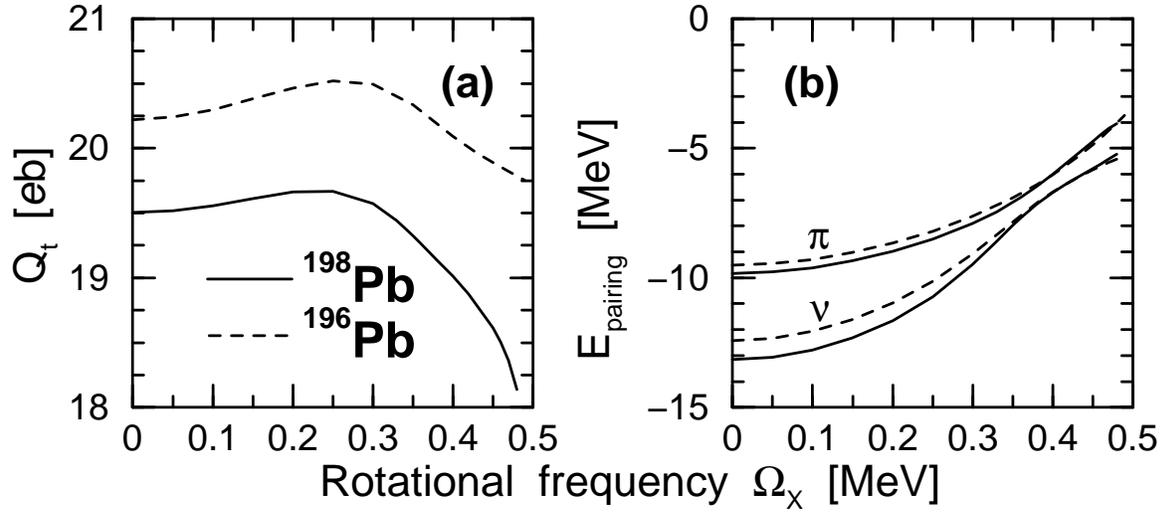}
\caption{Calculated transition quadrupole moments $Q_t$ and 
proton and neutron pairing energies $E_{pairing}$ of the 
lowest SD configuration in $^{198}$Pb. For comparison,
the same quantities obtained for the lowest SD 
configuration in $^{196}$Pb are shown by dashed 
lines.}
\label{pb98-qt-epair}
\end{figure}

 The results of the calculations for the lowest SD band in 
$^{198}$Pb are shown in Figs.\ \ref{pb98-j2j1} and 
\ref{pb98-qt-epair}. The calculated kinematic moment of 
inertia agrees reasonably well with experiment up to 
$\Omega_x\sim 0.32$ MeV, while considerable disagreement is seen 
at higher frequencies. The increase of calculated $J^{(1)}$ as a 
function of $\Omega_x$ is larger in $^{198}$Pb than in 
$^{196}$Pb (Fig.\ \ref{pb98-j2j1}b). As a result, the calculated 
dynamic moment of inertia in $^{198}$Pb is larger than the one 
in $^{196}$Pb at $\Omega_x \geq 0.2$ MeV and smaller at 
$\Omega_x \leq 0.2$ MeV. On the contrary, the experimental data 
show the opposite trend for the dynamic moment of inertia of the 
$^{198}$Pb band having a much smaller increase as a function of rotational 
frequency (Fig.\ \ref{pb98-j2j1}a). Thus the results of the calculations 
for $^{198}$Pb do not reproduce neither absolute rotational properties
of SD band nor their relative properties with respect to SD band in 
$^{196}$Pb. A similar problem with the reproduction of the properties 
of the $^{198}$Pb band exists also in the cranked Nilsson-Strutinsky 
Lipkin-Nogami calculations presented in Refs.\ \cite{WS.94,Pb198}. The 
investigation of neighbouring odd nuclei within the CRHB+LN theory 
is needed in order to understand better the origin of these problems. 
The calculated values of $Q_t$ for SD band in $^{198}$Pb are by 
$\approx 1$ $e$b smaller than in $^{196}$Pb 
(see Fig.\ \ref{pb98-qt-epair}a) which is in agreement with the 
decrease of $Q_t$ with increasing $N$ seen in the lighter Pb 
isotopes (see Fig.\ 4 in Ref.\ \cite{A190}). Precise measurements 
of the relative transition quadrupole moments of the yrast SD 
bands in $^{196,198}$Pb nuclei and a theoretical study of neighboring
odd nuclei can be useful for the understanding of the present problems 
with the description of $J^{(1)}$ and $J^{(2)}$. Pairing energies in 
$^{198}$Pb are somewhat larger than in $^{196}$Pb 
(Fig.\ \ref{pb98-qt-epair}b) which agrees with the trend of the 
increase of pairing correlations within the isotopic chain with 
increasing neutron number $N$ (see Section \ref{PbHg}).

%%%%%%%%%%%%%%%%%%%%%%%%%%%%%%%%%%%%%
\subsection{The nucleus $^{198}$Po}
%%%%%%%%%%%%%%%%%%%%%%%%%%%%%%%%%%%%%
  
\begin{figure}
\epsfxsize 16.0cm
\epsfbox{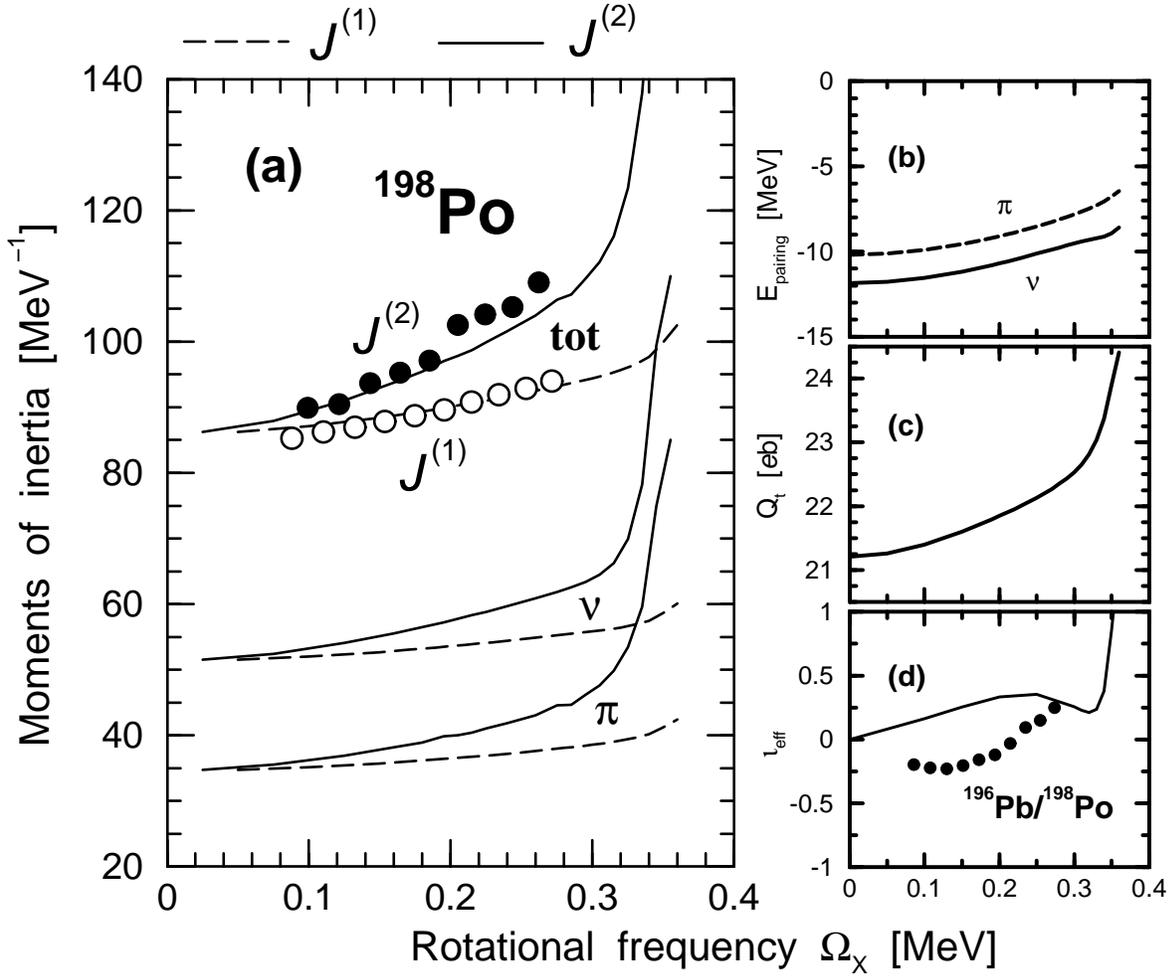}
\caption{The results of the calculations for the lowest SD 
configuration in $^{198}$Po. Dynamic and kinematic moments 
of inertia, proton and neutron pairing energies $E_{pairing}$, 
transition quadrupole moment $Q_t$ and effective alignment $i_{eff}$
in the $^{196}$Pb/$^{198}$Po pair are shown in panels (a), (b), (c) and 
(d) respectively. The experimental data for the kinematic and dynamic 
moments of inertia are taken from Ref.\ \protect\cite{Po198} and
shown by unlinked open and solid circles in panel (a). In panel (d), 
the experimental effective alignment is shown by unlinked solid circles.}
\label{fig-po98}
\end{figure}

\begin{figure}
\epsfxsize 16.0cm
\epsfbox{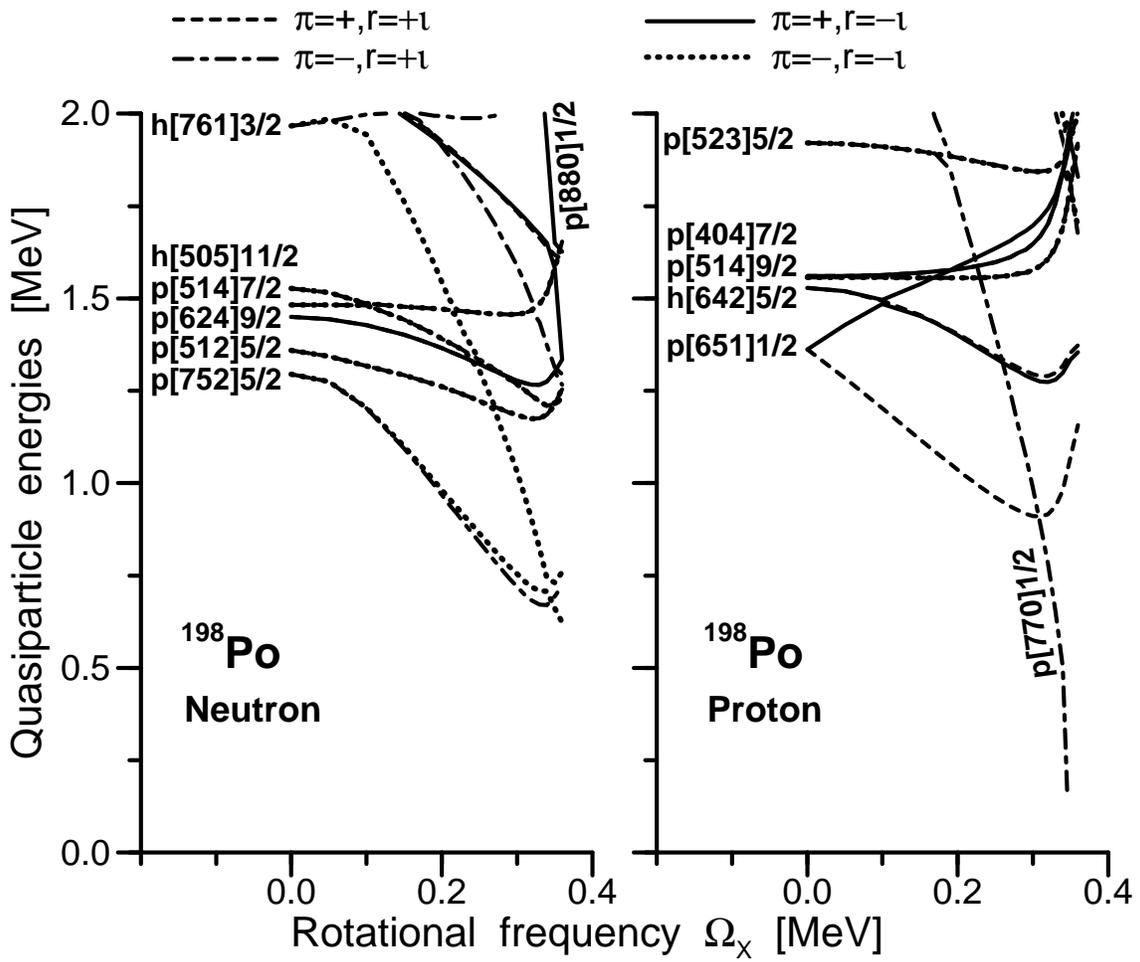}
\caption{Quasiparticle routhians corresponding to the 
lowest SD configuration in $^{198}$Po, see caption 
of Fig.\ \protect\ref{qpe-routh} for details.}
\label{qpe-po98}
\end{figure}

  The CRHB+LN calculations for the lowest SD configuration 
in $^{198}$Po very well describe experimental dynamic and 
kinematic moments of inertia of the yrast SD band in this nucleus
(Fig.\ \ref{fig-po98}a). At $\Omega_x \sim 0.34$ MeV, the 
calculations predict sharp increase in dynamic moment of 
inertia caused mainly by the alignment of the lowest proton 
hyperintruder $\pi [770]1/2$ orbital (see Fig.\ \ref{qpe-po98}). 
The absolute values
of proton and neutron pairing energies and their behavior
as a function of rotational frequency (Fig.\ \ref{fig-po98}b) 
are similar to the ones seen in other nuclei. A specific feature 
of this nucleus, which has not been observed in other nuclei, 
is considerable increase of the transition quadrupole moment 
$Q_t$ (by $\sim 3$ $e$b) with increasing rotational frequency 
(Fig.\ \ref{fig-po98}c). At $\Omega_x=0.0$ MeV, the $Q_t$ 
values in $^{198}$Po are larger than the ones in isotonic
$^{196}$Pb by $\approx 1.0$ $e$b (see Fig.\ \ref{fig-po98}c
and Fig.\ 4 in Ref.\ \cite{A190}).  One should note that the 
results of the GCM+GOA calculations based on the Gogny force 
also show the same feature \cite{LGD.99}. Finally, the experimental 
effective alignment in the $^{196}$Pb/$^{198}$Po pair is reasonably 
well described in the calculations (Fig.\ \ref{fig-po98}d).

%%%%%%%%%%%%%%%%%%%%%%%%%%%%%%%%%%%%%%%%%%%%%%%%%
\subsection{Mass hexadecupole moments $Q_t$}
%%%%%%%%%%%%%%%%%%%%%%%%%%%%%%%%%%%%%%%%%%%%%%%%%

\begin{figure}
\epsfxsize 16.0cm
\epsfbox{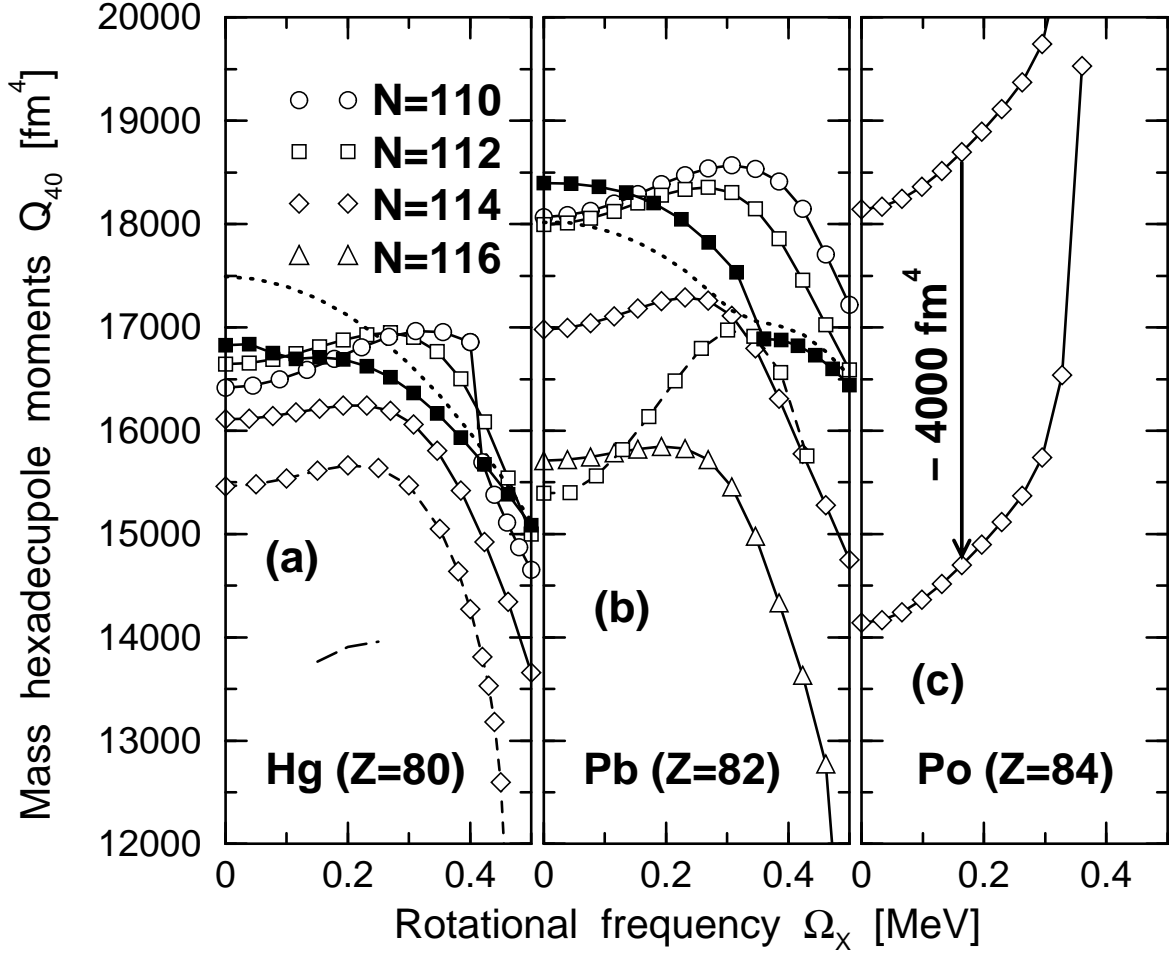}
\caption{Calculated mass hexadecupole moments $Q_{40}$. 
Circles, squares, diamonds and triangles are used
for the nuclei with $N=110$, 112, 114 and 116, 
respectively. Open symbols are used for the 
results of the calculations with APNP(LN), while 
filled symbols for the ones without APNP(LN). In
panels (a) and (b), the results of the calculations
without pairing are shown for $^{192}$Hg and 
$^{194}$Pb by dotted lines. Dashed lines are used
to indicate the results obtained with the NL3 force, 
while long-dashed line in panel (a) shows the
ones obtained in $^{194}$Hg with the NLSH force. The lower
curve in panel (c) is shifted down by $-4000$ fm$^4$
relative to original (upper) curve in order to show the 
results of the calculations at high rotational frequencies.
}
\label{h40-sys}
\end{figure}

 Experimental information on mass hexadecupole moments 
$Q_{40}$ is not available so far for SD bands 
in any mass region. Thus only the results of the calculations 
for this quantity are presented in Fig.\ \ref{h40-sys}. 
These results have been obtained with the NL1 force for the 
RMF Lagrangian and the set D1S for the Gogny force if it is 
not specified otherwise. Note 
that we do not show the results of the calculations obtained with either 
different parametrizations or different scalings of the 
Gogny force. The $Q_{40}$ values calculated with no pairing
(dotted lines in Figs.\ \ref{h40-sys}a,b) show a gradual 
decrease with an increase of rotational frequency. A similar
trend is also seen in the results of the calculations with no 
APNP(LN) (solid lines with solid squares in Figs.\ 
\ref{h40-sys}a,b). The sharp change of the slope of the 
$Q_{40}(\Omega_x)$ curve seen in $^{194}$Pb at $\Omega_x \sim 0.35$ 
MeV (Fig.\ \ref{h40-sys}b) is related to a sharp 
band crossing associated with the 
collapse of proton pairing correlations (see Sect.\ \ref{sect-pair}). 
The results of the calculations with APNP(LN) (open symbols in Fig.\ 
\ref{h40-sys}) show a somewhat different trend. With the exception 
of $^{198}$Po, the calculated $Q_{40}$ values stay nearly constant 
or smoothly increase with increasing rotational frequency up to 
$\Omega_x \approx 0.35$ MeV. Above this frequency
they decrease with increasing $\Omega_x$. In $^{198}$Po, the 
$Q_{40}$ values increase with increasing $\Omega_x$ in the whole
calculated frequency range (Fig.\ \ref{h40-sys}c). This increase 
is especially pronounced in the band crossing region. Within the  
isotopic chain the increase of neutron number $N$ leads to the 
decrease of $Q_{40}$. The $Q_{40}$ values increase within isotonic 
chain with increasing proton number $Z$. The results of the
calculations with NL3 and NLSH lead to smaller values of $Q_{40}$ 
compared with the ones obtained with NL1 force (Fig.\
\ref{h40-sys}a,b). The features discussed above are very similar 
to the ones obtained for the transition quadrupole moment $Q_t$, 
see the discussion in previous sections.

%%%%%%%%%%%%%%%%%%%%%%%%%%%%%%%%%%%%%%%%%%%%%%%%%%%%%%%%%%%%%%%%%%
\subsection{Particle number fluctuation $\langle(\hat{\Delta N})^2\rangle$}
%%%%%%%%%%%%%%%%%%%%%%%%%%%%%%%%%%%%%%%%%%%%%%%%%%%%%%%%%%%%%%%%%%

\begin{figure}
\epsfxsize 16.0cm
\epsfbox{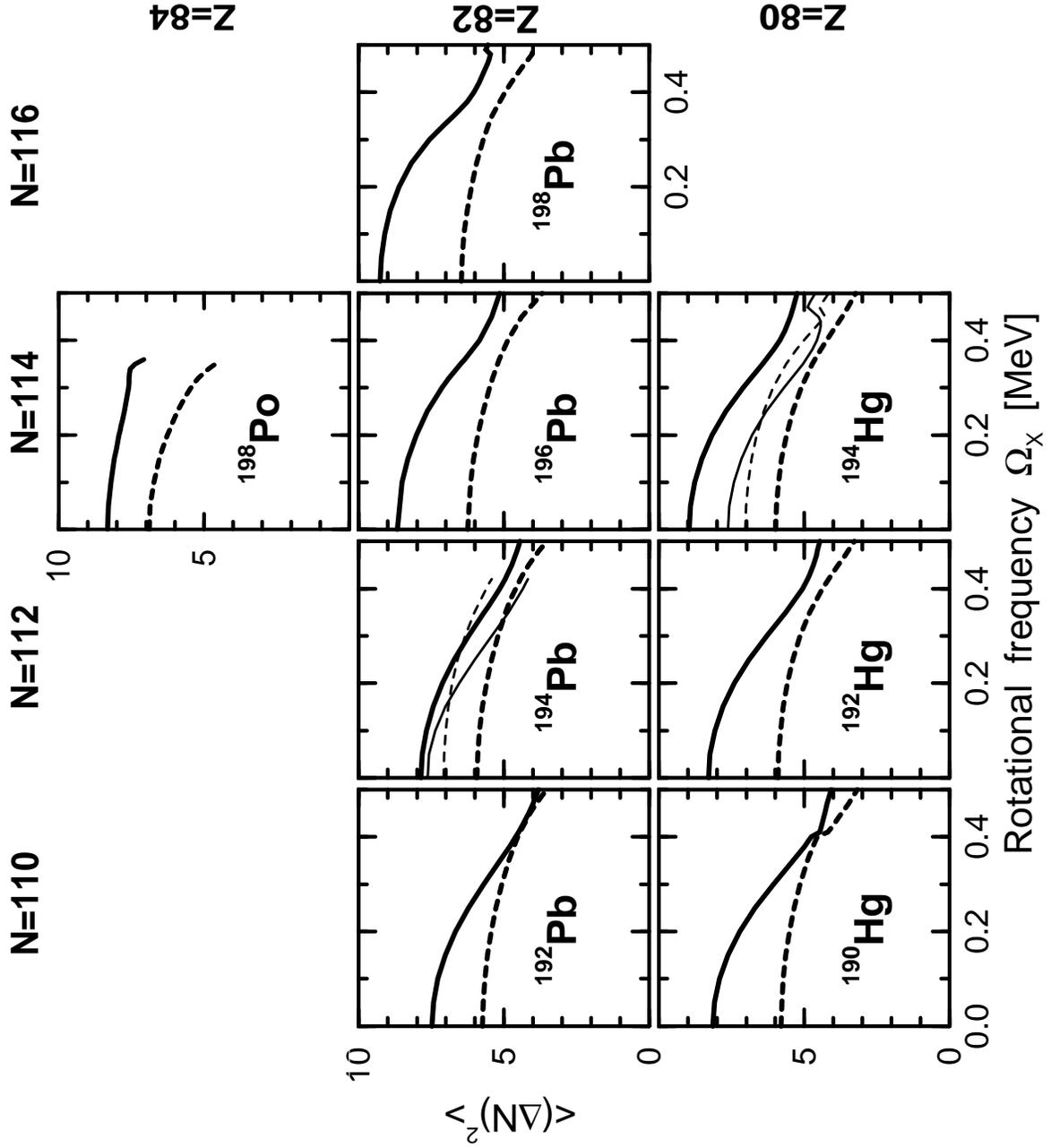}
\caption{Particle number fluctuations $\langle (\hat{N})^2 \rangle$ 
as a function of rotational frequency. Solid and dashed lines are 
used for neutrons and protons, respectively. Thick and thin lines 
are used for the values obtained with the forces NL1 and NL3, 
respectively.}
\label{fig-dn2}
\end{figure}

   The basis assumption behind the Kamlah expansion to second order 
\cite{Ka68} used in the derivation of the Lipkin-Nogami method is that 
the system is well pair-correlated which means that the particle 
number fluctuation in the unprojected wave function $\langle (\Delta
\hat{N})^2\rangle$ is large. These quantities obtained 
in the CRHB+LN calculations with the NL1 and NL3 forces for the RMF 
Lagrangian and the D1S set for the Gogny force are shown in Fig.\ 
\ref{fig-dn2} for all nuclei studied in the present manuscript. The 
particle number fluctuations decrease more or less smoothly with 
increasing rotational frequency indicating the quenching of pairing 
correlations due to the Coriolis antipairing effect. One should note that 
even at the highest rotational frequencies these fluctuations remain 
reasonably 
large thus indicating that the approximate particle number
projection by means of the Lipkin-Nogami method still remains
within the applicability range of the Kamlah expansion. The 
values of $\langle (\Delta \hat{N})^2 \rangle$ in neutron and 
proton subsystems correlate with the pairing energies $E_{pairing}$ calculated 
in these subsystems. For example, in the calculations with
the NL1 force the particle number fluctuations are larger
for neutrons than for protons (see Fig.\ \ref{fig-dn2})
which correlates with the fact that the absolute values of
pairing energies are larger for neutrons (see Figs.\ 
\ref{sys-pair}, \ref{pb98-qt-epair}b and \ref{fig-po98}b).
The situation is somewhat different in the calculations
with the NL3 force, where at medium and high rotational frequencies
the proton subsystem is more pair-correlated than the
neutron one as reflected in the particle number fluctuations
(Fig.\ \ref{fig-dn2}) and pairing energies (Fig.\ \ref{qtep-nl3}a,b).

%%%%%%%%%%%%%%%%%%%%%%%%%%%%%%%%%%%%%%%%%%%%%%%%%%%%%%
\section{Conclusions}
\label{conclusions}
%%%%%%%%%%%%%%%%%%%%%%%%%%%%%%%%%%%%%%%%%%%%%%%%%%%%%%

  The formalism of the Cranked Relativistic Hartree-Bogoliubov
theory with and without approximate particle number projection
before variation by means of the Lipkin-Nogami method is 
presented in detail. The relativistic mean field theory is
used in the particle-hole channel of this theory, while
a non-relativistic finite range two-body force of Gogny 
type is employed in the particle-particle (pairing) 
channel. Considering that the pairing is a genuine 
non-relativistic effect which plays a role only in 
the vicinity of the Fermi surface, the use of the best 
non-relativistic force in the pairing channel seems
well justified. 
 
  Its applicability to the description of rotating
nuclei and the main features of this theory have been 
studied on the example of the yrast superdeformed bands 
observed in even-even nuclei of the $A\sim 190$ 
mass region. The main conclusions emerging from this 
study are the following:

{\bf (i)} The calculations without particle number 
projection do provide only a poor description of 
experimental rotational features such as the kinematic 
$J^{(1)}$ and the dynamic $J^{(2)}$ 
moments of inertia. The calculated kinematic moments of 
inertia are larger than the experimental values. The same 
is also true for the dynamic moments of inertia at low and 
medium rotational frequencies. The calculations without
particle number projection lead to a unphysical collapse
of pairing correlations as has been seen in the proton
subsystem of $^{192}$Hg and $^{194}$Pb. As was shown by
subsequent calculations with particle number projection
these problems are related to the poor treatment of 
pairing correlations.
  
{\bf (ii)} Approximate particle number projection by means 
of the Lipkin-Nogami method considerably improves an agreement 
with experiment. The correlations induced by the Lipkin-Nogami 
method produce the desired effects, namely, {\bf (a)} to 
increase the strength
of the pairing correlations, {\bf (b)} to diminish the 
kinematic moments of inertia at all frequencies, {\bf (c)} 
to diminish the dynamic moments of inertia at low and medium 
rotational frequencies and {\bf (d)} to delay proton and 
neutron alignments to higher frequencies. In addition, 
there is no collapse of pairing correlations in the whole 
rotational frequency range under investigation. Systematic
calculations with the NL1 force for the RMF Lagrangian and 
the D1S set for the Gogny force have been performed
for all even-even nuclei in which SD bands have been 
observed so far. With an exception of $^{198}$Pb,
an excellent description of the rotational properties of 
yrast SD bands such as the dynamic and kinematic moments 
of inertia is obtained in a way free from adjustable 
parameters. It was concluded that the investigation of the 
structure of SD bands in neighboring odd nuclei is needed 
in order to better understand the problems seen in this 
nucleus. 
 
{\bf (iii)} It has been investigated how much 
the results of calculations
with approximate particle number projection by means of the 
Lipkin-Nogami method depend
on the parametrization of the RMF Lagrangian 
and the Gogny force. It was found that 
the combination of the NL1 set for the RMF Lagrangian and the 
D1S set for the Gogny force produce very good agreement with 
experimental rotational properties. The D1 and D1P sets of 
the parameters for the Gogny force produce too strong 
pairing correlations and thus fail to describe properly the 
rotational properties of SD bands.
An unexpected result of the present investigation is the fact that 
the NL3 force, which is believed to be the best RMF force for 
the description of nuclear properties far from beta-stability 
region, provides a less accurate description of the rotational 
properties of SD bands in the $A\sim 190$ mass region compared 
with the force NL1. This is most likely related to the single-particle 
spectra produced by this force in the SD minimum. It is possible 
that the improvement of the description of the isospin properties 
far from beta-stability region obtained in the NL3 force as compared 
with the NL1 force is reached at the cost of some worsening of the 
description of the single-particle spectra close to beta-stability
region.

{\bf (iv)} The dependence of the results on 
the strength of the Gogny force in the pairing channel has been 
studied in the calculations with approximate particle number 
projection by means of the Lipkin-Nogami 
method. It was found that a change of the strength by $\pm10$\% 
has significant impact on the rotational properties of SD bands such
as the kinematic and dynamic moment of inertia. Contrary to previous 
investigations within 
the Relativistic Hartree-Bogoliubov theory at no 
rotation and with no particle number projection, 
where the strength of the Gogny force has been 
increased by factor 1.15, here a very good 
description of rotational properties has been obtained 
with no modification of the strength of the Gogny force.
 
{\bf (v)} The results of the calculations with particle number
projection indicate the general trend of a decrease of the 
average transition quadrupole ($Q_t$) and mass hexadecupole 
$(Q_{40}$) moments in the isotopic chain with increasing neutron 
number $N$ and an increase of these quantities within the isotonic 
chain with increasing proton number $Z$.

{\bf (vi)} Neutron and proton pairing energies in all the calculated 
nuclei decrease with increasing rotational frequency reflecting 
the quenching of pairing correlations due to the Coriolis anti-pairing 
effect. The pairing energies $E_{pairing}$ calculated in the 
NL1+D1S+LN scheme do not show considerable variations as a function 
of $Z$ and $N$: the maximum difference between pairing energies 
calculated in two different nuclei is around 16\%. A smooth increase 
of absolute values of pairing energies with increasing $N$ is 
observed in the isotopic Pb and Hg chains which correlates with
the decrease of the transition quadrupole moments $Q_t$.
The size of the pairing energies strongly depends on the
parametrization of the Gogny force and the RMF Lagrangian, as well 
as on the strength of the Gogny force.

{\bf (vii)} The difference between the experimental dynamic moments 
of inertia of two bands and the effective alignment $i_{eff}$ between
these bands is reasonably well reproduced in most of the cases.
Further investigation of neighboring odd nuclei is needed in order
to find the origin of the remaining discrepancies.

  Finally, the present work should be considered as one of the first 
steps in the investigation of rotating nuclei in the pairing regime 
within the framework of the RMF theory. Different tasks 
definitely lie ahead. For example, an investigation of the rotational 
bands based on one- and multi-quasiparticle configurations in the 
$A\sim 190$ mass region is mandatory in order to see how the 
present theory can reproduce the effects connected with the blocking 
of one or several quasiparticle orbitals. One can expect that the 
Lipkin-Nogami method is a reasonably good approximation to the exact 
particle number projection in the regimes of strong pairing
correlations as it holds in the case of the $A\sim 190$ mass region.
It remains to investigate if this method is also a good approximation 
in the regimes of weak pairing correlations typical at high 
rotational frequencies in SD bands of other regions of the periodic 
table, such as the $A\sim 60$, $A\sim 130$ and $A\sim 150$ mass 
regions.

\section{Acknowledgments}

    A.V.A. acknowledges support from the Alexander von
Humboldt Foundation. This work is also supported in part
by the Bundesministerium f{\"u}r Bildung und Forschung
under the project 06 TM 875.

\newpage
%------------------------------------------------------------
%-------------------------------------------------------------

%------------------------------------------------------------
%-------------------------------------------------------------

%------------------------------------------------------------
%-------------------------------------------------------------

%------------------------------------------------------------
%-------------------------------------------------------------

%------------------------------------------------------------
%-------------------------------------------------------------

%------------------------------------------------------------
%-------------------------------------------------------------

\newpage
%------------------------------------------------------------
%-------------------------------------------------------------

%------------------------------------------------------------
%-------------------------------------------------------------

%------------------------------------------------------------
%-------------------------------------------------------------

%------------------------------------------------------------
%-------------------------------------------------------------

%------------------------------------------------------------
%-------------------------------------------------------------

%------------------------------------------------------------
%-------------------------------------------------------------

%------------------------------------------------------------
%-------------------------------------------------------------

%------------------------------------------------------------
%-------------------------------------------------------------

%------------------------------------------------------------
%--------------------------------------------------------------

%------------------------------------------------------------
%-------------------------------------------------------------

%------------------------------------------------------------
%-------------------------------------------------------------

%------------------------------------------------------------
%-------------------------------------------------------------

%------------------------------------------------------------
%-------------------------------------------------------------

%------------------------------------------------------------
%-------------------------------------------------------------

%------------------------------------------------------------
%-------------------------------------------------------------

%------------------------------------------------------------
%-------------------------------------------------------------

%------------------------------------------------------------
%-------------------------------------------------------------

%------------------------------------------------------------
%-------------------------------------------------------------

\newpage
%%%%%%%%%%%%%%%%%%%%%%%%%%%%%%%%%%%%%%%%%%%%%%%%%%%%%%%%%%%%%%%%%%%%
\begin{table}[h]
\caption{The values of the parameters of different
sets for the finite range Gogny forces.}
\vspace{0.5cm}
\begin{tabular}{|c|c|c|c|c|c|c|}
Parameter       &  \multicolumn{2}{c|}{  D1S  (Ref.\ \protect\cite{D1S})} & 
                    \multicolumn{2}{c|}{  D1P (Ref.\ \protect\cite{D1P})}  & 
                     \multicolumn{2}{c|}{  D1 (Ref.\ \protect\cite{D1}) } \\ \hline  
$\mu_i$ (fm)    & $0.7$     &    1.2  &    0.9  &   1.44 &    0.7  &   1.2   \\  
$W_i$           & $-1720.30$ &  103.64 & $-372.89$ &  34.62 & $-402.4$  & $-21.30$   \\  
$B_i$           &  1300.00 & $-163.48$ &   62.69 & $-14.08$ & $-100.0$  & $-11.77$   \\  
$H_i$           & $-1813.53$ &  162.81 & $-464.51$ &  70.95 & $-496.2$
                    &  37.27    \\  
$M_i$           &  1397.60 & $-223.93$ & $-31.49$ & $-20.96$ &
$-23.56$ & $-68.81$  \\  
\end{tabular}
\label{table-Gog}
\end{table}
%%%%%%%%%%%%%%%%%%%%%%%%%%%%%%%%%%%%%%%%%%%%%%%%%%%%%%%%%%%%%%%%%%%%%%%

%%%%%%%%%%%%%%%%%%%%%%%%%%%%%%%%%%%%%%%%%%%%%%%%%%%%%%%%%%%%%%%%%%%%%%%
\begin{table}[h]
\caption{ \sf The non-linear parameter sets NL1, NLSH and
NL3. The masses are given in MeV, the parameter $g_2$ in
$fm^{-1}$, while the rest of the parameters are
dimensionless. The nuclear matter properties, predicted
with these effectives forces, namely, the baryon density
$\rho_0$ (in units $fm^{-3}$), the binding energy per
particle $E/A$ (in MeV), the incompressibility $K$ (in MeV), the
effective mass $m^*/m$ and the asymmetry parameter $J$ (in
MeV) are also shown.}
\vspace{0.5cm}   
\begin{center}
\begin{tabular}{|c|c|c|c|} 
Parameter       &   NL1      &  NLSH     & NL3  \\ \hline
\multicolumn{4}{|c|}{\bf Masses} \\ \hline
$m_N$           & 938.0      & 939.0     & 939.0    \\
$m_{\sigma}$    & 492.25     & 526.059   & 508.194  \\
$m_{\omega}$    & 795.36    & 783.0     & 782.501  \\
$m_{\rho}$      & 763.0      & 763.0     & 763.0    \\ \hline
\multicolumn{4}{|c|}{\bf Coupling constants}        \\ \hline
$g_{\sigma}$     & 10.138     & 10.4444    & 10.217   \\
$g_2$           & $-12.172$    & $-6.9099$   & $-10.431$  \\
$g_3$           & $-36.265$    & $-15.8337$  & $-28.885$  \\
$g_{\omega}$     & 13.285     & 12.945    & 12.868   \\
$g_{\rho}$       & 4.976     & 4.383     & 4.474    \\ \hline
\multicolumn{4}{|c|}{\bf Nuclear matter properties } \\ \hline
$\rho_0$        & 0.153      & 0.146     & 0.148   \\
$E/A$           & $-16.488$    & $-16.346$   & $-16.299$   \\
$K$             & 211.29      & 355.36     & 271.76   \\
$m^*/m$         & 0.57      & 0.60     & 0.60     \\
$J$             & 43.7       & 36.1      & 37.4     \\
\end{tabular}
\end{center}
\label{RMFsets}
\end{table}
%%%%%%%%%%%%%%%%%%%%%%%%%%%%%%%%%%%%%%%%%%%%%%%%%%%%%%%%%%%%%%%%%%%

%%%%%%%%%%%%%%%%%%%%%%%%%%%%%%%%%%%%%%%%%%%%%%%%%%%%%%%%%%%%%%%%%%%
\begin{table}[h]
\caption{The spin assignment for the lowest state $I_0$ of 
the unlinked or tentatively linked [$^{192}$Pb] yrast SD bands. 
The lowest transition energies 
$E_{\gamma}(I_0+2\rightarrow I_0)$ are also shown.}
\vspace{0.5cm}
\begin{center}
\begin{tabular}{|c|c|c|c|} 
Nucleus      & Ref. & $E_{\gamma}(I_0+2 \rightarrow I_0)$ [keV] & $I_0$ \\ \hline
$^{190}$Hg   & \protect\cite{Hg190}  &  316.9 & $12^+$ \\
$^{192}$Hg   & \protect\cite{Hg192b} &  214.4 & $8^+$  \\
$^{192}$Pb   & \protect\cite{Pb192c} &  214.8 & $8^+$  \\
$^{196}$Pb   & \protect\cite{Pb196a} &  171.5 & $6^+$  \\
$^{198}$Pb   & \protect\cite{Pb198}  &  305.1 & $12^+$ \\
$^{198}$Po   & \protect\cite{Po198}  &  175.9 & $6^+$  \\ 
\end{tabular}
\end{center}
\label{table-I0}
\end{table}
%%%%%%%%%%%%%%%%%%%%%%%%%%%%%%%%%%%%%%%%%%%%%%%%%%%%%%%%%%%%%%%%%%%%

\end{document}